\documentclass[aps,prb,preprint,superscriptaddress]{revtex4}

\usepackage{graphicx}
\usepackage{amssymb,amsmath}
\usepackage{bm}

\begin{document}



\title{Topology of Andreev Bound States with Flat Dispersion} 


\author{Masatoshi Sato}
\affiliation{The Institute for Solid State Physics, The University of
Tokyo, Chiba, 277-8581, Japan}
\author{Yukio Tanaka}
\affiliation{
Department of Applied Physics, Nagoya University, Nagoya, 464-8603,
Japan
}
\author{Keiji Yada}
\affiliation{
Venture Business Laboratory, Nagoya University, Nagoya, 464-8603,
Japan
}
\author{Takehito Yokoyama}
\affiliation{
Department of Physics, Tokyo Institute of Technology, Tokyo, 152-8551,
Japan
}



\date{\today}

\begin{abstract}
A theory of dispersionless Andreev bound states on
surfaces of time-reversal invariant unconventional superconductors is presented.
The generalized criterion for the dispersionless Andreev bound state is
derived from the bulk-edge correspondence, and the chiral spin structure of
the dispersionless Andreev bound states is argued from which the Andreev
 bound state is stabilized.
Then we summarize the criterion in a form of index theorems.
The index theorems are proved in a general framework to certify the
 bulk-edge correspondence.
As concrete examples, we discuss (i) $d_{xy}$-wave superconductor (ii)
 $p_x$-wave superconductor, and (iii)
 noncentrosymmetric superconductors. 
In the last example, we find a peculiar time-reversal invariant Majorana
 fermion.
The time-reversal invariant Majorana fermion shows an
unusual response to the Zeeman magnetic field, which can be used to
 identify it experimentally.
\end{abstract}

\pacs{}


\maketitle

\section{Introduction}
In unconventional superconductors, 
due to the sign change of the pair potential on the Fermi surface 
it is known that an Andreev bound state (ABS)
via Andreev reflection \cite{Andreev64,BTK82,Bruder90} 
is generated at a surface of superconductors
\cite{BZ81,HN86,Hu94,KT00,LSW01,ATK04}.
The ABS shows up as a zero bias 
conductance peak of tunneling spectroscopy \cite{Hu94,TK95,KT00,LSW01}, 
and through the study of tunneling spectroscopy of 
high $T_{c}$ cuprates \cite{KTKTK95,KTTKUATTK98,CAPGXZM97,ATKTITKK97,WYGS98,IWYTK00,BFQSBG02,CSDSKKT05,CDDHKKASH06,CSH08}, it has been established that 
a dispersionless zero energy ABS \cite{KT00} is generated 
for $d_{xy}$-wave superconductors.  
The existence of the ABS influences seriously 
interface and surface properties 
\cite{MS95,Nagato95,TKPRB96,Tanuma98,TanumaJ99,Tanuma99}.   
Great amount of  unconventional quantum phenomena  
appear in  
quasiparticle tunneling \cite{TK95,Barash97,Tanuma01},
Josephson effect \cite{TK96,TK97,BBR96,Yakovenko,TK99,TK00}, 
spin transport phenomena \cite{KTYB99,HTYAIK03},  
Meissner effect \cite{Higashitani97,BKK00,WPSKAHBRS98,TAGK05} 
and macroscopic quantum tunneling  \cite{Kawabata04,Kawabata05,Yokoyama07}.  \par

Dispersionless ABSs appear not only in spin-singlet $d_{xy}$-wave
superconductors, but also in spin-triplet $p_x$-wave ones
\cite{BZ81,HN86,KT00}.
However, it has been shown that 
the proximity effect of the ABS
is completely different in these
superconductors \cite{TNS03,TNGK04,TK04,TKY05,ATK06}.
In the latter case, the ABS can penetrate into a diffusive normal 
metal (DN) attached to the superconductor, while it is not possible 
in the former one. 
The underlying physics is the existence of odd-frequency
pairings \cite{TG07}:
Odd-frequency pairings are induced on boundaries of
superconductors due to the breakdown of the  translational
invariance \cite{ELCCS07}.
From 
the requirement of the Fermi statistics,
the ABS in $d_{xy}$-wave superconductors
is expressed by an odd-frequency spin-singlet odd-parity
state, but the ABS in $p_{x}$-wave superconductors
an odd-frequency spin-triplet even-parity one \cite{TGKU07,TTG07}. 
Then the ABS in $p_x$-wave superconductors can penetrate
in DN since the odd-frequency spin-triplet $s$-wave component in
$p_x$-wave superconductors is robust against
impurity scattering
\cite{TG07,TGKU07,TAGK05,ATGK07,Higashitani09}.

Another important character of ABSs is that they realize Majorana fermions
in condensed matter systems.
For time-reversal breaking superconductors, ABSs with linear
dispersion have been studied
in the context of the chiral $p$-wave superconductor such as
Sr$_{2}$RuO$_{4}$ \cite{MHYNFBL94,MS99,HS98,YTK97}. 
In this case, the ABSs are an analogous state to 
the edge modes of quantum Hall system (QHS) \cite{Volovik97,RG00},
which induce a spontaneous charge current along the edge \cite{GI98,FMS01}. 
The main difference between the edge state of QHS and the ABS is that the
latter excitation is a Majorana fermion realizing a non-Abelian anyon,
which can be used in a fault tolerant topological quantum
computation \cite{Kitaev03, FKLW03}.
In addition to spin-triplet superconductors \cite{RG00,
Ivanov01,SF09, Sato10}, now it
has been known that spin-singlet superconducting states may support
a non-Abelian anyon \cite{Sato03,FK08,STF09,STF10,SLTD10,Alicea10,SF10}, 
and the various unusual transport properties have been explored
\cite{RG00,Ivanov01,FK08,FK09,NAB08,ANB09,STF09,STF10,TYN09,LLN09,LTYSN10a,LTYSN10b,SF10,Alicea10,SLTD10,SRM10,Flensberg10,SFN10,MZ10,NOF10}. 
For time-reversal invariant superconductors, 
the ABS with linear dispersion have been studied mainly in the context of
non-centrosymmetric (NCS) superconductors
\cite{BHMPSGSNSR04,TBNOTH04,NIZ05,HQC09,RTCKHRSKRJGMTM07}.
One of the remarkable features of the NCS superconductors is that the
superconducting pair potential becomes a
mixture of  spin-singlet even-parity and spin-triplet odd-parity
ones \cite{GR01},
and for CePt$_3$Si \cite{FAKS04},
the ABS and the relevant charge transport properties 
\cite{YTJ05,IHSYMTS07,EIT10,VVE08,TYBN09,Sato06,SF09,LY09,SBMT10}  
have been studied based on $s+p$-wave 
pairing.
The resulting ABS has two linear dispersions
analogous to helical edge modes\cite{SF09}, in Quantum spin Hall systems
(QSHS)\cite{KM05a,KM05b,BZ06,BHZ06},
thus instead of charge current, spin current is spontaneously generated 
along the edge 
The ABS forms a Kramers pair of Majorana modes called a
helical Majorana edge mode, and
several new features of  spin and charge transport
stemming from these helical Majorana edge modes 
have been predicted \cite{VVE08,TYBN09,Sato06,SF09,LY09,ATN10}. 

It has been found recently that Majorana fermions are possible also 
for ABSs with flat dispersion \cite{TMYYS10,YSTY10}.
Although $s+p$-wave superconducting states have been
studied intensively, 
other types of NCS pairing symmetry may appear
in strongly correlated systems. 
Microscopic calculations have shown that a parity mixed pairing
state between the spin-singlet $d_{x^{2}-y^{2}}$-wave 
pairing and the spin-triplet $f$-wave one is realized in the Hubbard model
near the half filling \cite{YOT07a,YOT07b, YOT08a, YOT08b}.
Moreover,
a gap function which consists 
of the spin-singlet $d_{xy}$-wave component and spin-triplet 
$p$-wave one has been proposed,\cite{YOTI09} as a possible
candidate of superconductivity generated at heterointerface
LaAlO$_3$/SrTiO$_3$
\cite{RTCKHRSKRJGMTM07}. 
Examining the ABSs in these NCS
superconductors,\cite{TMYYS10,YSTY10} we found a  
new type of ABS in the $d_{xy}+p$-wave NCS superconductor \cite{dxy}: 
Due to the Fermi surface splitting by the spin-orbit coupling,
there appears a single branch of Majorana edge state with flat dispersion 
preserving the time reversal symmetry. 
A similar ABS with flat band in NCS superconductors was also discussed
in Ref.\cite{SR10}.

In this paper, we will address topological properties of the
dispersionless ABSs on surfaces of time-reversal invariant superconductors.
For the ABSs with linear dispersion, it has been known that
their presence is
guaranteed by the topological invariance defined 
in bulk Hamiltonian.
For instance, for full
gapped two dimensional time-reversal breaking (invariant)
superconductors, a non-zero Chern
number\cite{TKNN82, Kohmoto85} (non-trivial ${\bm Z}_2$ topological number
\cite{KM05a,KM05b}) ensures the existence of the Majorana
fermions\cite{Hatsugai93,Volovik97,RG00,FK06,QHRZ09,Roy08,SRFL08,Sato09,FF10},
and for nodal superconductors, the parity of Chern number ensures the
existence of Majorana fermions \cite{SF10}.  
Furthermore, for spin-triplet superconductors, 
the ABSs on the boundary can be predicted from the topology of the Fermi
surface \cite{Sato09,Sato10,FB10}.
However, Majorana edge mode  with flat dispersion and its relevance 
to the symmetry of Hamiltonian 
has not been clarified yet. 
It is an interesting issue to clarify the relevance of the 
topological property to the wave function 
of ABS with flat dispersion.

The organization of this paper as follows.
In section \ref{sec:topological criterion}, we will define a
topological number starting from the bulk Bogoliubov de Gennes (BdG)
Hamiltonian, and present a topological criterion for dispersionless
ABSs.
In section \ref{sec:chirality}, we discuss topological stability of the
ABSs using the chiral symmetry of time-reversal invariant BdG Hamiltonians. 
Then, we propose index theorems for the dispersionless ABSs. 
In section \ref{sec:2x2}, the relation between the sign change of gap
function and topological criterion will be discussed. 
We will show that 
the sign change of the pair potential is directly relevant to 
the index theorem we proposed. 
As concrete examples, we will discuss the ABSs in spin-singlet
$d_{xy}$-wave superconductors and spin 
triplet $p_{x}$-wave ones, respectively. 
By constructing the wave function of the ABS explicitly, 
the index theorems are confirmed.
We also find that the topological structures of the ABSs are consistent
with those of the odd-frequency pairings.
In section \ref{sec:4x4}, we will apply our theory to the ABSs in
noncentrosymmetric superconductors, in which a Majorana fermion with
flat dispersion is realized.
The wave function of the ABS for $d_{xy}+p$-wave
superconductors will be derived, and the relevant topological
number will be discussed. 
It will be also found that the Majorana fermion has an anisotropic response to 
Zeeman magnetic field, which can be explained by  
a $Z_{2}$ topological number. 
In section \ref{sec:indextheorems}, we will finally provide a general
framework to certify the bulk-edge correspondence, and by using it the index
theorems will be proved. 
In section \ref{sec:conclusion}, we will
summarize our results, and discuss possible application of our theory to
other systems.
Throughout this paper, we use a convention of $\hbar=1$ except
in Sec.\ref{sec:indextheorems}.

\section{topological criterion for dispersionless Andreev bound state}
\label{sec:topological criterion}

We start with a general BdG Hamiltonian for
superconducting states. 
\begin{eqnarray}
{\cal H}=\frac{1}{2}\sum_{{\bm k}\alpha \alpha'}
\left(c_{{\bm k}\alpha}^{\dagger},c_{-{\bm k}\alpha}
\right)
{\cal H}({\bm k})
\left(
\begin{array}{c}
c_{{\bm k}\alpha'} \\
c_{-{\bm k}\alpha'}^{\dagger}
\end{array}
\right), 
\end{eqnarray} 
with 
\begin{eqnarray}
{\cal H}({\bm k})=
\left(
\begin{array}{cc}
\hat{\cal E}({\bm k})_{\alpha\alpha'} & \hat{\Delta}({\bm k})_{\alpha\alpha'} \\
\hat{\Delta}^{\dagger}({\bm k})_{\alpha\alpha'} &
-\hat{\cal E}^{T}(-{\bm k})_{\alpha\alpha'}
\end{array}
\right), 
\end{eqnarray}
where $c_{{\bm k}\alpha}^{\dagger}$ ($c_{{\bm k}\alpha}$) denotes the
creation (annihilation) operator of electron with momentum ${\bm k}$.
The suffix $\alpha$ labels other degrees of freedom for electron such as
spin, orbital degrees of freedom, sublattice indices, and so on. 
$\hat{\cal E}({\bm k})$ is the Hermitian matrix describing the normal
dispersion of electron, and $\hat{\Delta}({\bm k})$ the gap function
which satisfies $\hat{\Delta}^{T}(-{\bm k})=-\hat{\Delta}({\bm k})$.

In the following, we suppose time-reversal invariant superconducting
states.
The time-reversal operation for electron is given by the
antiunitary operator ${\cal T}$ in the form,
\begin{eqnarray}
{\cal T}=UK, 
\end{eqnarray}
where $U$ is a constant unitary matrix acting on the suffix $\alpha$ of
electron, and $K$ the complex conjugate operator.
$K$ also reverses the sign of the momentum ${\bm k}$.
$U$ satisfies $UU^{*}=-1$ since ${\cal T}^2=-1$ for electron. 
Then, for time-reversal invariant superconductors, the BdG Hamiltonian
${\cal H}({\bm k})$ satisfies,
\begin{eqnarray}
\Theta {\cal H}({\bm k})\Theta^{-1}={\cal H}^{*}(-{\bm k}),
\quad
\Theta=
\left(
\begin{array}{cc}
U_{\alpha\alpha'} & 0\\
0 & U^{*}_{\alpha\alpha'}
\end{array}
\right),
\label{eq:time-reversal}
\end{eqnarray}
which implies 
\begin{eqnarray}
U\hat{\cal E}({\bm k})U^{\dagger}=\hat{\cal E}^{*}(-{\bm k}),
\quad
U\hat{\Delta}({\bm k})U^{T}=\hat{\Delta}^{*}(-{\bm k}). 
\end{eqnarray}
In addition, the BdG Hamiltonian ${\cal H}({\bm k})$ has the following
particle-hole symmetry,
\begin{eqnarray}
{\cal C} {\cal H}({\bm k}){\cal C}^{-1}=-{\cal H}^{*}(-{\bm k}),
\quad
{\cal C}=\left(
\begin{array}{cc}
0 & \delta_{\alpha\alpha'}\\
\delta_{\alpha\alpha'} & 0
\end{array}
\right). 
\label{eq:particle-hole}
\end{eqnarray}
Therefore, combining with (\ref{eq:time-reversal}) and
(\ref{eq:particle-hole}),  we obtain the so-called chiral symmetry 
\begin{eqnarray}
\{\Gamma, {\cal H}({\bm k})\}=0,
\quad
\Gamma=-i{\cal C}\Theta=
\left(
\begin{array}{cc}
0 & -iU^{*}_{\alpha\alpha'}\\
-iU_{\alpha\alpha'} & 0
\end{array}
\right),
\label{eq:chiral}
\end{eqnarray} 
which is the central ingredient of our theory.
For later convenience, we perform here the following unitary transformation
by which $\Gamma$ is diagonalized 
\begin{eqnarray}
U_{\Gamma}^{\dagger}\Gamma U_{\Gamma}= 
\left(
\begin{array}{cc}
\delta_{\alpha\alpha'} & 0 \\
0 &-\delta_{\alpha\alpha'}
\end{array}
\right),
\quad
U_{\Gamma}=\frac{1}{\sqrt{2}}\left(
\begin{array}{cc}
\delta_{\alpha\alpha'} & iU^{\dagger}_{\alpha\alpha'}\\
-iU_{\alpha\alpha'} & -\delta_{\alpha\alpha'}
\end{array}
\right).
\label{eq:gammadiag}
\end{eqnarray}
Then the BdG Hamiltonian becomes off-diagonal,
\begin{eqnarray}
U_{\Gamma}^{\dagger}{\cal H}({\bm k})U_{\Gamma}
=\left(
\begin{array}{cc}
0 & \hat{q}({\bm k})\\
\hat{q}^{\dagger}({\bm k}) & 0
\end{array}
\right), 
\end{eqnarray}
with
\begin{eqnarray}
\hat{q}({\bm k})=i\hat{\cal E}({\bm k})U^{\dagger}-\hat{\Delta}({\bm k}).
\end{eqnarray}

\begin{figure}[h]
\begin{center}
\includegraphics[width=8cm]{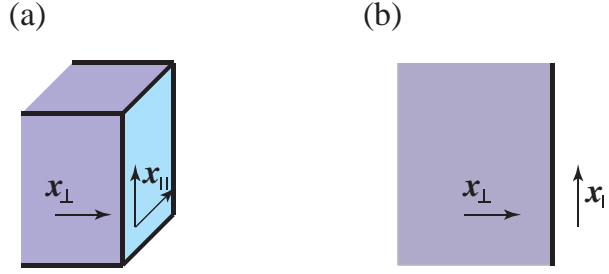}
\caption{(color online) Surfaces of superconductors. (a) two dimensional surface. (b) 1
 dimensional edge. 
The conjugate momenta of ${\bm x}_{\perp}$ and ${\bm x}_{\parallel}$
are ${\bm k}_{\perp}$ and ${\bm
 k}_{\parallel}$, respectively.}
\label{fig:surface}
\end{center}
\end{figure}

Now consider the ABS on a surface of the superconductor.
The relevant topological number for the flat dispersion (or
dispersionless) ABS is 
\begin{eqnarray}
W({\bm k}_{\parallel})&=&
-\frac{1}{4\pi i}\int d{\bm k}_{\perp}
{\rm tr}[\Gamma {\cal H}^{-1}({\bm k})
\partial_{{\bm k}_{\perp}}{\cal H}({\bm k})]
\nonumber\\
&=& 
\frac{1}{4\pi i}
\int d{\bm k}_{\perp}{\rm tr}[\hat{q}^{-1}({\bm k})
\partial_{{\bm k}_{\perp}} \hat{q}({\bm k})
-\hat{q}^{\dagger -1}({\bm k})
\partial_{{\bm k}_{\perp}}\hat{q}^{\dagger}({\bm k})]
\nonumber\\
&=&\frac{1}{2\pi}{\rm Im}\left[
\int d{\bm k}_{\perp}\partial_{{\bm k}_{\perp}}\ln
{\rm det}\hat{q}({\bm k})
\right],
\label{eq:windingnumber}
\end{eqnarray} 
where ${\bm k}_{\parallel}$ (${\bm k}_{\perp}$) is the momentum parallel
(perpendicular) to the surface we consider, and the integration is along
the ${\bm k}_{\perp}$-direction with fixing the value of ${\bm k}_{\parallel}$ 
(Fig.\ref{fig:surface}). 
For instance, for the surface perpendicular to the $x$ direction,  ${\bm
k}_{\perp}=k_x$ and ${\bm k}_{\parallel}=(k_y,k_z)$.

\begin{figure}[h]
\begin{center}
\includegraphics[width=8cm]{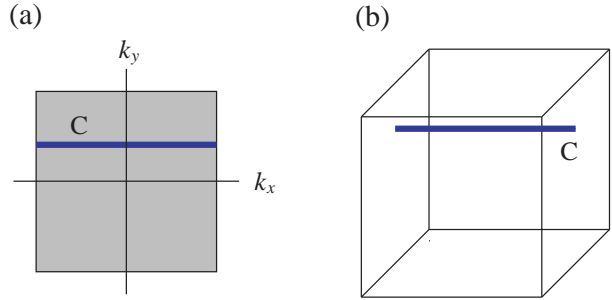}
\caption{(color online) The integral path $C$ in the first Brillouin zone. 
(a) two dimensional  case. (b) three dimensional case.  
On the path $C$, the momentum ${\bm k}$ has a fixed ${\bm
 k}_{\parallel}$.
}
\label{fig:loop}
\end{center}
\end{figure}

For superfluids or continuum models of superconductors, 
the line integral in (\ref{eq:windingnumber}) is
performed from ${\bm k}_{\perp} =-\infty$ to ${\bm k}_{\perp}=\infty$.
Far apart from the Fermi surfaces, we can regulate the gap function as
$\hat{\Delta}({\bm k})\rightarrow 0$ at ${\bm k}_{\perp}=\pm\infty$
and neglect off-diagonal terms in 
${\cal E}({\bm k})$ without changing the physics. 
With this
regularization, ${\rm det}\hat{q}({\bm
k})$ becomes identical at ${\bm k}_{\perp}=\pm \infty$. 
As a result, we can show that the topological number $W({\bm
k}_{\parallel})$ takes an integer value. 
On the other hand, for lattice models of superconductors, the line integral in
(\ref{eq:windingnumber}) should be performed in the noncontractable
closed loop $C$ in the Brillouin zone as illustrated in Fig.\ref{fig:loop}.
In this case, the periodicity of the Hamiltonian with respect to a
reciprocal vector ${\bm G}$,  ${\cal H}({\bm
k})={\cal H}({\bm k}+{\bm G})$, ensures the quantization of $W({\bm
k}_{\parallel})$.

From the bulk-edge correspondence, we obtain the following simple criterion for
the ABS \cite{Sato10b,TMYYS10,YSTY10,SR10}:
\begin{itemize}
 \item {\it When the topological number $W({\bm k}_{\parallel})$ takes a
       non-zero integer, 
a dispersionless ABS exists on the surface.}
\end{itemize}
We notice here that the resultant ABS has flat
dispersion: 
This is because the topological number $W({\bm
k}_{\parallel})$ is nonzero in a finite region of ${\bm
k}_{\parallel}$ since it cannot change unless the integration path
intersects a gap node.
From the above criterion, this implies that
the zero energy ABS
also exists in a finite region of ${\bm k}_{\parallel}$. In other words,  
the ABS obtained from the above criterion has flat
dispersion.
On the other hand, if the integration path intersects a gap node, ${\rm
det}\hat{q}({\bm k})$ becomes zero, thus a
discontinuous change of  $W({\bm k}_{\parallel})$ becomes possible.
(Note that at a gap node ${\bm k}_0$, ${\rm det}{\cal H}({\bm k}_0)\propto|{\rm
det}\hat{q}({\bm k}_0)|^2=0$.)
Therefore, we find that
\begin{itemize}
 \item {\it the flat dispersion ABSs are terminated in a
       gap node in the surface momentum ${\bm k}_{\parallel}$ space}.
\end{itemize}
These results are confirmed in concrete examples in Secs.\ref{sec:2x2}
and \ref{sec:4x4}, and proved eventually in Sec.\ref{sec:indextheorems}.

\section{Chirality and topological stability of Andreev bound state}
\label{sec:chirality}

In the previous section, we discussed the topological number constructed
from the bulk BdG Hamiltonian ${\cal H}({\bm k})$ which
ensures the existence of the dispersionless ABS.
In this section, we present an alternative
argument on the stability of the flat dispersion ABS
from a viewpoint of the boundary.
This argument reveals a peculiar spin structure of the dispersionless
ABSs, which we call chirality, from which the scattering between them is
much suppressed.

In order to argue the boundary state, we 
perform the Fourier transformation of the BdG Hamiltonian
${\cal H}({\bm k})$ with respect to ${\bm k}_{\perp}$, and denote   
the resultant BdG Hamiltonian as ${\cal H}({\bm x}_{\perp}, {\bm
k}_{\parallel})$, where ${\bm x}_{\perp}$ is the conjugate coordinate of
${\bm k}_{\perp}$.
Then consider the semi-infinite superconductor on ${\bm x}_{\perp}>0$
where the surface is located on ${\bm x}_{\perp}=0$.
The corresponding BdG equation is given by 
\begin{eqnarray}
{\cal H}({\bm x}_{\perp}, {\bm k}_{\parallel})
|u({\bm x}_{\perp},{\bm k}_{\parallel})\rangle
=E({\bm k}_{\parallel})
|u({\bm x}_{\perp},{\bm k}_{\parallel})\rangle, 
\label{eq:boundaryBdG}
\end{eqnarray}
with the boundary condition
\begin{eqnarray}
|u({\bm x}_{\perp},{\bm k}_{\parallel})\rangle=0,
\quad
\label{eq:boundarycond}
\end{eqnarray}
at ${\bm x}_{\perp}=0$.
The zero energy ABS satisfies (\ref{eq:boundaryBdG})
and (\ref{eq:boundarycond}) with $E({\bm k}_{\parallel})=0$.

In our argument, it is convenient to use the following equation
\begin{eqnarray}
{\cal H}^2({\bm x}_{\perp},{\bm k}_{\parallel})|v({\bm x}_{\perp},{\bm
 k}_{\parallel})\rangle=
E^2({\bm k}_{\parallel})|v({\bm x}_{\perp},{\bm k}_{\parallel})\rangle.
\label{eq:BdGsquare}
\end{eqnarray}
instead of the original BdG equation (\ref{eq:boundaryBdG}).
As is shown in Appendix \ref{sec:appendixa}, it is 
found that
there exists one to one correspondence between the solutions of the BdG equation
(\ref{eq:boundaryBdG}) and those of (\ref{eq:BdGsquare}).
In particular, the zero energy state 
$|u_0({\bm x}_{\perp},{\bm k}_{\parallel})\rangle$
of (\ref{eq:boundaryBdG}) is exactly the same as the state
$|v_0({\bm x}_{\perp}, {\bm k}_{\parallel})\rangle$ of
(\ref{eq:BdGsquare}) with $E^2({\bm k}_{\parallel})=0$. 
Therefore, we consider (\ref{eq:BdGsquare}) instead of (\ref{eq:boundaryBdG})
in the following.

\begin{figure}[h]
\begin{center}
\includegraphics[width=8cm]{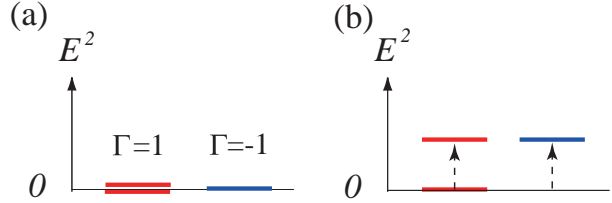}
\caption{(color online) Zero energy states of ${\cal H}^2({\bm x}_{\perp}, {\bm
 k}_{\parallel})$. (a) $N_0^{(+)}=2$ and $N_0^{(-)}=1$. (b)
 $N_0^{(+)}=1$ and $N_0^{(-)}=0$. By perturbation, some of the zero
 modes in (a) might become non-zero modes as in (b), however,
 $N_0^{(+)}$ $- N_0^{(-)}$ does not change.} 
\label{fig:chirality}
\end{center}
\end{figure}

The starting point of our argument is the chiral symmetry
(\ref{eq:chiral}) of the
BdG Hamiltonian. From (\ref{eq:chiral}), we have 
\begin{eqnarray}
\{{\cal H}({\bm x}_{\perp},{\bm k}_{\parallel}), \Gamma \}=0, 
\label{eq:boundarychiral}
\end{eqnarray}
thus, ${\cal H}^2({\bm x}_{\perp},{\bm k}_{\parallel})$
and $\Gamma$ commute with each other,
\begin{eqnarray}
[{\cal H}^2({\bm x}_{\perp},{\bm k}_{\parallel}), \Gamma]=0.
\end{eqnarray}
Therefore, the solution $|v({\bm x}_{\perp},{\bm k}_{\parallel})\rangle$
of (\ref{eq:BdGsquare}) can be an eigenstate of $\Gamma$ at the same time.
Then we denote  the solution with the eigenvalue $\Gamma=\pm 1$ as
$|v^{(\pm)}({\bm x}_{\perp},{\bm k}_{\parallel})\rangle$. 
From the argument in Appendix \ref{sec:appendixa}, we find the following
properties of $|v^{(\pm)}({\bm x}_{\perp},{\bm k}_{\parallel})\rangle$. 
\begin{enumerate}
\item For a solution with nonzero $E^2({\bm k}_{\parallel})\neq 0$, 
the state 
$|v^{(+)}({\bm x}_{\perp},{\bm k}_{\parallel})\rangle$ is always paired
      with $|v^{(-)}({\bm x}_{\perp},{\bm k}_{\parallel})\rangle$ . 
In other words, for nonzero energy solutions, 
the number of the $\Gamma=1$ states is equal to that of the $\Gamma=-1$ states.
\item On the other hand, for zero energy solutions, the number of the
      $\Gamma=1$ states is not always the same as that of the
      $\Gamma=-1$ states. 
\end{enumerate}
If we denote the number of the zero energy state with $\Gamma=\pm 1$ as
$N_0^{(\pm)}$, 
these properties imply that $N_0^{(+)}-N_0^{(-)}$ does not change its value
against perturbation preserving the chiral symmetry.
As illustrated in Fig.\ref{fig:chirality},
some of the zero energy states might acquire non-zero energy by the
perturbation, however, they always form a pair with
opposite chirality $\Gamma$. So the difference $N_0^{(+)}-N_0^{(-)}$ does not
change as a result.
This result implies that 
the existence of the zero
energy state is robust against the perturbation once $N_0^{(+)}-N_0^{(-)}$
becomes nonzero.

As was shown in the previous section, 
from the bulk-edge correspondence, we find that
the nonzero
$W({\bm k}_{\parallel})$ implies the existence of the
zero energy ABS. 
At the same time, in this section, we find that the nonzero
$N_0^{(+)}-N_0^{(-)}$ also ensures the
robustness of the existence of the zero energy state.   
Therefore, it is naturally conjectured that these two quantities,
$W({\bm k}_{\parallel})$ and $N_0^{(+)}-N_0^{(-)}$, should be equaled.
Since there is a sign ambiguity to equal these two quantities,  we
have two possible equations in a form of index theorems,
\begin{eqnarray}
W({\bm k}_{\parallel})=N^{(-)}_0-N^{(+)}_0,
\label{eq:index}
\end{eqnarray}
or
\begin{eqnarray}
W({\bm k}_{\parallel})=N^{(+)}_0-N^{(-)}_0.
\label{eq:index2}
\end{eqnarray}
In the following examples, 
we will show that the conjecture of the index theorem (\ref{eq:index}) or
(\ref{eq:index2}) indeed holds. 
Then they will be proved eventually in Sec.\ref{sec:indextheorems}.
Here note that there are two possible choices of the surfaces of the
superconductor which is
perpendicular to ${\bm x}_{\perp}$, i.e. the surface of the
semi-infinite superconductor on ${\bm x}_{\perp}>0$ or that on ${\bm
x}_{\perp}<0$. 
It will be found that these two possible choices of the surface exactly
correspond to the two possible equalities (\ref{eq:index}) and (\ref{eq:index2}).

\section{sign change of gap function and topological criterion in
 superconductor preserving $S_z$ \cite{Sato10b}}
\label{sec:2x2}

We first consider the simplest case where the Cooper pair preserves
spin in a certain direction, say $S_z$.
As is shown below, the gap function consists of a
real single component, and  
the BdG Hamiltonian reduces to a $2\times 2$
matrix in this case.
Since the particle-hole symmetry (\ref{eq:particle-hole}) and the
time-reversal invariance (\ref{eq:time-reversal}) have different forms in the
$2\times 2$ BdG Hamiltonian, it needs a special care to consider this case.
In particular, the particle-hole symmetry (\ref{eq:particle-hole}) is
not manifest in the $2\times 2$ BdG Hamiltonian. Nevertheless,  we find a
chiral symmetry in the $2\times 2$ BdG Hamiltonian in the following, 
and using it, we will present topological criteria similar to
(\ref{eq:index}) and (\ref{eq:index2}).

In this case, it also has been known that a sign change of the gap function
implies the existence of the zero energy ABS.\cite{KT00,Hu94}
Below, we will show that our topological criterion reproduces this result
if the Fermi surface has a simple sphere-like shape.
In addition, we consider a general Fermi surface,
where these two criteria are not coincident with each other.
It will be shown that our topological
criterion excellently agrees with the details of the zero energy
ABS, while the previous one is not.

Let us consider a time-reversal invariant superconductor described by a
single-band electron.
We also assume that the spin component of a certain direction is
preserved. 
Without loss of generality, we can select the preserved spin axis as
$z$, and the gap function is given by
\begin{eqnarray}
\hat{\Delta}({\bm k})=\left\{
\begin{array}{ll}
i\psi({\bm k})\sigma_y, & \mbox{for spin-singlet superconductor}\\ 
id_z({\bm k})\sigma_z\sigma_y, &
\mbox{for spin-triplet superconductor}
\end{array}
\right.,
\end{eqnarray}
where $\psi({\bm k})$ and $d_z({\bm k})$ are real functions.
Under this assumption, the Hamiltonian ${\cal H}$ reduces to
\begin{eqnarray}
{\cal H}= \sum_{{\bm k}}\left(c^{\dagger}_{{\bm k}\uparrow}, 
c_{-{\bm k}\downarrow}\right)
{\cal H}_{2\times 2}({\bm k})
\left(
\begin{array}{c}
c_{{\bm k}\uparrow} \\
c^{\dagger}_{-{\bm k}\downarrow}
\end{array}
\right),
\end{eqnarray}
with
\begin{eqnarray}
{\cal H}_{2\times 2}({\bm k}) 
=\left(
\begin{array}{cc}
\varepsilon({\bm k}) & \Delta({\bm k})\\
\Delta({\bm k}) & -\varepsilon({\bm k})
\end{array}
\right).
\label{eq:BdG2x2}
\end{eqnarray}
Here $\Delta({\bm k})$ is given by
\begin{eqnarray}
\Delta({\bm k})=
\left\{
\begin{array}{ll}
\psi({\bm k})& \mbox{for spin-singlet superconductor}\\
d_z({\bm k}) &\mbox{for spin-triplet superconductor}
\end{array}
\right. .
\end{eqnarray}
Note that the BdG Hamiltonian is now reduced to the $2\times 2$ matrix
(\ref{eq:BdG2x2}).

As has been mentioned above, the particle-hole symmetry
(\ref{eq:particle-hole}) is not manifest in (\ref{eq:BdG2x2}). 
However, we find the reduced Hamiltonian
(\ref{eq:BdG2x2}) has the
following chiral symmetry, 
\begin{eqnarray}
\left\{\gamma, {\cal H}_{2\times 2}({\bm k}) \right\}=0,
\quad \gamma=\sigma_y.
\label{eq:chiral2x2}
\end{eqnarray}
Indeed, we can show that the chiral symmetry
(\ref{eq:chiral2x2}) is indeed a remnant
of the original chiral symmetry (\ref{eq:chiral}).

In a similar manner as Sec.\ref{sec:chirality}, we can define the
topological number by using the chiral symmetry (\ref{eq:chiral2x2}).
By using the unitary transformation $u_{\gamma}$ which diagonalizes
$\gamma$ as 
\begin{eqnarray}
u_{\gamma}^\dagger\gamma u_{\gamma}=
\left(
\begin{array}{cc}
1 &0 \\
0 & -1
\end{array}
\right),
\quad
u_{\gamma}=\frac{1}{\sqrt{2}}
\left(
\begin{array}{cc}
1 & -i \\
i & -1
\end{array}
\right), 
\end{eqnarray}
${\cal H}_{2\times 2}({\bm k})$ is recast into
\begin{eqnarray}
u_{\gamma}^{\dagger}{\cal H}_{2\times 2}({\bm k})u_{\gamma}=
\left(
\begin{array}{cc}
0 & q({\bm k})\\
q^{*}({\bm k}) & 0
\end{array}
\right), 
\end{eqnarray}
with $q({\bm k})=-i\varepsilon({\bm k})-\Delta({\bm k})$.
Now we introduce the topological number $w(k_y)$ as
\begin{eqnarray}
w(k_y)&=&-\frac{1}{4\pi i}\int dk_x {\rm tr}\left[
\gamma {\cal H}_{2\times 2}^{-1}({\bm k})\partial_{k_x}{\cal H}_{2\times
2}({\bm k})
\right]
\nonumber\\
&=&
\frac{1}{2\pi }{\rm Im}\left[
\int dk_x \partial_{k_x}\ln q({\bm k})  
\right].
\label{eq:wdef}
\end{eqnarray}
Here the line integral (\ref{eq:wdef})
is performed in a manner similar to (\ref{eq:windingnumber}).
As is shown in Appendix \ref{sec:appendixb1}, it is found that this
integral can be rewritten as the following simple summation,
\begin{eqnarray}
w(k_y)=\frac{1}{2}\sum_{\varepsilon({\bm k})=0}
{\rm sgn}[\partial_{k_x}\varepsilon({\bm k})]\cdot{\rm sgn}[\Delta({\bm k})],
\label{eq:formula}
\end{eqnarray}
where the summation is taken for $k_x$ satisfying $\varepsilon({\bm
k})=0$ with a fixed $k_y$.
The formula (\ref{eq:formula}) makes it easy to evaluate the topological
number $w(k_y)$.
Then from the bulk-edge correspondence, we can say that if $w({k_y})\neq
0$, there exists a dispersionless ABS on a surface of
the superconductor which is perpendicular to the $x$-direction.

Furthermore, by using an argument similar to that in
Sec.\ref{sec:chirality}, it is found that the ABS
with flat dispersion is an eigenstate of
the chirality operator $\gamma$ again. 
Then, in a similar manner as (\ref{eq:index}), we can conjecture that the number
$n_0^{(\pm)}$ of the zero energy ABS with chirality
$\gamma=\pm 1$ satisfies
\begin{eqnarray}
w(k_y)=n_0^{(-)}-n_0^{(+)}, 
\label{eq:index2x2}
\end{eqnarray}
or
\begin{eqnarray}
w(k_y)=n_0^{(+)}-n_0^{(-)}.
\label{eq:index2x2_2}
\end{eqnarray}
These give a criterion for the ABS with flat
dispersion.

\begin{figure}[h]
\begin{center}
\includegraphics[width=10cm]{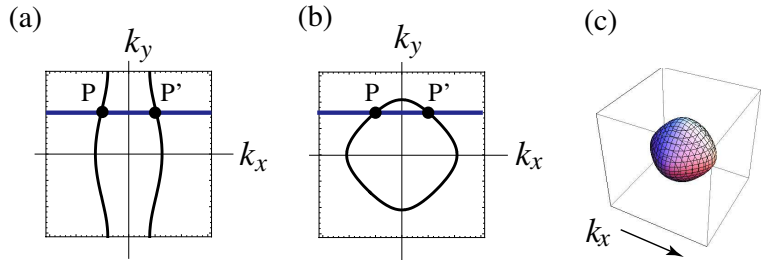}
\caption{(color online) Fermi surfaces with simple topology in 
(a) quasi-one dimensional system, (b) quasi-two dimensional one,     
and (c) three dimensional one. 
The thick blue lines denote the integral path of $w$. 
For simplicity, we illustrate the integral path only in (a) and (b).
For each case, the integral path gets across the Fermi surface only twice at
$k_x=\pm k_x^0$. In (a) and (b), P and P' denote the intersection point
 $(-k_x^0,k_y)$ and $(k_x^0, k_y)$, respectively. 
}
\label{fig:simpleFS}
\end{center}
\end{figure}

We notice here that our topological criterion of dispersionless ABS
includes the criterion proposed previously. 
In the case where the topology of the Fermi surface is simple as
illustrated in Fig.\ref{fig:simpleFS}, it has been known that
if the gap function satisfies 
\begin{eqnarray}
\Delta(k_x,k_y)\Delta(-k_x,k_y)<0 
\label{eq:signchange}
\end{eqnarray} 
then a zero energy ABS exists on the boundary
perpendicular to $x$-direction.\cite{KT00,Hu94}
In other words, 
a sign change of the gap function with respect to $k_x \rightarrow -k_x$
implies the existence of the zero energy ABS on a
surface perpendicular to the $x$-direction.
Our topological criterion reproduces this result correctly:
The formula (\ref{eq:formula}) leads to
\begin{eqnarray}
w(k_y)&=&\frac{1}{2}\left[
{\rm sgn}[\partial_{k_x}\varepsilon(-k_x^0,k_y)]\cdot
{\rm sgn}[\Delta(-k_x^0,k_y)] 
+{\rm sgn}[\partial_{k_x}\varepsilon(k_x^0,k_y)]\cdot
{\rm sgn}[\Delta(k_x^0,k_y)] 
\right],
\end{eqnarray}
where $(\pm k_x^0, k_y)$ denote the intersection points between the integral
path of $w(k_y)$ and the Fermi surface. See Fig. \ref{fig:simpleFS}. 
Noticing that ${\rm sgn}[\partial_{k_x}\varepsilon(k_x^0,k_y)]
=-{\rm sgn}[\partial_{k_x}\varepsilon(-k_x^0,k_y)]$, we can rewrite this as
\begin{eqnarray}
w(k_y)&=&\frac{1}{2}
{\rm sgn}[\partial_{k_x}\varepsilon(-k_x^0,k_y)]
\left[
{\rm sgn}[\Delta(-k_x^0,k_y)]-
{\rm sgn}[\Delta(k_x^0,k_y)] 
\right].
\end{eqnarray}
Thus the topological number $w(k_y)$ becomes nonzero only when the gap
function satisfies (\ref{eq:signchange}), which means that our
topological criterion reproduces the previous one in this particular
simple case.

We would like to emphasis here that our topological criterion does not
merely reproduce the known criterion, but is more informative.
First, the chirality of the zero energy ABS is determined in
a manner consistent with the formulas
(\ref{eq:index2x2}) and (\ref{eq:index2x2_2}). 
Second, our formula is also applicable to more
complicated cases in which the previous criterion does not work. 
We will see them in Secs. \ref{sec:dxy}, \ref{sec:px} and \ref{sec:px2}.

\subsection{$d_{xy}$-wave superconductor}
\label{sec:dxy}

Here we consider the $d_{xy}$-wave superconductor where $\varepsilon({\bm k})$ and 
$\Delta({\bm k})$ in (\ref{eq:BdG2x2}) are given by
\begin{eqnarray}
\varepsilon({\bm k})=\frac{{\bm k}^2}{2m}-\mu,
\quad 
\Delta({\bm k})=\Delta_0 \frac{k_x k_y}{{\bm k}^2}.
\label{eq:epsilon}
\end{eqnarray}
Here $\Delta_0$ is a positive constant.
From (\ref{eq:formula}), the topological number $w(k_y)$ is evaluated as
\begin{eqnarray}
w(k_y)=
\left\{
\begin{array}{rl}
1, & \mbox{for $0<k_y< k_{\rm F}$}\\
-1,& \mbox{for $0>k_y> -k_{\rm F}$}\\
0, & \mbox{for $|k_y|>k_{\rm F}$}
\end{array}
\right., 
\label{eq:wdxy}
\end{eqnarray}
where $k_{\rm F}=\sqrt{2m\mu}$ is the Fermi momentum. 
Thus our topological criterion (\ref{eq:index2x2}) implies the existence
of the zero energy ABS for $|k_y|<k_{\rm F}$. 

Now, 
let us solve the BdG equation for the semi-infinite $d_{xy}$ superconductor 
on $x>0$ with the boundary condition 
$|u(x=0,k_y)\rangle=0$.
For $|k_y|<k_{\rm F}$,  the following zero energy ABS on $x=0$ is found
\cite{Hu94}
\begin{eqnarray}
|u_0(x)\rangle=C
\left(
\begin{array}{c}
1 \\
-i{\rm sgn}k_y
\end{array}
\right)e^{ik_y y}\sin (k_x x) e^{-x/\xi},
\end{eqnarray}
where $C$ is a normalization constant, $k_x=\sqrt{k_{\rm F}^2-k_y^2}$
and $\xi^{-1}=m\Delta_0 k_y/k^2_{\rm F}$. 
Since the ABS
is an eigenstate of $\gamma(=\sigma_y)$ with eigenvalue $\gamma=-1$
($\gamma=1$) for $0<k_y<k_{\rm F}$ ($0>k_y>-k_{\rm F}$), 
it is found that $n_0^{(+)}=0$ and $n_0^{(-)}=1$ for $0<k_y<k_{\rm F}$ 
($n_0^{(+)}=1$ and
$n_0^{(-)}=0$ for $0>k_y>-k_{\rm F}$).
On the other hand, for $|k_y|>k_{\rm F}$, we do not find any zero energy
ABS, thus $n_0^{(+)}=n_0^{(-)}=0$.
We summarize these results in Table \ref{table:dxy} (a), which
shows that the index theorem (\ref{eq:index2x2})
holds in this case.

If we choose the
semi-infinite $d_{xy}$ superconductor on $x<0$, the zero energy ABS on
the surface at $x=0$
is given by
\begin{eqnarray}
|u_0(x)\rangle=C
\left(
\begin{array}{c}
1 \\
i{\rm sgn}k_y
\end{array}
\right)e^{ik_y y}\sin (k_x x) e^{x/\xi}.
\end{eqnarray}
Thus $n_0^{(+)}$ and $n_0^{(-)}$ are summarized as Table \ref{table:dxy} (b).
The index theorem (\ref{eq:index2x2_2}) holds in this case.

\begin{table}
\begin{center} 
\begin{tabular}[t]{|c|c|c|c|c|}
\hline
\hline
\multicolumn{5}{c}{(a) $d_{xy}$-wave superconductor on $x>0$} \\ 
 \hline
$k_y$ 
& $n_0^{(+)}$& $n_0^{(-)}$& $n_0^{(+)}-n_0^{(-)}
$ & $w(k_y)$ \\ 
\hline 
$0<k_y<k_{\rm F}$ 
&0 &1 &-1 &1 \\ 
$0>k_y>-k_{\rm F}$
&1 &0& 1 & -1 \\
$|k_y|>k_{\rm F}$
&0 &0 &0 & 0 \\
\hline
\multicolumn{5}{c}{}\\
\multicolumn{5}{c}{(b) $d_{xy}$-wave superconductor on $x<0$} \\ 
 \hline
$k_y$ 
& $n_0^{(+)}$& $n_0^{(-)}$& $n_0^{(+)}-n_0^{(-)}
$ & $w(k_y)$ \\ 
\hline 
$0<k_y<k_{\rm F}$ 
&1 &0 &1 &1 \\ 
$0>k_y>-k_{\rm F}$
&0 &1& -1 & -1 \\
$|k_y|>k_{\rm F}$
&0 &0 &0 & 0 \\
\hline
\multicolumn{5}{c}{}\\
\hline
\hline
 \end{tabular} 
\end{center}
\caption{The number $n_0^{(\pm)}$ of the zero energy ABSs with the
 chirality $\gamma=\pm 1$ for (a) the semi-infinite $d_{xy}$-wave
superconductor on $x>0$ and (b) that on $x<0$. 
For comparison, we also show the topological number $w(k_y)$ given in
 (\ref{eq:wdxy}). The index theorem (\ref{eq:index2x2}) and
 (\ref{eq:index2x2_2}) hold in (a) and (b), respectively.}
\label{table:dxy}
\end{table}

\subsection{$p_x$-wave superconductor}
\label{sec:px}

Now consider the semi-infinite $p_x$-wave superconductor on $x>0$.
The gap function is $\Delta({\bm k})=\Delta_0 k_x/k$ with $k=\sqrt{{\bm
k}^2}$ and $\varepsilon({\bm k})$ is the same as that in (\ref{eq:epsilon}).
For the $p_x$-wave superconductor, we have
\begin{eqnarray}
w(k_y)=
\left\{
\begin{array}{rl}
1, & \mbox{for $|k_y|<k_{\rm F}$}\\
0, & \mbox{for $|k_y|>k_{\rm F}$}
\end{array}
\right., 
\label{eq:wpx}
\end{eqnarray}
from (\ref{eq:formula}).
Correspondingly, if $|k_y|<k_{\rm F}$, we obtain the following zero
energy ABS on $x=0$ 
\begin{eqnarray}
|u(x,k_y)\rangle=
\left(
\begin{array}{c}
1\\
-i
\end{array}
\right)e^{ik_y y}\sin(k_x x)e^{-x/\xi_{p}}, 
\end{eqnarray}
for the semi-infinite $p_x$-wave superconductor on $x>0$, and 
\begin{eqnarray}
|u(x,k_y)\rangle=
\left(
\begin{array}{c}
1\\
i
\end{array}
\right)e^{ik_y y}\sin(k_x x)e^{x/\xi_{p}}, 
\end{eqnarray}
for the semi-infinite $p_x$-wave  superconductor on $x<0$.
Here $k_x=\sqrt{k_{\rm F}^2-k_y^2}$ and $\xi_p^{-1}=m\Delta_0 /k_{\rm
F}$.
It is also found that these solutions are the eigenstates of $\gamma$ with
the eigenvalue $\gamma=-1$ and $\gamma=1$, respectively. 
Thus $n_0^{(+)}$ and $n_0^{(-)}$ are summarized as Table \ref{table:px}
(a) and (b).
We confirm the relations (\ref{eq:index2x2}) and (\ref{eq:index2x2_2}),
respectively again. 

\begin{table}
\begin{center} 
\begin{tabular}[t]{|c|c|c|c|c|}
\hline
\hline
\multicolumn{5}{c}{(a) $p_{x}$-wave superconductor on $x>0$} \\ 
 \hline
$k_y$ 
& $n_0^{(+)}$& $n_0^{(-)}$& $n_0^{(+)}-n_0^{(-)}
$ & $w(k_y)$ \\ 
\hline 
$|k_y|<k_{\rm F}$ 
&0 &1 &-1 &1 \\ 
$|k_y|>k_{\rm F}$
&0 &0 &0 & 0 \\
\hline
\multicolumn{5}{c}{}\\
\multicolumn{5}{c}{(b) $p_{x}$-wave superconductor on $x<0$} \\ 
 \hline
$k_y$ 
& $n_0^{(+)}$& $n_0^{(-)}$& $n_0^{(+)}-n_0^{(-)}
$ & $w(k_y)$ \\ 
\hline 
$|k_y|<k_{\rm F}$ 
&1 &0 &1 &1 \\ 
$|k_y|>k_{\rm F}$
&0 &0 &0 & 0 \\
\hline
\multicolumn{5}{c}{}\\
\hline
\hline
 \end{tabular} 
\end{center}
\caption{The number $n_0^{(\pm)}$ of the zero energy ABSs with the
 chirality $\gamma=\pm 1$ for (a) the semi-infinite $p_{x}$-wave
 superconductor on $x>0$ and (b) that on $x<0$, respectively.
For comparison, we also show the topological number $w(k_y)$ given in
 (\ref{eq:wpx}). The index theorem (\ref{eq:index2x2}) and
 (\ref{eq:index2x2_2}) hold in (a) and (b), respectively. }
\label{table:px}
\end{table}

\subsection{Zero energy Andreev bound state and odd-frequency pairing}
\label{sec:zero-odd}

In this subsection, we discuss the ABS in $d_{xy}$-wave and
$p_x$-wave superconductors from a viewpoint of odd-frequency pairing.
It will be shown that the odd-frequency pairing has a
topological structure similar to that of the ABSs discussed in 
Secs.\ref{sec:dxy} and \ref{sec:px}.

Odd-frequency pairing 
is 
the pair function (pairing amplitude) that 
changes a sign when exchanging the time coordinates of two electrons.
\cite{Berezinskii74} 
Thus the Fourier transformed pairing function is an
odd function of frequency. 
Near the surface of superconductor,
due to the breakdown of translational
invariance, the  pair potential 
acquires a spatial dependence leading to the coupling between the even and
odd-parity pairing states.
From the Fermi-Dirac statistics,
the induced 
pair amplitude at the interface
should be odd in frequency. 
%
It has been established recently that 
zero energy dispersionless ABSs induce odd-frequency pairings at the
surface of the superconductors.\cite{TG07,ELCCS07,TGKU07,TTG07}
%

First, we consider a two-dimensional semi-infinite superconductor in
$x>0$ where the surface is located at $x=0$.
As shown in the appendix  \ref{sec:appendixd}, 
the pair amplitude at the surface of 
$d_{xy}$-wave superconductor is given by 
\begin{equation}
f(k_{x},k_{y})=
\frac{i{\rm sgn}(k_{y}) \Delta_{0}}{\omega_{n}}
\frac{\mid k_{x} \mid \mid k_{y} \mid}{{\bm k}^{2}} 
\label{zeroodd1}
\end{equation} 
in the Matsubara representation.   
Thus the odd-parity odd-frequency pairing is realized at the surface. 
On the other hand, for 
$p_{x}$-wave pair potential, we obtain 
\begin{equation}
f(k_{x},k_{y})=\frac{i \Delta_{0}}{\omega_{n}}
\frac{\mid k_{x} \mid}{\sqrt{ {\bm k}^{2}}},
\label{zeroodd2}
\end{equation}
which means the realization of even-parity odd-frequency pairing in this case. 
As well as the wave function of zero energy ABS of $d_{xy}$ and 
$p_{x}$-wave superconductor in Secs.\ref{sec:dxy} and \ref{sec:px}, the
factor ${\rm{sgn}}k_{y}$ exists only for $d_{xy}$-wave case. 
This factor decides the difference of the parity 
of induced Cooper pair. 
The difference of the parity results in a
serious difference when we consider proximity effect 
into DN attached to 
superconductor \cite{TG07,TGKU07,TTG07}. 
In DN, only $s$-wave even parity 
pairing is possible. 
Thus, odd-parity odd-frequency pairing amplitude
cannot penetrate into DN. 
This implies that 
ABS in $d_{xy}$-wave superconductor cannot enter into DN since it is
expressed by odd-frequency spin-singlet odd-parity state. 
On the other hand, for $p_{x}$-wave 
superconductor, 
ABS can enter into DN since it is expressed by
odd-frequency spin-singlet even-parity state including $s$-wave channel 
\cite{TNS03,TNGK04,TK04,TKY05,ATK06}. 

Now we consider a two-dimensional semi-infinite superconductor 
in $x<0$. 
The corresponding pair amplitude at surface ($x=0$) is 
given by 
\begin{equation}
f(k_{x},k_{y})=
-\frac{i{\rm sgn}(k_{y}) \Delta_{0}}{\omega_{n}}
\frac{\mid k_{x} \mid \mid k_{y} \mid}{{\bm k}^{2}} 
\label{zeroodd3}
\end{equation}
for spin-singlet $d_{xy}$-wave superconductor and 
\begin{equation}
f(k_{x},k_{y})=-\frac{i \Delta_{0}}{\omega_{n}}
\frac{\mid k_{x} \mid}{\sqrt{ {\bm k}^{2}}}
\label{zeroodd4}
\end{equation}
for spin-triplet $p_{x}$-wave one, respectively.  
Comparing
Eq. (\ref{zeroodd1}) [(\ref{zeroodd2})] with 
(\ref{zeroodd3})  [(\ref{zeroodd4})], we find the 
difference between them is 
the presence of $-$ sign. 
The present $-$ sign exactly corresponds to the different values 
of $n_{0}^{+}$ and $n_{0}^{-}$ between case (a) and 
case (b) in Tables \ref{table:dxy} and \ref{table:px}. \par

\subsection{Dispersionless Andreev bound states in superconductors with multiple Fermi surfaces}
\label{sec:px2}

\begin{figure}[h]
\begin{center}
\includegraphics[width=5cm]{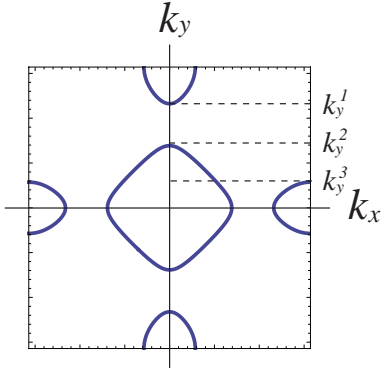}
\caption{(color online) The multiple Fermi surfaces in the first Brillouin zone. We take
 $t=t'=1$ and $\mu=-2.5$ in (\ref{eq:multiFSdispersion}).
}
\label{fig:multiFS}
\end{center}
\end{figure}

In the above two examples, we assume a simple Fermi surface
obtained in (\ref{eq:epsilon}). 
Here we consider superconducting states with more complicated Fermi
surfaces. 
To realize multiple Fermi surfaces, we consider a model in the
square lattice. 
By taking into account the second next nearest-neighbor hopping, the
normal dispersion $\varepsilon({\bm k})$ is
given by 
\begin{eqnarray}
\varepsilon({\bm k})=-2t[\cos k_x+\cos k_y]
-2t'[\cos 2k_x+\cos 2k_y]-\mu.
\label{eq:multiFSdispersion}
\end{eqnarray}
As illustrated in Fig.\ref{fig:multiFS}, if we choose the parameters
$t$, $t'$ and $\mu$ properly, multiple Fermi surfaces are realized in the
first Brillouin zone.

For the gap function, we consider (a) $d_{xy}$-wave pairing with
$\Delta({\bm k})=\Delta_0 \sin k_x\sin k_y$, (b) $p_{x}$-wave pairing with
$\Delta({\bm k})=\Delta_0 \sin k_x$ and (c) $p_x$-wave pairing with $\Delta({\bm
k})=\Delta_0\sin 2k_x$.
Since the condition (\ref{eq:signchange}) is satisfied in all the cases in
the above,
the sign change criterion suggests the existence of zero energy ABS on a surface
perpendicular to the $x$-direction.
However, we find that zero energy ABSs do not always appear.

To study the edge state, we consider the model above in the square lattice. 
The BdG Hamiltonian in the lattice space is given by
\begin{eqnarray}
&&{\cal H}={\cal H}_{\rm kin}+{\cal H}_{\rm s},
\nonumber\\
&&{\cal H}_{\rm kin}=-t\sum_{\langle {\bm i},{\bm j}\rangle,\sigma}
c^{\dagger}_{{\bm i}, \sigma}c_{{\bm j}, \sigma}  
-t'\sum_{\langle\langle {\bm i},{\bm j}\rangle\rangle,\sigma}
c^{\dagger}_{{\bm i}, \sigma}c_{{\bm j}, \sigma} 
-\mu\sum_{{\bm i},\sigma}c^{\dagger}_{{\bm i}, \sigma}c_{{\bm i}, \sigma},
\end{eqnarray}
where ${\bm i}=(i_x,i_y)$ denotes a site on the square lattice,
$c^{\dagger}_{{\bm i},\sigma}$ ($c_{{\bm k},\sigma}$) the creation
(annihilation) operator of an electron with spin $\sigma$ at site ${\bm
i}$. The summation $\langle {\bm i}, {\bm j}\rangle$ ($\langle\langle
{\bm i}, {\bm j}\rangle\rangle$) is taken for the nearest-neighbor
(the second next nearest-neighbor) sites. 
${\cal H}_s$ is the pairing term depending on the symmetry of
the Cooper pair.
For the $d_{xy}$-wave pairing with $\Delta({\bm k})=\Delta_0\sin k_x \sin k_y$, 
it is given by
\begin{eqnarray}
{\cal H}_s=-\frac{\Delta_0}{4} \sum_{\bm i}
(c^{\dagger}_{{\bm i}+\hat{x}+\hat{y},\uparrow}c^{\dagger}_{{\bm i},\downarrow} 
+c^{\dagger}_{{\bm i}-\hat{x}-\hat{y},\uparrow}c^{\dagger}_{{\bm i},\downarrow} 
-c^{\dagger}_{{\bm i}+\hat{x}-\hat{y},\uparrow}c^{\dagger}_{{\bm i},\downarrow} 
-c^{\dagger}_{{\bm i}-\hat{x}+\hat{y},\uparrow}
c^{\dagger}_{{\bm i},\downarrow}) +\mbox{h.c.}.
\end{eqnarray}
For the $p_x$-wave pairing with $\Delta({\bm k})=\Delta_0\sin k_x$, 
\begin{eqnarray}
{\cal H}_s=\frac{\Delta_0}{2i} \sum_{\bm i}
(c^{\dagger}_{{\bm i},\uparrow}c^{\dagger}_{{\bm i}+\hat{x},\downarrow} 
-c^{\dagger}_{{\bm i},\uparrow}c^{\dagger}_{{\bm i}-\hat{x},\downarrow}) 
+\mbox{h.c.},
\end{eqnarray}
and for the $p_x$-wave pairing with $\Delta({\bm k})=\Delta_0\sin 2k_x$, 
\begin{eqnarray}
{\cal H}_s=\frac{\Delta_0}{2i} \sum_{\bm i}
(c^{\dagger}_{{\bm i},\uparrow}c^{\dagger}_{{\bm i}+2\hat{x},\downarrow} 
-c^{\dagger}_{{\bm i},\uparrow}c^{\dagger}_{{\bm i}-2\hat{x},\downarrow}) 
+\mbox{h.c.}.
\end{eqnarray}
Suppose that the system has two open boundary edges at $i_x=0$ and
$i_x=N_x$, and impose the periodic boundary condition in the $y$-direction.
Solving numerically the energy spectrum as a function of the momentum
$k_y$, we examine the edge states.

\begin{figure}[h]
\begin{center}
\includegraphics[width=15cm]{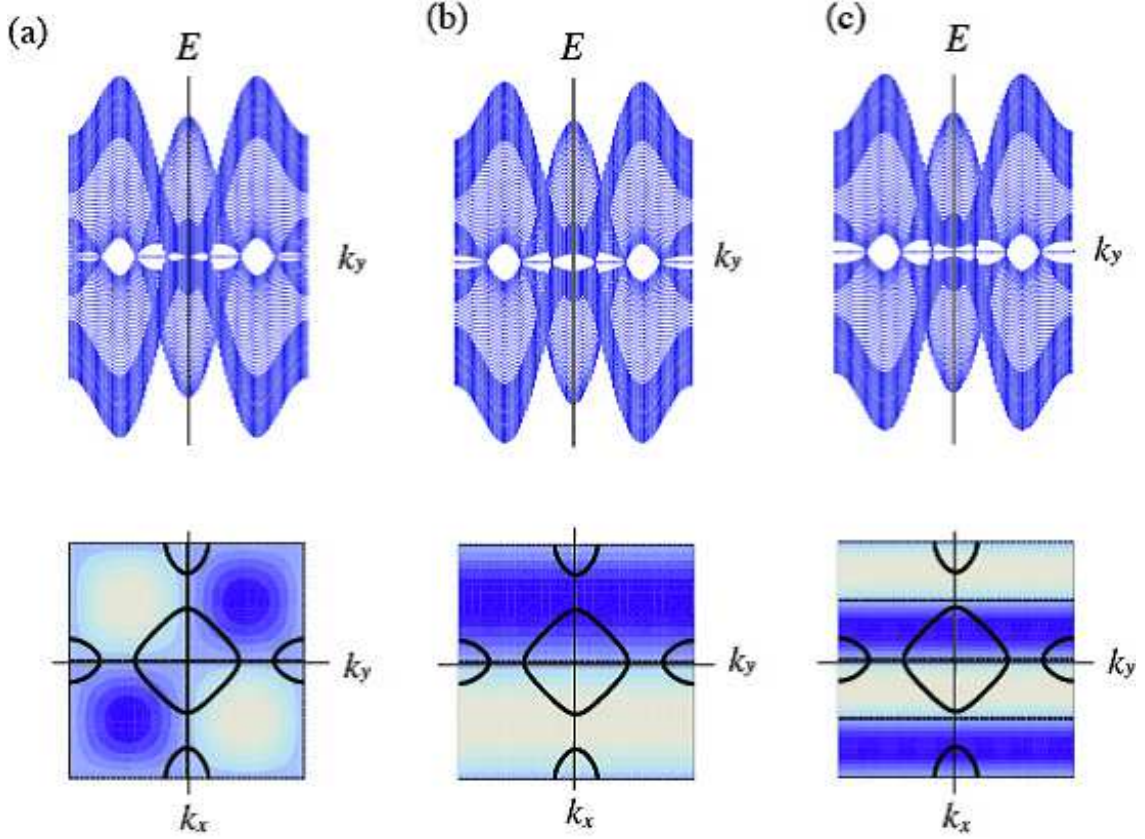}
\caption{(color online) The energy spectra with open edges at $i_x=0$ and $i_x=100$ (upper panels) and the corresponding
gap functions in the first Brillouin zone (lower panels) for
 superconductors with multiple Fermi surfaces. 
(a) $d_{xy}$-wave superconductor with $\Delta({\bm
 k})=\Delta_0\sin k_x\sin k_y$, (b) $p_{x}$-wave superconductor with 
$\Delta({\bm  k})=\Delta_0\sin k_x$, and (c) $p_x$-wave superconductor with
 $\Delta({\bm  k})=\Delta_0\sin 2k_x$. 
We take $t=t'=1$ and $\mu=-2.5$.  $\Delta_0$ is chosen as $\Delta_0=1$
 for (a) and $\Delta_0=0.5$ for (b) and (c), respectively.
In the upper panels, $k_y$ is the momentum in the $y$-direction, and $k_y\in[-\pi,\pi]$.
The ABSs appear as zero energy states with flat dispersion.  
In the lower panels, the solid curves denotes the Fermi surfaces.
The darker areas denotes the negative gap functions, the lighter areas
 the positive ones, and the gap functions vanish at the dashed lines.
As a matter of convenience, we take the horizontal axis as the
 $k_y$-direction in the lower panels.
}
\label{fig:edgestate}
\end{center}
\end{figure}

In Fig.\ref{fig:edgestate}(a)-(c), we illustrate the energy spectra
for each case.
In spite that zero energy ABSs appear on the edges, 
the previous criterion does not explain the detailed
structures.
Indeed, in the cases (a) and (b), the dispersionless ABSs disappear near
$k_y\sim 0$,  although the sign change condition (\ref{eq:signchange})
is still satisfied for both cases.   
Therefore, the sign change criterion does not work in these cases.
On the other hand, the topological criterion does work in all the cases.
From the formula (\ref{eq:formula}),  we obtain $w(k_y)$ in Table
\ref{table:multiFS}.
Comparing $w(k_y)$ in Table \ref{table:multiFS} with the zero energy
states in Fig. \ref{fig:edgestate}, we find an excellent agreement
between them.
In particular, for (a) and (b), we find that $w(k_y)=0$ at $k_y\sim 0$
(i.e. $k_y^3>k_y>-k_y^3$). 
Thus in contrast to the sign change criterion, our topological criterion
explains the detailed structures of the dispersionless ABSs.

\begin{table}
\begin{center} 
\begin{tabular}[t]{|c|c|c|c|}
\hline
\hline
\multicolumn{4}{c}{ } \\ 
 \hline
& \multicolumn{3}{c|}{$w(k_y)$} \\ 
\cline{2-4}
\raisebox{1.5ex}{$k_y$} & (a) &(b) & (c)\\ 
\hline 
$\pi>k_y>k_y^1$ 
&1 &1 & 1
\\ 
$k_y^1>k_y>k_y^2$
&0 &0 &0 
\\
$k_y^2>k_y>k_y^3$
&1 &1 &1 
\\
$k_y^3>k_y>-k_y^3$
&0 &0 &2 
\\
-$k_y^3>k_y>-k_y^2$
&-1 &1 &1 
\\
-$k_y^2>k_y>-k_y^1$
&0 &0 &0 
\\
-$k_y^1>k_y>-\pi$
&-1 &1 &1 
\\
\hline
\multicolumn{4}{c}{}\\
\hline
\hline
 \end{tabular} 
\end{center}
\caption{Topological number $w(k_y)$ for (a) the $d_{xy}$-wave
 superconductor with $\Delta({\bm k})=\Delta_0\sin k_x\sin k_y$, and
the $p_x$-wave superconductor with (b) $\Delta({\bm k})=\Delta_0\sin k_x$
 and (c) $\Delta({\bm k})=\Delta_0 \sin 2k_x$. We suppose that $t=t'=1$,
 $\mu=-2.5$ and $\Delta_0>0$. Here $k_y^i$ $(i=1,2,3)$
 are defined in Fig.\ref{fig:multiFS}.}
\label{table:multiFS}
\end{table}

\section{Noncentrosymmetric superconductor}
\label{sec:4x4}
In this section, we apply our topological criterion to time-reversal invariant
superconductors supporting multiple components of the gap function. 
For simplicity, we consider the superconductor in a single-band
description. 
In a single-band description, the general BdG Hamiltonian for
time-reversal invariant superconductor is given by
\begin{eqnarray}
{\cal H}({\bm k})=
\left(
\begin{array}{cc}
\varepsilon({\bm k})+{\bm g}({\bm k})\cdot{\bm \sigma} & \hat{\Delta}({\bm k})\\
\hat{\Delta}^{\dagger}({\bm k}) & -\varepsilon({\bm k})+{\bm g}({\bm k})\cdot
 {\bm \sigma}^{*}
\end{array}
\right), 
\label{eq:BdG4x4}
\end{eqnarray}
where $\varepsilon({\bm k})=\varepsilon(-{\bm k})$ is the kinetic energy
of electron 
measured from the Fermi energy, ${\bm g}({\bm k})=-{\bm g}(-{\bm k})$
the anti-symmetric spin-orbit interaction such as the Rashba spin-orbit
coupling, 
$\hat{\Delta}({\bm k})$ the gap function
$
\hat{\Delta}({\bm k})=i\psi({\bm k})\sigma_y+i{\bm d}({\bm k})\cdot{\bm
 \sigma}\sigma_y. 
$
This Hamiltonian has the particle-hole symmetry (\ref{eq:particle-hole}) and the
time-reversal invariance (\ref{eq:time-reversal}) with $U=i \sigma_y$.
Thus it also satisfies (\ref{eq:chiral}) with
\begin{eqnarray}
\Gamma= 
\left(
\begin{array}{cc}
0 & \sigma_y\\
\sigma_y & 0
\end{array}
\right).
\label{eq:gamma}
\end{eqnarray}

For the dispersionless ABS for the semi-infinite
superconductor on $x>0$, the topological number (\ref{eq:windingnumber}) is
given by
\begin{eqnarray}
W(k_y)=\frac{1}{2\pi }
{\rm Im}\left[
\int dk_x \partial_{k_x} \ln {\rm det}\hat{q}({\bm k})
\right],
\label{eq:Wdef}
\end{eqnarray}
with $\hat{q}({\bm k})=[\varepsilon({\bm k})-i\psi({\bm k})]\sigma_y
+[{\bm g}({\bm k})-i{\bm d}({\bm k})]\cdot{\bm \sigma}\sigma_y$.
As is shown in Appendix \ref{sec:appendixb2}, $W(k_y)$ is evaluated as
\begin{eqnarray}
W(k_y)=-\frac{1}{2}\sum_{\varepsilon({\bm k})^2-{\bm g}({\bm k})^2=0} 
{\rm sgn}\left[{\bm g}({\bm k})\cdot {\bm d}({\bm k})
-\varepsilon({\bm k})\psi({\bm k})\right]
\cdot {\rm sgn}\left[\partial_{k_x}\left(\varepsilon({\bm k})^2-{\bm
				    g}({\bm k})^2\right)\right],
\label{eq:formula2}
\end{eqnarray} 
where the summation is taken for $k_x$ satisfying $\varepsilon({\bm
k})^2-{\bm g}({\bm k})^2=0$ with a fixed $k_y$.

\subsection{two-dimensional Rashba superconductor}
\begin{figure}[h]
\begin{center}
\includegraphics[width=5cm]{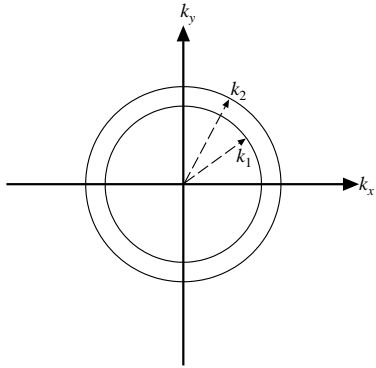}
\caption{(color online) Fermi surfaces of two-dimensional Rashba noncentrosymmetric SC.}
\label{fig:splitFS}
\end{center}
\end{figure}

Now consider two-dimensional Rashba noncentrosymmetric superconductors.
Here $\varepsilon({\bm k})$ is
given by (\ref{eq:epsilon}) and ${\bm g}({\bm k})=\lambda(\hat{\bm
 x}k_y-\hat{\bm y}k_x)$ where $\lambda$ is the coupling constant of
Rashba spin-orbit interaction.  
Due to the Rashba spin-orbit interaction, the Fermi surface is split
into two as illustrated in Fig.\ref{fig:splitFS}.
The Fermi momenta for the smaller and larger
Fermi surface are given by 
\begin{eqnarray}
k_1=-m\lambda + \sqrt{
(m\lambda)^{2} + 2m\mu },
\end{eqnarray}
\begin{eqnarray}
k_2=m
\lambda + \sqrt{ (m\lambda)^{2} +
2m\mu }, 
\end{eqnarray}
respectively.
The spin-orbit interaction also mixes the parity of the gap function
generally, so the spin-singlet component and the spin-triplet one
coexist in the gap function \cite{FAKS04,Fujimoto07a,Fujimoto07b}.
The spin-triplet component $\bm{d}(\bm{k})$
is aligned with the polarization vector of the Rashba spin orbit
coupling, $\bm{d}(\bm{k}) || \bm{g}(\bm{k})$ \cite{FAKS04}.
Then, the triplet component is given by
\begin{eqnarray}
\bm{d}(\bm{k})=\Delta_{t}f({\bm k})
(\hat{\bm{x}}k_{y}-\hat{\bm{y}}k_{x})/ k
\end{eqnarray}
with $k=\sqrt{{\bm k}^2}$ 
while the singlet component reads
\begin{eqnarray}
\psi(\bm{k})=\Delta_{s}f({\bm k}) 
\end{eqnarray}
with $\Delta_{t} \geq 0$ and 
$\Delta_{s} \geq 0$. 
The superconducting gaps are $\Delta_1=|\bar{\Delta}_1({\bm k})|$ and
$\Delta_2=|\bar{\Delta}_{2}({\bm k})|$ for the two spin split bands with
$\bar{\Delta}_1({\bm k})=(\Delta_t+\Delta_s)f({\bm k})$ and
$\bar{\Delta}_2({\bm k})=(\Delta_t-\Delta_s)f({\bm k})$. 

Using the formula (\ref{eq:formula2}), we evaluate the topological
number $W(k_y)$.
We obtain
\begin{eqnarray}
W(k_y)=&&
\frac{1}{2}{\rm sgn}[(\Delta_t-\Delta_s)f(-{\bm k}_2)] 
-\frac{1}{2}{\rm sgn}[(\Delta_t-\Delta_s)f({\bm k}_2)] 
\nonumber\\
&&-\frac{1}{2}{\rm sgn}[(\Delta_t+\Delta_s)f(-{\bm k}_1)] 
+\frac{1}{2}{\rm sgn}[(\Delta_t+\Delta_s)f({\bm k}_1)]
\label{eq:top1}
\end{eqnarray}
for $|k_y|<k_1$, 
\begin{eqnarray}
W(k_y)&=&\frac{1}{2}{\rm sgn}[(\Delta_t-\Delta_s)f(-{\bm k}_2)] 
-\frac{1}{2}{\rm sgn}[(\Delta_t-\Delta_s)f({\bm k}_2)]
\label{eq:top2}
\end{eqnarray}
for $k_1<|k_y|<k_2$, and $W(k_y)=0$ for $k_2<|k_y|$. 
Here, $\pm{\bm k}_i=(\pm \sqrt{k_i^2-k_y^2},k_y)$ $(i=1,2)$.

From Eqs.(\ref{eq:top1}) and (\ref{eq:top2}), it is found that
$W(k_y)$ is nonzero only when $f({\bm k})$ is an odd function with
respect to $k_x$.
Therefore, for example, for an $s+p$-wave or
a $d_{x^2-y^2}+f$-wave Rashba superconductor, where
$f({\bm k})$ is given by $f({\bm k})=1$ or $f({\bm
k})=(k_x^2-k_y^2)/{\bm k}^2$, respectively, $W(k_y)$ becomes zero.
Consistently, we find that they do not support the
dispersionless ABSs.
On the other hand, for a $d_{xy}+p$-wave superconductor, we have a
dispersionless ABS. 

\subsubsection{$d_{xy}+p$-wave superconductor}
For the $d_{xy}+p$-wave superconductor, we have $f({\bm k})=k_xk_y/{\bm k}^2$. 
Thus, $W(k_y)$ can be nonzero value. 
From Eqs. (\ref{eq:top1}) and (\ref{eq:top2}), 
we obtain
\begin{eqnarray}
W(k_y)= 
\left\{
\begin{array}{cl}
2{\rm sgn}k_y, &\mbox{for $|k_y|<k_1$} \\
{\rm sgn}k_y, & \mbox{for $k_1<|k_y|<k_2$}\\
0, & \mbox{for $|k_y|>k_2$}
\end{array}
\right.,
\label{eq:W}
\end{eqnarray}
for $\Delta_s>\Delta_t$, and
\begin{eqnarray}
W(k_y)= 
\left\{
\begin{array}{cl}
0, & \mbox{for $|k_y|<k_1$}\\
-{\rm sgn}k_y, &\mbox{for $k_1<|k_y|<k_2$}\\
0, & \mbox{for $|k_y|>k_2$}
\end{array}
\right.,
\label{eq:W2}
\end{eqnarray}
for $\Delta_t>\Delta_s$.

To confirm the relation (\ref{eq:index}), let us 
consider the two-dimensional semi-infinite $d_{xy}+p$-wave Rashba superconductor
on $x>0$ where the surface is located at $x=0$. 
The corresponding wave function 
is given by \cite{TYBN09,TMYYS10}
\begin{eqnarray}
|u(x,k_y)\rangle
&=&
\left[
c_{1}^{+}\psi_{1}^{+}\exp(iq^{+}_{1x}x) 
+ c_{1}^{-}\psi_{1}^{-}\exp(-iq^{-}_{1x}x)
+ c_{2}^{+}\psi_{2}^{+}\exp(iq^{+}_{2x}x) 
+ c_{2}^{-}\psi_{2}^{-}\exp(-iq^{-}_{2x}x)
\right]
\nonumber\\
&&\times
\exp(ik_y y), 
\label{wavefunction}
\end{eqnarray}
\begin{eqnarray}
q^{\pm}_{1(2)x}
=k^{\pm}_{1(2)x} \pm \frac{ k_{1(2)}}{k^{\pm}_{1(2)x}}
\sqrt{\frac{E^{2}-[ \bar \Delta_{1(2)}(\bm{k}_{1(2)}^{\pm})]^{2}}
{\lambda^{2} + 2\mu /m }},
\label{eq:qpm12}
\end{eqnarray}
with 
$k^{+}_{1(2)x}=k^{-}_{1(2)x}=\sqrt{k_{1(2)}^2-k_y^2}$ for $|k_{y}|
\leq k_{1(2)}$
and $k^{+}_{1(2)x}=-k^{-}_{1(2)x}=i \sqrt{k_y^2-k_{1(2)}}$ for $|k_{y}|
>k_{1(2)}$, and 
$\bm{k}_{1(2)}^{\pm}=(\pm k_{1(2)x}^{\pm},k_{y})$. 
Here $\psi^{\pm}_i$ $(i=1,2)$ are
given by 
\begin{eqnarray}
^T\psi_{1}^{\pm} =\left(
1,-i\alpha_{1\pm}^{-1},i\alpha_{1\pm}^{-1}\Gamma_{1\pm},\Gamma_{1\pm} \right), 
\end{eqnarray}
\begin{eqnarray}
^T\psi_{2}^{\pm} =\left(1,i\alpha_{2\pm}^{-1},
i \alpha_{2\pm}^{-1}\Gamma_{2\pm},-\Gamma_{2\pm} \right)
\end{eqnarray}
with
\begin{eqnarray}
\Gamma_{1(2)\pm}=
\frac{\bar{\Delta}_{1(2)}(\bm{k}_{1(2)}^{\pm})}
{ E \pm \sqrt{E^2 - [\bar{\Delta}_{1(2)}(\bm{k}_{1(2)}^{\pm}) ] ^2 } },
\label{eq:gamma12pm}
\end{eqnarray}
and $\alpha_{1(2)\pm}=(\pm k^{\pm}_{1(2)x}-ik_{y})/k_{1(2)}$.
$E$ is the
quasiparticle energy measured from the Fermi energy.
For $E^2<[\bar{\Delta}_{1(2)}({\bm k}^{\pm}_{1(2)})]^2$, the branch of
the square root in Eqs.(\ref{eq:qpm12}) and (\ref{eq:gamma12pm}) is
chosen so as the wave function (\ref{wavefunction}) is normalized
(i.e. $|u(x,k_y)\rangle \rightarrow 0$ as $x\rightarrow \infty$).
When $E=0$, we find 
\begin{eqnarray}
\Gamma_{1\pm}=
\left\{
\begin{array}{cl}
-i{\rm sgn}k_y,  & \mbox{for $|k_y|<k_1$}\\
\pm i {\rm sgn}k_y, & \mbox{for $|k_y|>k_1$}
\end{array}
\right.,
\end{eqnarray}
and 
\begin{eqnarray}
\Gamma_{2\pm}=
\left\{
\begin{array}{cl}
-i{\rm sgn}(\Delta_t-\Delta_s){\rm sgn}k_y,  & \mbox{for $|k_y|<k_2$}\\
\pm i {\rm sgn}(\Delta_t-\Delta_s){\rm sgn}k_y, & \mbox{for $|k_y|>k_2$}
\end{array}
\right. .
\end{eqnarray}
Thus, it is found that $\psi_{1\pm}$ and $\psi_{2\pm}$ are
eigenstates of $\Gamma$ in (\ref{eq:gamma}),
\begin{eqnarray}
\Gamma \psi_{1}^{\pm}=
\left\{
\begin{array}{cl}
-{\rm sgn}k_y\psi_{1}^{\pm}, & \mbox{for $|k_y|<k_1$}\\
\pm {\rm sgn}k_y \psi_{1}^{\pm}, & \mbox{for $|k_y|>k_1$}
\end{array}
\right.,
\end{eqnarray}
\begin{eqnarray}
\Gamma \psi_{2}^{\pm}=
\left\{
\begin{array}{cl}
{\rm sgn}(\Delta_t-\Delta_s){\rm sgn}k_y\psi_{2}^{\pm}, 
& \mbox{for $|k_y|<k_2$}\\
\mp{\rm sgn}(\Delta_t-\Delta_s) {\rm sgn}k_y \psi_{2}^{\pm}, 
& \mbox{for $|k_y|>k_2$}
\end{array}
\right. .
\end{eqnarray}
In Table \ref{table:ABS}, we summarize the chirality of $\psi^{\pm}_{1(2)}$.

To construct the dispersionless ABS, we put the boundary
condition (\ref{eq:boundarycond}) on the wave function above.
Then, we have
\begin{eqnarray}
c_1^{+}\psi_1^{+}+c_1^{-}\psi_1^{-}+c_2^{+}\psi_2^{+}+c_2^{-}\psi_2^{-}=0. 
\label{eq:c_boundary}
\end{eqnarray}
The dispersionless ABS is obtained if there exist
non-zero $c_i^{\pm}$s satisfying (\ref{eq:c_boundary}).
Applying the chiral projection operator $P_{\pm}=(1\pm \Gamma)/2$ on the
both side, we can divide (\ref{eq:c_boundary}) into the sector with $\Gamma=1$
and that with $\Gamma=-1$. 
For example, when $\Delta_s>\Delta_t$ and $k_2>k_y>k_1$, we obtain
\begin{eqnarray}
c_1^{+}\psi_1^{+}=0 
\label{eq:ex1}
\end{eqnarray}  
in the $\Gamma=1$ sector, and 
\begin{eqnarray}
c_1^{-}\psi_1^-+c_2^{+}\psi_2^{+}+c_2^{-}\psi_2^{-}=0, 
\label{eq:ex2}
\end{eqnarray}  
in the $\Gamma=-1$ sector.
Then, we solve (\ref{eq:c_boundary}) in each chiral sector. 
In the above example, we find that $c_1^+$ in (\ref{eq:ex1}) is
identically zero, but there exists a single non-zero solution of
$(c_1^{-},c_2^{+},c_2^{-})$ satisfying (\ref{eq:ex2}).
To see this, let us notice that only two components of $\psi_i^{\pm}$ 
are independent when $\psi_i^{\pm}$ is an eigenstate of
$\Gamma$.
Hence, we obtain two independent linear equations from (\ref{eq:ex2}). 
Solving these linear equations, we have a unique solution
$(c_1^{-},c_2^+,c_2^-)$ of (\ref{eq:ex2}) up to a overall normalization factor.
This result means that no ABS exists in the $\Gamma=1$ sector but a
single ABS in the $\Gamma=-1$ sector. 
Thus, $N_0^{(+)}=0$ and $N_0^{(-)}=1$. 
In a similar manner, we can solve (\ref{eq:c_boundary}) for other $k_y$s'.
In general, we find that if one of the chiral sectors consists of three (four)
wave functions, we have a single non-trivial solution (a pair of non-trivial
solutions) of (\ref{eq:c_boundary}).  
In Table \ref{table:ABS}, we summarize the number $N_0^{(\pm)}$ of
dispersionless ABS in each sector obtained in this manner.
Comparing with the topological number $W(k_y)$ given in (\ref{eq:W}) and
(\ref{eq:W2}),
we find that our result agrees with the index theorem (\ref{eq:index})
exactly. 

In a similar manner, we can also construct the zero energy ABSs on the
surface of the two-dimensional semi-infinite $d_{xy}+p$ wave Rashba
superconductor on $x<0$. In comparison with the ABS of the semi-infinite
superconductor on $x>0$, the chirality of the ABS is found to be
opposite. Thus we find that the index theorem (\ref{eq:index2}) holds in
this case.

We notice here that the zero energy bound state for $k_2>|k_{y}|>k_{1}$ 
is a Majorana edge mode.
The wave function for the zero energy edge state 
$\Psi_{m}(k_{y})$ can be written as  
 $^{T}\Psi_{m}(k_{y}) =
 (u_{1}(k_{y}),u_{2}(k_{y}),v_{1}(k_{y}),v_{2}(k_{y}))$ where
\begin{eqnarray}
u_{1}(k_{y})=-i\sigma v_{2}(k_{y})
=\frac{(\alpha f_{1}-\beta_{1}f_{2})\exp(ik_{y}y-i\frac{\pi}{4})}
{\sqrt{\sigma \alpha}} 
\\ 
u_{2}(k_{y})=i\sigma v_{1}(k_{y})
=\frac{(f_{1}+\beta_{2}f_{2})\exp(ik_{y}y-i\frac{\pi}{4})}
{\sqrt{\sigma \alpha}}
\end{eqnarray}
with $\alpha=(k_{y}-\sqrt{k_{y}^{2}-k_{1}^{2}})/k_{1}$, 
$\beta_{1}=(\alpha k_{y}/k_{2} +1)$, 
$\beta_{2}=(\alpha + k_{y}/k_{2})$ and $\sigma={\rm sgn}(k_{y})$.  
The functions $f_{1}$ and $f_{2}$ decays exponentially as a function of $x$ and are even function of $k_{y}$. 
The Bogoliubov quasiparticle creation operator for this state is constructed in the usual way as $\gamma^\dagger(k_{y}) 
= u_1(k_y) c_\uparrow^\dagger(k_y) + u_2(k_y) c_\downarrow^\dagger(k_y) 
+ v_1(k_y) c_\uparrow(-k_y) + v_2(k_y) c_\downarrow(-k_y)$. 
Since $u_{1}(k_{y})=v_{1}^{*}(-k_{y})$ and 
$u_{2}(k_{y})=v_{2}^{*}(-k_{y})$ are satisfied, it is possible to verify that 
$\gamma^{\dagger}(k_{y})=\gamma(-k_{y})$.  
This means the generation of Majorana bound state at the edge for 
$k_2>|k_{y}| > k_{1}$. 
A similar Majorana bound state also 
appears for $\Delta_{s}>\Delta_{t}$ and $k_2>|k_{y}| > k_{1}$. \par

Unlike Majorana Fermions studied before, 
the present single branch of Majorana bound
state is realized with time reversal symmetry. 
The TRI Majorana edge mode has the following three characteristics.
a) It has a unique flat dispersion: To be consistent with the
time-reversal invariance, the single branch of zero mode should be
symmetric under $k_y\rightarrow -k_y$.
Therefore, by taking into account the particle-hole symmetry as well,
the flat dispersion is required. 
On the other hand, the conventional time-reversal breaking Majorana
has a linear dispersion.
b) The spin-orbit coupling is
necessary to obtain the TRI Majorana edge mode.
Without spin-orbit coupling, the TRI Majorana edge mode vanishes. 
c) The TRI Majorana edge mode is topologically stable under small
deformations of the Hamiltonian (\ref{eq:BdG4x4}).
The topological stability is ensured by the topological invariant $W(k_y)$.

\begin{table}
\begin{center} 
\begin{tabular}[t]{|c|c|c|c|c|c|c|c|}
\hline
\hline
\multicolumn{6}{c}{(a) $\Delta_s>\Delta_t$} \\ 
 \hline
$k_y$ &$\Gamma=1$ sector & $\Gamma=-1$ sector 
& $N_0^{(+)}$& $N_0^{(-)}$& $N_0^{(+)}-N_0^{(-)}
$ & $W(k_y)$ & $(-1)^{\nu(k_y)}$ \\ 
\hline 
$k_y>k_2$& $\psi_1^{+}$, $\psi_2^{+}$ &$\psi_1^{-}$, $\psi_2^{-}$ 
&0 &0 &0 &0 &1\\ 
$k_2>k_y>k_1$
& $\psi_1^{+}$ & $\psi_1^{-}$, $\psi_2^{+}$, $\psi_2^{-}$ 
&0 &1& -1 & 1 & -1\\
 $k_1>k_y>0$
& - &$\psi_1^{+}$, $\psi_1^{-}$, $\psi_2^{+}$, $\psi_2^{-}$ 
&0 &2 &-2 & 2 & 1\\
 $0>k_y>-k_1$ 
&$\psi_1^{+}$, $\psi_1^{-}$, $\psi_2^{+}$, $\psi_2^{-}$ & -
& 2 & 0 & 2 &-2 & 1\\
 $-k_1>k_y>-k_2$ 
&$\psi_1^{-}$, $\psi_2^{+}$, $\psi_2^{-}$ & $\psi_1^{+}$
&1& 0 & 1 &-1&-1\\
 $-k_2>k_y$ 
&$\psi_1^{-}$, $\psi_2^{-}$ & $\psi_1^{+}$, $\psi_2^{+}$
&0& 0 & 0 &0&1\\
\hline
\multicolumn{6}{c}{}\\
\multicolumn{6}{c}{(b) $\Delta_t>\Delta_s$} \\ 
 \hline
$k_y$ &$\Gamma=1$ sector & $\Gamma=-1$ sector 
&$N_0^{(+)}$ & $N_0^{(-)}$& $N_0^{(+)}-N_0^{(-)}$ & $W(k_y)$ & $(-1)^{\nu(k_y)}$ \\ 
\hline 
$k_y>k_2$
&$\psi_1^{+}$, $\psi_2^{-}$& $\psi_1^{-}$, $\psi_2^{+}$ 
&0 &0 &0 &0 & 1\\ 
$k_2>k_y>k_1$
&$\psi_1^{+}$, $\psi_2^{+}$, $\psi_2^{-}$ & $\psi_1^{-}$ 
& 1 & 0 &1 &-1&-1\\
 $k_1>k_y>0$
&$\psi_2^{+}$, $\psi_2^{-}$ & $\psi_1^{+}$, $\psi_1^{-}$
& 0 & 0 &0 &0&1\\
 $0>k_y>-k_1$ 
&$\psi_1^{+}$, $\psi_1^{-}$ & $\psi_2^{+}$, $\psi_2^{-}$
& 0 & 0 & 0 & 0 & 1\\
 $-k_1>k_y>-k_2$ 
&$\psi_1^{-}$ & $\psi_1^{+}$, $\psi_2^{+}$, $\psi_2^{-}$
& 0& 1 & -1 &1& -1\\
 $-k_2>k_y$ 
&$\psi_1^{-}$, $\psi_2^{+}$ & $\psi_1^{+}$, $\psi_2^{-}$
& 0& 0 & 0 &0&1\\
\hline
\multicolumn{6}{c}{}\\
\hline
\hline
 \end{tabular} 
\end{center}
\caption{The zero energy ABSs of the semi-infinite $d_{xy}+p$-wave Rashba
 superconductor on $x>0$. 
In (a), we consider the spin-singlet dominant
 Cooper pair, $\Delta_s>\Delta_t$, and in (b) the spin-triplet dominant
 one, $\Delta_t>\Delta_s$.
In the second and third columns, the chirality of each wave function
 $\psi_i^{\pm}$ in
 (\ref{wavefunction}) is summarized.
As is explained in the text, 
the numbers $N_0^{(\pm)}$ of the zero energy ABSs with $\Gamma=\pm 1$
are determined from the chirality.
For comparison, we also show the topological number $W(k_y)$ given in
 (\ref{eq:W}) and (\ref{eq:W2}). In both of the case (a) and (b), the
 results agree with the index theorem (\ref{eq:index}) excellently. }
\label{table:ABS}
\end{table}

\subsubsection{$Z_2$ topological number and anisotropic response to the Zeeman magnetic field}
In this section, we would like to point out that 
the time-reversal invariant Majorana fermion is also characterized
by another topological number taking a ${\bm Z}_2$ value.
The ${\bm Z}_2$ topological number explains the anisotropic response of
the Majorana fermion to the Zeeman magnetic field found in
Ref.\cite{TMYYS10,YSTY10}. 

For a $d_{xy}+p$-wave superconductor, the BdG Hamiltonian has the
following symmetry,
\begin{eqnarray}
{\cal C}'{\cal H}(k_x,k_y){\cal C'}^{-1}=-{\cal H}^{*}(-k_x,k_y),
\quad
{\cal C}'=
\left(
\begin{array}{cc}
0 & -i\sigma_y\\
i\sigma_y & 0
\end{array}
\right). 
\label{eq:particle-hole2}
\end{eqnarray}
Regarding $k_y$ as a parameter,  we can consider this as
the particle-hole symmetry in one dimension.
Thus the ${\bm Z}_2$ topological number can be introduced in a similar
manner as shown in Ref.\cite{Sato10}. 
As seen from  Appendix \ref{sec:appendixc}, the ${\bm Z}_2$ topological
number is given by $(-1)^{\nu(k_y)}$ with
\begin{eqnarray}
\nu(k_y)=\frac{1}{\pi}\int_0^{\infty}dk_x A_x({\bm k}). 
\end{eqnarray}
Here $A_x({\bm k})$ is the ``gauge field'' defined by the bulk wave
function $|u_n({\bm k})\rangle$,
\begin{eqnarray}
A_x({\bm k})=i\sum_{n}\langle u_n({\bm k})|\partial_{k_x} u_n({\bm k})\rangle.
\end{eqnarray}
Then the integral can be evaluated as 
\begin{eqnarray}
(-1)^{\nu(k_y)}=
{\rm sgn}\left[(k_y^2/2m-\mu)^2-(\lambda k_y)^2\right]. 
\end{eqnarray}
(See Eq.(\ref{eq:z21}).)
From this, it is found that the ${\bm Z}_2$ topological number is
non-trivial, i.e. $(-1)^{\nu(k_y)}=-1$, in the region $k_1<|k_y|<k_2$
where the dispersionless Majorana fermion exists. Therefore, in addition
to $W(k_y)$,
the ${\bm Z}_2$ topological number $(-1)^{\nu(k_y)}$ also ensures the
topological stability of the dispersionless Majorana fermion.

The merit of the ${\bm Z}_2$ topological invariant is evident if we
apply a Zeeman magnetic field.
In the presence of Zeeman magnetite field, the chiral symmetry
(\ref{eq:chiral}) is broken since it is a combination of the
particle-hole symmetry and the time-reversal symmetry, and the Zeeman
magnetic field breaks the time-reversal invariance.
This implies that the topological protection of the gapless state
discussed in Sec.\ref{sec:chirality} does not work in the presence of
a Zeeman magnetic field.  
In addition, the topological number $W(k_y)$ cannot be defined
without the chiral symmetry.

On the other hand, the symmetry (\ref{eq:particle-hole2}) survives even
in the presence of the Zeeman magnetic field if we
apply it in the $y$-direction.
Indeed, for the $d_{xy}+p$-wave superconductor, we can show that the BdG
Hamiltonian
\begin{eqnarray}
{\cal H}({\bm k})=
\left(
\begin{array}{cc}
\varepsilon({\bm k})+{\bm g}({\bm k})\cdot{\bm \sigma}-\mu_{\rm B}H_y\sigma_y
 &\hat{\Delta}({\bm k}) \\
\hat{\Delta}^{\dagger}({\bm k}) &
-\varepsilon({\bm k})+{\bm g}({\bm k})\cdot{\bm \sigma}^*+\mu_{\rm
B}H_y\sigma_y^*
\end{array}
\right) 
\end{eqnarray}
in the presence of the Zeeman magnetic field $H_y$ in the $y$-direction
satisfies (\ref{eq:particle-hole2}).
In a similar manner, we can introduce the ${\bm Z}_2$ invariant,
and we obtain 
\begin{eqnarray}
(-1)^{\nu(k_y)}={\rm sgn}\left[(k_y^2/2m-\mu)^2-(\lambda k_y)^2-(\mu_{\rm
			  B}H_y)^2\right]. 
\end{eqnarray}
(See.Eq. (\ref{eq:z22}).)
Thus, the ${\bm Z}_2$ number remains
non-trivial in the region where the dispersionless Majorana fermion
exists. 
In fact, the region of $k_y$ in which the ${\bm Z}_2$ topological number
is non-trivial is extended in the presence of $H_y$, which suggests that
the magnetic field in the $y$-direction stabilizes the dispersionless
Majorana fermion.
This is a peculiar property that is not seen in other dispersionless
ABSs. 
From the
bulk-edge correspondence, we can conclude that the time-reversal invariant
dispersionless Majorana fermion survives even in the presence of the
Zeeman magnetic field if we apply it in  the $y$-direction.

Here we notice that the ${\bm Z}_2$ number
is very sensitive to the direction of the Zeeman magnetic field we
considered. While it is well defined in the presence of a magnetic
field in the $y$-direction, it becomes meaningless if we apply a
magnetic field in the other directions since in the latter case
the symmetry (\ref{eq:particle-hole2}) is broken.
Therefore, the dispersionless Majorana fermion is also very sensitive to
the direction of a magnetic field, which is consistent with the
surface density of states calculated in Ref.\cite{TMYYS10,YSTY10}.

\section{Index theorems: A proof of bulk-edge correspondence}
\label{sec:indextheorems}
Finally, we would like to prove the index theorems, (\ref{eq:index}),
({\ref{eq:index2}}), (\ref{eq:index2x2}) and (\ref{eq:index2x2_2}).

\subsection{Strategy}
\label{sec:strategy}
Before going into the details, we would like to outline our
strategy.
In order to prove the index theorems, we introduce an adiabatic parameter
$a$ in the Plank constant
$\hbar$,
\begin{eqnarray}
\hbar =a \hbar_0, 
\end{eqnarray}
where $\hbar_0$ denotes an original value of the Plank constant. When 
$a\rightarrow 0$, we have a classical limit of $\hbar\rightarrow 0$, and 
when $a=1$, the system returns to the original.

We will first consider the semi-classical limit $\hbar<<1$ ($a\ll 1$). 
Using the WKB approximation, we prove
the index theorems in the following sections. ((\ref{eq:2x2indexwkb}) and
(\ref{eq:2x2indexwkb2}) in Sec.\ref{sec:index2x2}, and (\ref{eq:indexwkb})
and (\ref{eq:index2wkb}) in Sec.\ref{sec:indexgeneral}, respectively.)

Then we adiabatically increase $a\rightarrow 1$ until $\hbar$
goes back to the original value $\hbar_0$.
From the argument in Sec.\ref{sec:chirality}, we notice here that
$N_0^{(+)}-N_0^{(-)}$ cannot change in this process. Indeed, 
in order to change $N_0^{(+)}-N_0^{(-)}$, we need a continuum mode which closes
the gap at the corresponding ${\bm k}_{\parallel}$, but we have a gap in
the bulk from
the assumption.  
A new state might appear near the boundary, but 
it should have a discrete spectrum for a fixed ${\bm k}_{\parallel}$
since it is localized near the boundary.
Thus, the boundary state cannot change the value of
$N_0^{(+)}-N_0^{(-)}$ as well. 
Therefore, the index theorems in the
semi-classical limit remain to hold even when $\hbar$ goes back to the
original value.

We would like to emphasis here that our strategy adapted here provides a general
framework to prove the bulk-edge correspondence:
For any topological state (of non-interacting systems), 
a mismatch of the topological number on the boundary results in a gap closing
point near the boundary, in the classical limit ($a=0$). 
Then, by using the WKB quantization of the gap closing
point, an gapless edge state is obtained semi-classically $(a\ll 1)$.
From the existence of a gap in the bulk,  
the edge state is stable against an adiabatic
change of $a$,  thus we can conclude the existence of the edge state in
the original theory $(a=1)$.

\subsection{2 $\times$ 2 BdG Hamiltonian}
\label{sec:index2x2}
In this subsection, we prove  (\ref{eq:index2x2}) and (\ref{eq:index2x2_2}).

\begin{figure}[h]
\begin{center}
\includegraphics[width=9cm]{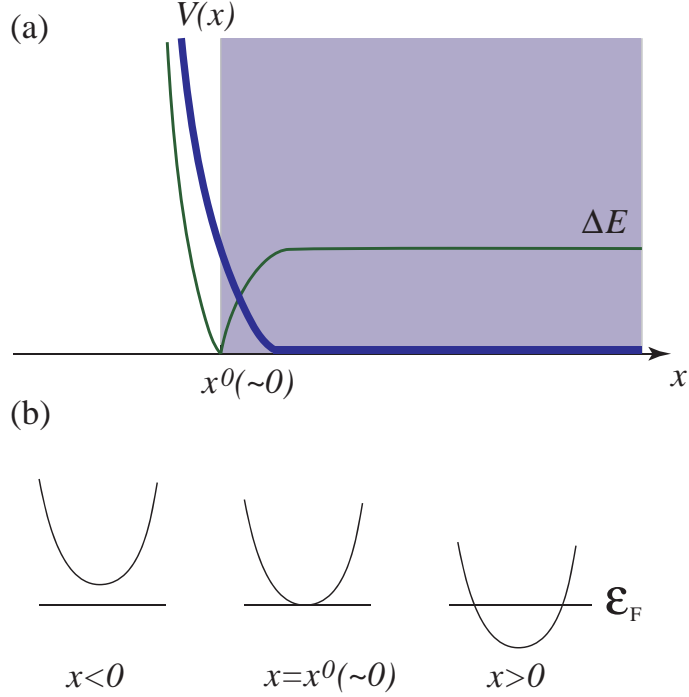}
\caption{(color online) (a) A semi-infinite superconductor on $x>0$. The
 thick curve denotes the confining potential $V(x)$, and the thin curve
 a gap of the system in the classical limit. If Eq. (\ref{eq:gapcloseedge}) 
 is satisfied, the gap $\Delta E$ in the classical limit closes near the
 edge at $x=0$.
(b) The corresponding normal dispersion of electrons.
Inside the superconductor $(x>0)$, the system supports the Fermi
 surface, but outside the superconductor $(x<0)$, the system becomes a band
 insulator.
Correspondingly, 
$\Delta E$ is a superconducting gap for $x>0$, and 
$\Delta E$ is a band gap ($\sim V(x)$) for $x<0$.}
\label{fig:confining_potential}
\end{center}
\end{figure}

To make a boundary of a superconductor, let us introduce a confining
potential $V(x)$ illustrated in Fig.\ref{fig:confining_potential},
\begin{eqnarray}
{\cal H}_{2\times 2}({\bm k})\rightarrow {\cal H}_{2\times 2}({\bm k},x)
=\left(
\begin{array}{cc}
\varepsilon({\bm k})+V(x) &\Delta({\bm k}) \\
\Delta({\bm k}) & -\varepsilon({\bm k})-V(x)
\end{array}
\right). 
\label{eq:2x2+c}
\end{eqnarray}
We assume that the confining potential $V(x)$ is steep enough near the edge. 
Now we will prove the index theorem (\ref{eq:index2x2}) for the BdG
Hamiltonian (\ref{eq:2x2+c}).

First, we use the WKB approximation to count the zero energy ABSs.
For this purpose, it is convenient to solve ${\cal H}_{2\times 2}({\bm
k},x)^2 v=E^2 v$ instead of the original BdG equation ${\cal H}_{2\times
2}({\bm k},x)u=Eu$. The zero energy states for these two equations are
the same as is shown in
Appendix.\ref{sec:appendixa}.

In the presence of the confining potential, ${\cal H}_{2\times 2}({\bm
k},x)^2$ is given by 
\begin{eqnarray}
{\cal H}_{2\times 2}({\bm k},x)^2= 
\left(
\begin{array}{cc}
(\varepsilon({\bm k})+V(x))^2+\Delta({\bm k}) & [V(x),\Delta({\bm k})]\\
-[V(x),\Delta({\bm k})] & (\varepsilon({\bm k})+V(x))^2+\Delta({\bm k})^2
\end{array}
\right).
\end{eqnarray} 
Here $k_x$ should be treated as $-i\hbar\partial_x$ in
the above.
Then noting that
\begin{eqnarray}
[V(x),\Delta({\bm k})]= 
\frac{\partial V}{\partial x}[x,k_x]
\frac{\partial \Delta}{\partial k_x}+O(\hbar^2)
=
i\hbar 
\frac{\partial V}{\partial x}
\frac{\partial \Delta}{\partial k_x}+O(\hbar^2). 
\end{eqnarray}
we have
\begin{eqnarray}
{\cal H}_{2\times 2}({\bm k},x)^2=[(\varepsilon({\bm
 k})+V(x))^2+\Delta({\bm k})^2]{\bm 1}_{2\times 2}-\hbar\frac{\partial
 V}{\partial x}\frac{\partial \Delta}{\partial k_x}\sigma_y +O(\hbar^2).
\label{eq:2x2wkb}
\end{eqnarray}

In the classical limit ($\hbar\rightarrow 0$) of the WKB
approximation, the energy spectrum is given by the first term of the
right hand side of (\ref{eq:2x2wkb}),
\begin{eqnarray}
E^2=(\varepsilon({\bm k})+V(x))^2+\Delta({\bm k})^2, 
\end{eqnarray} 
where ${\bm k}$ and $x$ should be considered as $c$-numbers.
For a fixed ${\bm k}_{\parallel},$
we have a zero energy state if $(k_x, x)=(k_x^0,x^0)$ satisfies 
\begin{eqnarray}
\varepsilon(k_x^0, {\bm k}_{\parallel})+V(x^0)=0,
\quad
\Delta(k_x^0,{\bm k}_{\parallel})=0.
\label{eq:gapcloseedge}
\end{eqnarray}

Let us now take into account the leading order correction of $\hbar$. 
Near the zero $(k_x^0,x^0)$ satisfying (\ref{eq:gapcloseedge}), we
obtain 
\begin{eqnarray}
&&\varepsilon({\bm k})+V(x)=\left(
\frac{\partial V(x)}{\partial x}\right)_{x=x^0}(x-x^0)+\cdots, 
\nonumber\\
&&\Delta({\bm k})=\left(\frac{\partial \Delta({\bm k})}{\partial k_x}
\right)_{k_x=k_x^0}
(k_x-k_x^0)
+\cdots, 
\end{eqnarray}
thus the first term of (\ref{eq:2x2wkb}) is evaluated as a harmonic oscillator,
\begin{eqnarray}
[(\varepsilon({\bm k})+V(x))^2+\Delta({\bm k})^2]= 
\left(\frac{\partial V}{\partial x}\right)_{x=x^0}^2(x-x^0)^2
+\left(\frac{\partial \Delta}{\partial k_x}\right)_{k_x=k_x^0}^2
(k_x-k_x^0)^2+\cdots.
\end{eqnarray}
From the Bohr-Sommerfeld quantization condition, 
it leads to
\begin{eqnarray}
[(\varepsilon({\bm k})+V(x))^2+\Delta({\bm k})^2]= 
2\hbar\left|\frac{\partial V}{\partial x}\right|_{x=x^0}
\left|\frac{\partial \Delta}{\partial k_x}\right|_{k_x=k_x^0}
\left(n+\frac{1}{2}\right)+O(\hbar^2),
\label{eq:wkb1}
\end{eqnarray}
with $n=0,1,2,\cdots$. 
The second term in (\ref{eq:2x2wkb}) is evaluated as 
the expectation value for the WKB wave function,
\begin{eqnarray}
-\hbar\left\langle\frac{\partial V}{\partial x}
\frac{\partial \Delta}{\partial k_x}
\right\rangle_0 \sigma_y 
=-\hbar\left(\frac{\partial V}{\partial x}\right)_{x=x^0} 
\left(\frac{\partial \Delta}{\partial k_x}\right)_{k_x=k_x^0}\sigma_y,
\label{eq:wkb2}
\end{eqnarray}
where we have used the fact that the WKB wave function has a sharp peak
at $x=x^0$ and $k_x=k_x^0$.
Combining (\ref{eq:wkb1}) and (\ref{eq:wkb2}), we obtain 
\begin{eqnarray}
{\cal H}_{2\times 2}({\bm k},x)^2=  
2\hbar\left|\frac{\partial V}{\partial x}\right|_{x=x^0}
\left|\frac{\partial \Delta}{\partial k_x}\right|_{k_x=k_x^0}
\left(n+\frac{1}{2}\right){\bm 1}_{2\times 2}
-\hbar\left(\frac{\partial V}{\partial x}\right)_{x=x^0} 
\left(\frac{\partial \Delta}{\partial k_x}\right)_{k_x=k_x^0}\sigma_y.
\label{eq:hamiltonianwkb}
\end{eqnarray}
For a semi-infinite superconductor on $x>0$, the confining potential
satisfies $(\partial V/\partial x)_{x=x^0}<0$. 
Therefore Eq.(\ref{eq:hamiltonianwkb}) yields a zero energy solution with
chirality $\gamma$($\equiv$ an eigenvalue of $\sigma_y$) given by
\begin{eqnarray}
\gamma=-{\rm sgn}\left(\frac{\partial \Delta}{\partial
		  k_x}\right)_{k_x=k_x^0}.  
\end{eqnarray}
Counting the zero energy states from all $(k_x^0,x^0)$s satisfying
(\ref{eq:gapcloseedge}),
we obtain 
\begin{eqnarray}
n_0^{(+)}-n_0^{(-)}=-\sum  
{\rm sgn}\left(\frac{\partial \Delta}{\partial
		  k_x}\right)_{k_x=k_x^0},
\label{eq:2x2indexwkb}
\end{eqnarray}
where the summation is taken for $(k_x^0,x^0)$ satisfying
(\ref{eq:gapcloseedge}).

Here we can show that (\ref{eq:2x2indexwkb}) is nothing but the index theorem
(\ref{eq:index2x2}):
First, we note that if and only if $\varepsilon({\bm k})<0$, there
exists a $x$ satisfying $\varepsilon({\bm k})+V(x)=0$.
Thus we can rewrite (\ref{eq:2x2indexwkb}) as
\begin{eqnarray}
n_0^{(+)}-n_0^{(-)}=-\sum_{\Delta(k_x^0,{\bm k}_{\parallel})=0,
 \varepsilon(k_x^0, {\bm k}_{\parallel})<0}  
{\rm sgn}\left(\frac{\partial \Delta}{\partial
		  k_x}\right)_{k_x=k_x^0},
\end{eqnarray}
As is shown in Sec.\ref{sec:appendixb1}, the right hand side of the
above equation is equal to $-w({\bm k}_{\parallel})$, thus we have the
index theorem
\begin{eqnarray}
n_0^{(-)}-n_0^{(+)}=w({\bm k}_{\parallel}).
\label{eq:2x2indexwkb2}
\end{eqnarray}
in the leading order in the WKB approximation.

While we have derived Eq. (\ref{eq:2x2indexwkb2}) in the
leading order of the WKB approximation, we can find that no
further correction exists:
As was discussed in Sec.\ref{sec:chirality}, any small perturbation
preserving the chiral symmetry cannot change $n_0^{(+)}-n_0^{(-)}$
unless the bulk energy gap closes. 
Therefore, a higher order correction might change each of $n_0^{(+)}$ or
$n_0^{(-)}$, but it cannot change their difference $n_0^{(+)}-n_0^{(-)}$.  
This means that Eq.(\ref{eq:2x2indexwkb2}) is exact.
(See also the discussions in Sec.\ref{sec:strategy}.)

In a similar manner, we can prove the index theorem (\ref{eq:index2x2_2}) for 
a semi-infinite superconductor on $x<0$. 
Following the same argument above, we obtain (\ref{eq:hamiltonianwkb}) again.
However, in comparison with the semi-infinite superconductor on $x>0$,
the sign of $(\partial V/\partial x)_{x=x^0} $ is reversed for a semi-infinite
superconductor on $x<0$, so the chirality $\gamma$ of the zero energy
state is also reversed
\begin{eqnarray}
\gamma={\rm sgn}\left(\frac{\partial \Delta}{\partial
		  k_x}\right)_{k_x=k_x^0}.   
\end{eqnarray}
Thus we obtain 
\begin{eqnarray}
n_0^{(+)}-n_0^{(-)}=\sum  
{\rm sgn}\left(\frac{\partial \Delta}{\partial
		  k_x}\right)_{k_x=k_x^0},
\end{eqnarray}
where the summation is taken for $(k_x^0,x^0)$ satisfying
(\ref{eq:gapcloseedge}).
This leads to 
\begin{eqnarray}
n_0^{(+)}-n_0^{(-)}=w({\bm k}_{\parallel}),
\end{eqnarray}
which is exact again for the same reason above.  

\subsection{General case}
\label{sec:indexgeneral}

Now, we prove the index theorems (\ref{eq:index}) and
(\ref{eq:index2}) for general time-reversal invariant BdG Hamiltonian. 

First let us consider the index theorem (\ref{eq:index}) for a
semi-infinite superconductor on $x>0$.
To realize the semi-infinite superconductor, we introduce a confining
potential $V(x)$ in a similar manner as Sec.\ref{sec:index2x2} (See also
Fig.\ref{fig:confining_potential}),
\begin{eqnarray}
{\cal H}({\bm k})\rightarrow 
{\cal H}({\bm k},x)=
\left(
\begin{array}{cc}
\hat{\cal E}({\bm k})_{\alpha\alpha'}+V(x)\delta_{\alpha\alpha'} 
& \hat{\Delta}({\bm k})_{\alpha\alpha'}\\
\hat{\Delta}^{\dagger}({\bm k})_{\alpha\alpha'} 
& -\hat{\cal E}^T(-{\bm k})_{\alpha\alpha'}-V(x)\delta_{\alpha\alpha'}
\end{array}
\right),
\end{eqnarray}
where $V(x)=0$ inside the superconductor $(x>0)$. We suppose a sharp
edge where the 
confining potential $V(x)$ is very steep near the edge $(x=0)$.

In the classical limit $\hbar\rightarrow 0$, $k_x$ and $x$ commute with
each other, thus we can treat them as $c$-numbers.
Then, considering $x$ as a parameter, we define the winding
number $W({\bm k}_{\parallel}, x)$ in a manner similar to
Eq.(\ref{eq:windingnumber}),
\begin{eqnarray}
W({\bm k}_{\parallel}, x)&=&-\frac{1}{4\pi i}\int d{\bm k}_{\perp}
{\rm tr}\left[\Gamma {\cal H}^{-1}({\bm k},x)\partial_{{\bm
	 k}_{\perp}}{\cal H}({\bm k},x)\right] 
\nonumber\\
&=&\frac{1}{2\pi}{\rm Im }\left[
\int d{\bm k}_{\perp}\partial_{{\bm k}_{\perp}}\ln \det \hat{q}({\bm k},x)
\right],
\end{eqnarray}
where $\hat{q}({\bm k}, x)=i[\hat{\cal E}({\bm
k})+V(x)]U^{\dagger}-\hat{\Delta}({\bm k})$ with $U$ in
Eq.(\ref{eq:time-reversal}),
and ${\bm k}_{\perp}=k_x$.
The line integral in the above is defined in the same way as
Eq.(\ref{eq:windingnumber}). 

Here we find that 
$W({\bm k}_{\parallel},x)$ is identical to $W({\bm k}_{\parallel})$
inside the superconductor $x>0$,
\begin{eqnarray}
W({\bm k}_{\parallel},x)=W({\bm k}_{\parallel}), 
\quad \mbox{for $x>0$, ($x\not\sim 0$)}, 
\end{eqnarray}
since $V(x)=0$ there.
We also find that $W({\bm k}_{\perp},x)$ becomes zero outside the
superconductor, 
\begin{eqnarray}
W({\bm k}_{\parallel},x)=0 
\quad \mbox{for $x<0$, ($x\not\sim 0$)}, 
\end{eqnarray}
since we have a vacuum there.
Therefore, if the bulk superconductor supports a non-zero winding number
$W({\bm k}_{\parallel})\neq 0$,  
then the value of $W({\bm k}_{\parallel},x)$ must be changed near the
edge $(x=0)$.
This immediately implies that when $W({\bm k}_{\parallel})\neq
0$, a gap of the system closes near the edge in the classical limit:
Since $W({\bm k}_{\parallel},x)$ is a topological number, it changes
only when the line integral is ill-defined. 
Then this occurs only when
$\det \hat{q}({\bm k},x)=0$, which means a gap closing of
the system in the classical limit.

To examine how $W({\bm k}_{\parallel},x)$ changes, we diagonalize $\hat{q}({\bm
k},x)$ as
\begin{eqnarray}
\hat{q}({\bm k},x)
=A({\bm k},x)\Lambda({\bm k},x)B^{\dagger}({\bm k},x),
\nonumber\\
\hat{q}^{\dagger}({\bm k},x)
=B({\bm k},x)\Lambda^*({\bm k},x)A^{\dagger}({\bm k},x), 
\label{eq:qdiag}
\end{eqnarray}
where $\Lambda({\bm k},x)={\rm diag}(\lambda_1,\lambda_2, \cdots,\lambda_N)$
with $\lambda_i$ the eigenvalue of $\hat{q}({\bm k},x)$, and  
$A({\bm k},x)$ and $B({\bm k},x)$ are $N\times N$ matrices satisfying
\begin{eqnarray}
A({\bm k},x)B^{\dagger}({\bm k},x)={\bm 1}_{N\times N}. 
\end{eqnarray}
(Here we assume that $\hat{q}({\bm k},x)$ is an $N\times N$ matrix. See
Appendix \ref{sec:diagq} for details.) 
Since Eq.(\ref{eq:qdiag}) yields that 
 $\det \hat{q}({\bm k},x)=\det\Lambda({\bm k},x)=\prod_{n}\lambda_n({\bm
 k},x)$, the winding number $W({\bm k},x)$ is recast into
\begin{eqnarray}
W({\bm k}_{\parallel},x)=\frac{1}{2\pi}{\rm Im}\left[
\sum_{n=1}^{N}
\int d{\bm k}_{\perp}\partial_{{\bm k}_{\perp}}\ln\lambda_n({\bm k},x)
\right]. 
\label{eq:windinglambda}
\end{eqnarray}
Thus the line integral of $W({\bm k}_{\parallel},x)$ is ill-defined,
when some of the eigenvalues $\lambda_n({\bm k},x)$ have a zero.

\begin{figure}[h]
\begin{center}
\includegraphics[width=5cm]{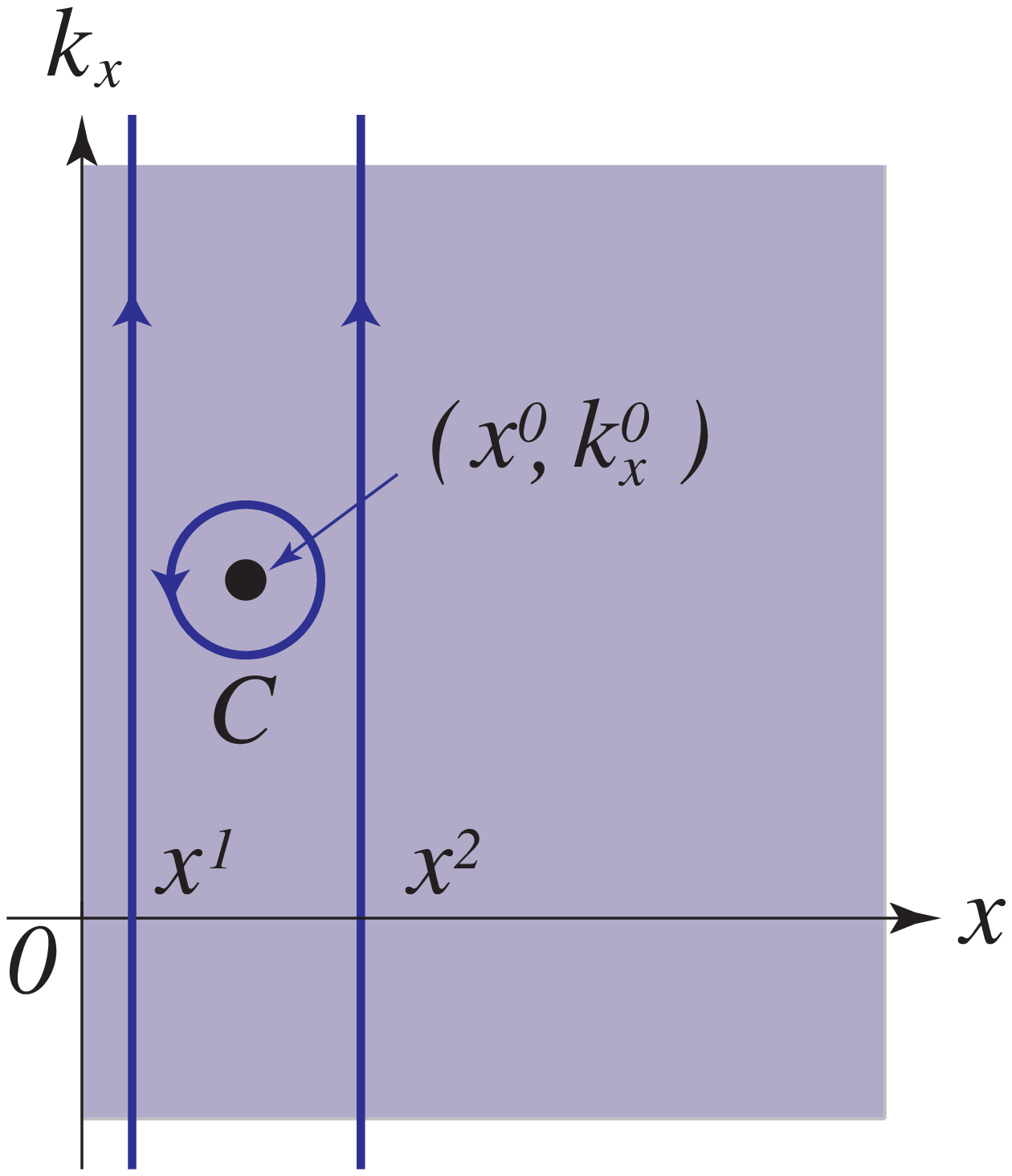}
\caption{(color online) A zero $(x^0,k_x^0)$ of $\lambda_n$ near the
 edge of a semi-infinite superconductor on $x>0$.}
\label{fig:contour}
\end{center}
\end{figure}

Now suppose that $\lambda_n({\bm k},x)$ has a zero at
$(x,k_x)=(x^0,k_x^0)$, 
\begin{eqnarray}
\lambda_n&=&\alpha e^{i\beta}(x-x^0)+\gamma e^{i\delta}(k_x-k_x^0)+\cdots,  
\nonumber\\
&=&\alpha e^{i\beta}\left[
(x-x^0)+(\gamma/\alpha) e^{i(\delta-\beta)}(k_x-k_x^0)
\right]
+\cdots,  
\end{eqnarray}
where $\alpha$, $\beta$, $\gamma$ and $\delta$ are real constants.
We notice here that $\alpha$ is very large since it is
proportional to $(\partial V/\partial x)_{x=x^0}$ and $V(x)$ is steep
near the edge.
Thus we can neglect the real part of
$(\gamma/\alpha)^{i(\delta-\beta)}(k_x-k_x^0)$, and obtain
\begin{eqnarray}
\lambda_n=\alpha e^{i\beta}\left[
(x-x^0)+i\eta(k_x-k_x^0)\right]+\cdots,
\end{eqnarray}
with $\eta=(\gamma/\alpha)\sin(\delta-\beta)$.
This zero changes the value of the winding number $W({\bm
k}_{\parallel},x)$ as follows.
Consider the line integrals near the zero along $x=x^i$ illustrated in
Fig \ref{fig:contour},
\begin{eqnarray}
\int_{x=x^i} dk_x \partial_{k_x}\ln \lambda_n({\bm k},x)
\quad 
(i=1,2)
\end{eqnarray}
with $x^1<x^0<x^2$.
The difference of these line integrals is evaluated as 
\begin{eqnarray}
&&\int_{x=x^2} dk_x \partial_{k_x}\ln \lambda_n({\bm k},x)
-\int_{x=x^1} dk_x \partial_{k_x}\ln \lambda_n({\bm k},x) 
\nonumber\\
&=&\oint_{C} dk_x \partial_{k_x}\ln \lambda_n({\bm k},x)
\nonumber\\
&=&2\pi i{\rm sgn}[\eta], 
\label{eq:diffline}
\end{eqnarray}
thus from Eq.(\ref{eq:windinglambda}), we have
\begin{eqnarray}
W({\bm k}_{\parallel},x^2)- W({\bm k}_{\parallel},x^1)={\rm sgn}[\eta].
\end{eqnarray}

In a similar manner, we can consider other zeros of the eigenvalues of
$\hat{q}({\bm k},x)$.
Then summing up all contribution of these zeros $(x^{0(i)},k_x^{0(i)})$
of $\lambda_{n(i)}$ $(i=1,\cdots,M)$, 
we have 
\begin{eqnarray}
W({\bm k}_{\parallel})=\sum_{i=1}^{M}{\rm sgn}[\eta_i], 
\label{eq:weta}
\end{eqnarray}
where $\eta_i$ is given by
\begin{eqnarray}
\lambda_{n(i)}=\alpha_i e^{i\beta_i}
\left[(x-x^{0(i)})+i\eta_i(k_x-k_x^{0(i)})\right]+\cdots, 
\end{eqnarray}
with real functions $\alpha_i$ and $\beta_i$. 
Since $\det {\cal H}({\bm k},x)$ around the zero $(x^{0(i)},k_x^{0(i)})$
is evaluated as
\begin{eqnarray}
\det {\cal H}({\bm k},x)&\propto & \det 
\hat{q}({\bm k},x)\hat{q}^{\dagger}({\bm
 k},x)
\nonumber\\
&=&
\prod_{n}|\lambda_n({\bm k},x)|^2
\nonumber\\
&\sim& [(x-x^{0(i)})^2+\eta_i^2(k_x-k_x^{0(i)})^2],
\end{eqnarray}
the classical spectrum around each zero is a harmonic oscillator. 

Let us now take into account the quantum corrections. 
In the leading order correction of $\hbar$, the classical spectrum of
a harmonic oscillator around each zero is quantized. 
To see whether the zero energy state survives or not after the
quantization, let us examine the BdG equation for the zero energy state,
\begin{eqnarray}
{\cal H}({\bm k},x)u_0=0,
\end{eqnarray}
which is equivalent to 
\begin{eqnarray}
\hat{q}({\bm k},x)\phi_0=0,  
\quad 
\hat{q}^{\dagger}({\bm k},x)\psi_0=0,
\label{eq:qzero}
\end{eqnarray}
with
\begin{eqnarray}
u_0=U_{\Gamma}
\left(
\begin{array}{c}
\psi_0 \\
\phi_0
\end{array}
\right). 
\label{eq:Uzero}
\end{eqnarray}
Substituting the following expression
\begin{eqnarray}
\phi_0=a_{n(i)}(k_x^{0(i)}, {\bm k}_{\parallel}, x^{0(i)})f(x),
\quad 
\psi_0=b_{n(i)}(k_x^{0(i)}, {\bm k}_{\parallel}, x^{0(i)})g(x),
\end{eqnarray}
into Eq.(\ref{eq:qzero}), we find that
\begin{eqnarray}
\left[
x-x^{0(i)}+i\eta_i(k_x-k_x^{0(i)})
\right]
f(x)=0,
\nonumber\\
\left[x-x^{0(i)}
-i\eta_i(k_x-k_x^{0(i)})
\right]g(x)=0,
\label{eq:zerowkb}
\end{eqnarray}
around the zero $(x^{0(i)},k_x^{0(i)})$.
Here $a_n({\bm k},x)$ and $b_n({\bm k},x)$ are the eigenfunctions of $q({\bm
k},x)$ and $q^{\dagger}({\bm k},x)$ in the classical limit, defined in
Appendix \ref{sec:diagq}.
Replacing $k_x$ with $-i\hbar\partial_x$, we can easily solve
Eq.(\ref{eq:zerowkb}), which reads
\begin{eqnarray}
f(x)=f_0 \exp \left[i\frac{k_x^{0(i)}}{\hbar}x\right]
\exp\left[\frac{-(x-x^{0(i)})^2}{2\eta_i\hbar}\right],
\nonumber\\
g(x)=g_0 \exp \left[i\frac{k_x^{0(i)}}{\hbar}x\right]
\exp\left[\frac{(x-x^{0(i)})^2}{2\eta_i\hbar}\right],
\end{eqnarray}
where $f_0$ and $g_0$ are constants. Since the normalization
condition requires that $f_0$ is zero if $\eta_i<0$, and $g_0$ is zero
if $\eta_i>0$, the zero energy state has $\Gamma=1$ for $\eta_i<0$ and
$\Gamma=-1$ for $\eta_i>0$, respectively.
(We have used Eq.(\ref{eq:gammadiag}) when applying $\Gamma$ to $u_0$ in
Eq.(\ref{eq:Uzero}).) 
In other words, in the leading order of the WKB quantization, there
exists a zero energy state for each zero $(x^{0(i)},k_x^{0(i)})$, and its
chirality $\Gamma$ is given by 
\begin{eqnarray}
\Gamma=-{\rm sgn}[\eta_i].
\label{eq:gammaeta}
\end{eqnarray}
From (\ref{eq:weta}), this leads to 
\begin{eqnarray}
W({\bm k}_{\parallel})=N_0^{(-)}-N_0^{(+)}. 
\label{eq:indexwkb}
\end{eqnarray} 
From the same argument as in Sec.\ref{sec:index2x2}, we can conclude
that no higher order correction of $\hbar$ exists in
Eq.(\ref{eq:indexwkb}) again.
Then using the argument in Sec.\ref{sec:strategy}, we conclude that
Eq.(\ref{eq:indexwkb}) is an exact result.

\begin{figure}[h]
\begin{center}
\includegraphics[width=5cm]{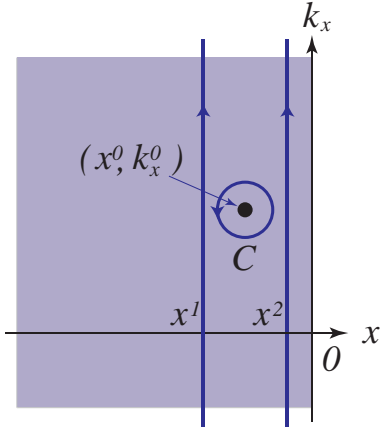}
\caption{(color online) A zero $(x^0,k_x^0)$ of $\lambda_n$ near the
 edge of a semi-infinite superconductor on $x<0$.}
\label{fig:contour2}
\end{center}
\end{figure}

In a similar manner, we can derive (\ref{eq:index2}) for a semi-infinite
superconductor on $x<0$.
As illustrated in Fig.\ref{fig:contour2}, for a zero
$(x^0,k_x^0)$ of $\lambda_n$ near the edge $x=0$, we have again
Eq.(\ref{eq:diffline}),
\begin{eqnarray}
\int_{x=x^2}dk_x\partial_{k_x}\ln\lambda_n 
-\int_{x=x^1}dk_x\partial_{k_x}\ln\lambda_n 
=2\pi i{\rm sgn}[\eta],
\end{eqnarray}
with $x^1<x^0<x^2$, but in contrast to the previous case, 
the line integral along $x=x^2$ is located at an outer region of the
superconductor. Thus the sign of the right hand side of Eq.(\ref{eq:weta})
is reversed, 
\begin{eqnarray}
W({\bm k}_{\parallel})=-\sum_i {\rm sgn }[\eta_i],
\end{eqnarray} 
and the following index theorem is obtained
\begin{eqnarray}
W({\bm k}_{\parallel})=N_0^{(+)}-N_0^{(-)}. 
\label{eq:index2wkb}
\end{eqnarray} 
From the argument in Sec.\ref{sec:strategy}, this result holds again
beyond the WKB approximation.

\section{Conclusion}
\label{sec:conclusion}

In this paper, we have discussed  the topology of the ABS with flat
dispersion at zero energy 
which appears on a boundary of time-reversal invariant
superconductors. 
Using the symmetry of the BdG Hamiltonian, we have introduced the
topological numbers $W({\bm k}_{\parallel})$ and $w({\bm
k}_{\parallel})$, and    
from the bulk-edge correspondence, topological criteria for the
dispersionless ABS  were obtained, which are summarized as 
the index theorems (\ref{eq:index}), (\ref{eq:index2}),
(\ref{eq:index2x2}) and (\ref{eq:index2x2_2}). 
We have shown that the index theorems
correctly predict the existence of the dispersionless ABSs.   
It has been also clarified that 
sign change of the gap function is directly related to
our topological criterion for superconductors preserving $S_z$ with
simple Fermi surfaces.
As concrete examples, we have also discussed (i) $d_{xy}$-wave superconductor, (ii) $p_x$-wave superconductor, and (iii) $d_{xy}+p$-wave
superconductor. 
All the examples confirms our topological criteria excellently.  
In the last part of this paper, we provide a general framework to
certify the bulk-edge correspondence, and we prove
the index theorems.

While we have considered only superconductors in this paper,
the index theorems we found can apply to other systems.
For example, graphene in the nearest neighbor tight-binding model has
chiral symmetry (or sublattice symmetry) similar to
(\ref{eq:chiral2x2}) \cite{RH02}, thus the
index theorems (\ref{eq:index2x2}) and (\ref{eq:index2x2_2}) hold.   
In general, our index theorems apply to any non-interacting Hamiltonian
with chiral symmetry.

\vspace{3ex}
\centerline{\bf Acknowledgment}

This work was supported in part by the Grant-in Aid for Scientific
Research from MEXT of Japan, "Topological Quantum Phenomena"
No. 22103005 (Y.T, M.S.), No. 20654030 (Y.T.) and No.22540383 (M.S.).

\appendix

\section{basic properties of system with chiral symmetry}
\label{sec:appendixa}

Here we summarize the properties of the system
with chiral symmetry.
Suppose that the Hamiltonian ${\cal H}$ has the chiral symmetry $\Gamma$
\begin{eqnarray}
\{{\cal H}, \Gamma\}=0, \quad \Gamma^2=1. 
\end{eqnarray}
Then consider the BdG equation
\begin{eqnarray}
{\cal H}|u_{E}\rangle =E|u_{E}\rangle 
\label{eq:chiralBdG}
\end{eqnarray}
First, we would like to show that there exists a one to one correspondence
between the solution of (\ref{eq:chiralBdG}) and that of the following
equation, 
\begin{eqnarray}
{\cal H}^2|v_{E^2}\rangle=E^2|v_{E^2}\rangle. 
\label{eq:chiralBdG2}
\end{eqnarray}
Since it is evident that the solution $|u_E\rangle$ of
(\ref{eq:chiralBdG}) satisfies (\ref{eq:chiralBdG2}), here we show only the
reverse. Namely, we can construct the solution $|u_E\rangle$ of
(\ref{eq:chiralBdG})  from the solution $|v_{E^2}\rangle$ of
(\ref{eq:chiralBdG2}):
For $E^2\neq 0$, if $({\cal H}+E)|v_{E^2}\rangle\neq 0$,
the solution $|u_E\rangle$ of
(\ref{eq:chiralBdG}) is obtained by
\begin{eqnarray}
|u_E\rangle =c({\cal H}+E)|v_{E^2}\rangle
\label{eq:uv}
\end{eqnarray}
with a normalization constant $c$.
Then if $({\cal H}+E)|v_{E^2}\rangle=0$, the solution is given by
\begin{eqnarray}
|u_E\rangle =\Gamma|v_{E^2}\rangle.
\label{eq:uv2}
\end{eqnarray}
It is easily found that $|u_E\rangle$ in (\ref{eq:uv}) and (\ref{eq:uv2}) has a
nonzero norm and satisfies (\ref{eq:chiralBdG}). 
For $E^2=0$, we have
\begin{eqnarray}
{\cal H}^2|v_0\rangle=0. 
\end{eqnarray}
Since this equation implies that
\begin{eqnarray}
\langle v_0|{\cal H}^2|v_0\rangle= || {\cal H}|v_0\rangle ||^2=0,
\end{eqnarray}
we obtain
\begin{eqnarray}
{\cal H}|v_0\rangle=0. 
\end{eqnarray}
Therefore, the zero energy solution $|u_0\rangle$ is obtained by
$|u_0\rangle=|v_0\rangle$.
Since there exists a one to one correspondence
between the solution of (\ref{eq:chiralBdG}) and that of
(\ref{eq:chiralBdG2}), we examine (\ref{eq:chiralBdG2}) instead of
(\ref{eq:chiralBdG}) in the following.

 Because ${\cal H}^2$ and $\Gamma$ commute with each other, we can take
 the solution of (\ref{eq:chiralBdG2}) as the eigen state of $\Gamma$,
 simultaneously. 
Then, it is found that for $E^2\neq 0$, the solution
 $|v_{E^2}^{+}\rangle$ satisfying $\Gamma
 |v_{E^2}^{+}\rangle=|v_{E^2}^{+}\rangle$ is constructed from the
 solution $|v_{E^2}^{-}\rangle$
satisfying $\Gamma|v_{E^2}^{-}\rangle=-|v_{E^2}^{-}\rangle$ by
 multiplying ${\cal H}$ from the left,
\begin{eqnarray}
|v_{E^2}^{+}\rangle =c' {\cal H}|v_{E^2}^{-}\rangle,
\label{eq:chiralpair}
\end{eqnarray}
where $c'$ is a normalization constant.
In a similar manner, we can construct $|v_{E^2}^{-}\rangle$ from
$|v_{E^2}^{+}\rangle$.
Therefore, for $E^2\neq 0$, the solution of (\ref{eq:chiralBdG2}) with
the chirality $\Gamma=1$ is always paired with the solution with $\Gamma=-1$. 

On the other hand, for $E^2=0$, the solution does not form a pair in
general. Indeed,  
as was shown in the above, the solution
$|v_0^{-}\rangle$ with $E^2=0$ satisfies ${\cal H}|v_0^{-}\rangle=0$, 
thus Eq.(\ref{eq:chiralpair}) leads to $|v_0^{+}\rangle=0$.
Thus we do not obtain the paired state in this case.

\section{Derivation of useful formulas}

\subsection{Formulas for $w(k_y)$}
\label{sec:appendixb1}
We first evaluate the integral (\ref{eq:wdef})
\begin{eqnarray}
w(k_y)=\frac{1}{2\pi }{\rm Im}\left[
\int dk_x \partial_{k_x}\ln q({\bm k})
\right], 
\end{eqnarray}
with $q({\bm k})=-i\varepsilon({\bm k})-\Delta({\bm k})$.
To calculate this integral, it is convenient to rewrite this as
\begin{eqnarray}
w(k_y)=-\frac{1}{2\pi}\int dk_x \epsilon^{ab}m_a({\bm
 k})\partial_{k_x}m_b({\bm k}), 
\label{eq:topappen2}
\end{eqnarray}
where 
\begin{eqnarray}
m_1({\bm k})=\frac{\varepsilon({\bm k})}{\sqrt{\varepsilon^2({\bm
 k})+\Delta^2({\bm k})}},
\quad 
m_2({\bm k})=\frac{\Delta({\bm k})}{\sqrt{\varepsilon^2({\bm
 k})+\Delta^2({\bm k})}}.
\end{eqnarray}
Then we apply a technique used in Ref.\cite{Sato09,Sato10b}.
Since the integral (\ref{eq:topappen2}) is a topological number, it can
not change its value even if we rescale $\Delta({\bm k})$ as
$a\Delta({\bm k})$ $(a\le 1)$, 
\begin{eqnarray}
m_1({\bm k})\rightarrow 
\frac{\varepsilon({\bm k})}
{\sqrt{\varepsilon^2({\bm k})+a^2\Delta^2({\bm k})}}, 
\quad
m_2({\bm k})\rightarrow 
\frac{a\Delta({\bm k})}
{\sqrt{\varepsilon^2({\bm k})+a^2\Delta^2({\bm k})}} 
\end{eqnarray} 
Then when $a\ll 1$, except the neighborhoods of the zero of
$\varepsilon({\bm k})$, $m_a({\bm k})$ becomes a constant
\begin{eqnarray}
m_1({\bm k})\sim \pm 1,
\quad
m_2({\bm k})\sim 0.
\end{eqnarray} 
Thus only the neighborhood of the zero of $\varepsilon({\bm k})$
contributes the integral. Expanding $\varepsilon({\bm k})$ and
$\Delta({\bm k})$ around the $k_x^0$ satisfying $\varepsilon(k_x^0,k_y)=0$,
\begin{eqnarray}
\varepsilon({\bm k})=\partial_{k_x}\varepsilon(k_x^0,k_y)(k_x-k_x^0)+\cdots, 
\quad
\Delta({\bm k})=\Delta(k_x^0,k_y)+\cdots,
\end{eqnarray}
we estimate the contribution near the $k_x^0$ as
\begin{eqnarray}
 \frac{1}{2}{\rm sgn}[\partial_{k_x}\varepsilon(k_x^0,k_y)]
{\rm sgn}\Delta(k_x^0,k_y).
\end{eqnarray}
Summing up the contribution from $k_x^0$'s, we obtain Eq.(\ref{eq:formula})
\begin{eqnarray}
w(k_y)= \frac{1}{2}\sum_{\epsilon(k_x^0,k_y)=0}{\rm
 sgn}[\partial_{k_x}\varepsilon(k_x^0,k_y)] {\rm sgn}\Delta(k_x^0,k_y).
\end{eqnarray}

\begin{figure}[h]
\begin{center}
\includegraphics[width=7cm]{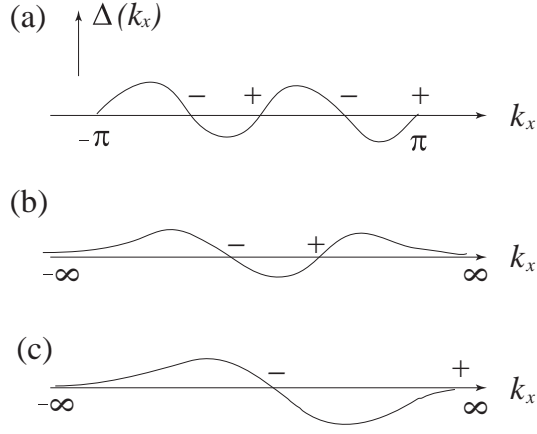}
\caption{Examples of $\Delta({\bm k})$ as a function of $k_x$. The
symbols $\pm$ represent the sign of $\partial_{k_x}\Delta(k_x)$ at $k_x=k_x^0$
with $\Delta(k_x^0)=0$. (a) $\Delta({\bm k})$ in a lattice model. From the
 periodicity in $k_x$,  Eq.(\ref{eq:deltaidentity}) holds. (b) and (c)
 $\Delta({\bm k})$ in a continuum model. Here we regulate $\Delta({\bm
 k})$ as $\Delta({\bm k})\rightarrow 0$ at $k_x=\pm \infty$. In the case
 (c), we need to take into account the contribution at $k_x=\infty$ to
 obtain Eq.(\ref{eq:deltaidentity}).
 }
\label{fig:deltaperiod}
\end{center}
\end{figure}

In a similar manner, 
we also obtain the following formula
by exchanging $\varepsilon({\bm k})$ with $\Delta({\bm k})$ in the above,  
\begin{eqnarray}
w(k_y)=-\frac{1}{2}\sum_{\Delta(k_x^0,k_y)=0}{\rm sgn}
\left[\partial_{k_y}\Delta(k_x^0,k_y)\right]{\rm sgn}\varepsilon(k_x^0,k_y).
\label{eq:wf2}
\end{eqnarray}
As illustrated in Fig.\ref{fig:deltaperiod}, the following identity
holds in general,
\begin{eqnarray}
\sum_{\Delta(k_x^0,k_y)=0}{\rm
 sgn}\left[\partial_{k_x}\Delta(k_x^0,k_y)\right]=0. 
\label{eq:deltaidentity}
\end{eqnarray}
For a lattice model, this equation comes from the periodicity in $k_x$,
and for a continuum model, this holds from the regularization in which
$\Delta({\bm k})$ at $k_x=\infty$ is identified with that at $k_x=-\infty$.
Since the left hand side of Eq.(\ref{eq:deltaidentity}) is rewritten as
\begin{eqnarray}
&&\sum_{\Delta(k_x^0,k_y)=0}
{\rm sgn}\left[\partial_{k_x}\Delta(k_x^0,k_y)\right]
\nonumber\\
&&=
\sum_{\Delta(k_x^0,k_y)=0,\,\varepsilon(k_x^0,k_y)>0}
{\rm sgn}\left[\partial_{k_x}\Delta(k_x^0,k_y)\right]
\nonumber\\
&&+
\sum_{\Delta(k_x^0,k_y)=0,\,\varepsilon(k_x^0,k_y)<0} 
{\rm sgn}\left[\partial_{k_x}\Delta(k_x^0,k_y)\right],
\end{eqnarray}
Eq.(\ref{eq:deltaidentity}) leads to
\begin{eqnarray}
\sum_{\Delta(k_x^0,k_y)=0,\,\varepsilon(k_x^0,k_y)>0}
{\rm sgn}\left[\partial_{k_x}\Delta(k_x^0,k_y)\right]
=
-\sum_{\Delta(k_x^0,k_y)=0,\,\varepsilon(k_x^0,k_y)<0} 
{\rm sgn}\left[\partial_{k_x}\Delta(k_x^0,k_y)\right]
\end{eqnarray}
From this equation, Eq.(\ref{eq:wf2}) is recast into
\begin{eqnarray}
w(k_y)&=&-\frac{1}{2}\sum_{\Delta(k_x^0,k_y)=0, \,
 \varepsilon(k_x^0,k_y)>0}{\rm sgn} 
\left[\partial_{k_x}\Delta(k_x^0,k_y)\right]
\nonumber\\
&+&\frac{1}{2}\sum_{\Delta(k_x^0,k_y)=0, \,
 \varepsilon(k_x^0,k_y)<0}{\rm sgn} 
\left[\partial_{k_x}\Delta(k_x^0,k_y)\right]
\nonumber\\
&=&
\sum_{\Delta(k_x^0,k_y)=0,\,{\rm sgn}
\varepsilon(k_x^0,k_y)<0} {\rm sgn}
\left[\partial_{k_x}\Delta(k_x^0,k_y)\right].
\end{eqnarray}
We will use this formula in Sec.\ref{sec:index2x2}.

\subsection{Formulas for $W(k_y)$}
\label{sec:appendixb2}
Here we evaluate the integral (\ref{eq:Wdef})
\begin{eqnarray}
W(k_y)=\frac{1}{2\pi }
{\rm Im}\left[
\int dk_x \partial_{k_x}\ln {\rm det}\hat{q}({\bm k})
\right],
\end{eqnarray}
with
\begin{eqnarray}
\hat{q}({\bm k})=[\varepsilon({\bm k})-i\psi({\bm k})]\sigma_y+
[{\bm g}({\bm k})-i{\bm d}({\bm k})]\cdot{\bm \sigma}\sigma_y. 
\end{eqnarray} 
Since the determinant of $\hat{q}({\bm k})$ is given by
\begin{eqnarray}
{\rm det}\hat{q}({\bm k})
=
-\varepsilon^2({\bm k})+{\bm g}^2({\bm k})+\psi^2({\bm  k})-{\bm d}^2({\bm k})
-2i\left[{\bm g}({\bm k})\cdot {\bm d}({\bm
			   k})-\varepsilon({\bm k})\psi({\bm k})\right], 
\end{eqnarray}
the above integral is recast into the same form as (\ref{eq:topappen2}) with
\begin{eqnarray}
m_1({\bm k})=\frac{
2({\bm g}({\bm k})\cdot {\bm d}({\bm k})-\varepsilon({\bm k})\psi({\bm
k}))
}
{\sqrt{[\varepsilon^2({\bm k})-{\bm g}^2({\bm k})
-\psi^2({\bm  k})+{\bm d}^2({\bm k})]^2+4[{\bm g}({\bm k})
\cdot {\bm d}({\bm k})-\varepsilon({\bm k})\psi({\bm k})]^2}},
\nonumber\\
m_2({\bm k})=\frac{
\varepsilon^2({\bm k})-{\bm g}^2({\bm
 k})-\psi^2({\bm  k})+{\bm d}^2({\bm k})
}
{\sqrt{[\varepsilon^2({\bm k})-{\bm g}^2({\bm
 k})-\psi^2({\bm  k})+{\bm d}^2({\bm k})]^2+4[{\bm g}({\bm k})
\cdot {\bm d}({\bm k})-\varepsilon({\bm k})\psi({\bm k})]^2
}}.
\end{eqnarray}
For a weak pairing superconductor, the energy scale of the gap functions
$\psi({\bm k})$ and $|{\bm d}({\bm k})|$ is much smaller than that of
$\varepsilon({\bm k})$ and $|{\bm g}({\bm k})|$.  Thus we can rescale
$\psi({\bm k})$ and ${\bm d}({\bm k})$ as $a\psi({\bm k})$ and $a{\bm
d}({\bm k})$ with a positive small constant $a$
without changing the value of $W(k_y)$.
Then for $a\ll 1$, it is found that only the neighborhood of the zeros of
$\varepsilon^2({\bm k})-{\bm g}^2({\bm k})$ contributes to $W(k_y)$. 
As a result, in a similar manner to Appendix \ref{sec:appendixb1}, we obtain
the formula (\ref{eq:formula2}).

\section{${\bm Z}_2$ topological number}
\label{sec:appendixc}
First consider the quasi-particle wave
function $|u_n({\bm k})\rangle$ with
\begin{eqnarray}
{\cal H}({\bm k})|u_n(k_x,k_y)\rangle= E_n({\bm k})|u_n(k_x,k_y)\rangle. 
\end{eqnarray}
From the symmetry (\ref{eq:particle-hole2}), we can say that if
$|u_n({\bm k})\rangle$ is a quasiparticle state with positive energy
$E_n({\bm k})>0$, then $C'|u^*_n(-k_x,k_y)\rangle$ is a quasiparticle
state with negative energy $-E_n(-k_x,k_y)<0$.  
(See Eq.(\ref{eq:particle-hole2}) for the definition of $C'$.)
Thus we use a positive (negative) $n$ for $|u_n(k_x,k_y)\rangle$ to represent
a positive (negative) energy quasiparticle state, and set
\begin{eqnarray}
|u_{-n}(k_x,k_y)\rangle =C'|u_n^*(-k_x,k_y)\rangle .
\label{eq:pn}
\end{eqnarray}
For the ground state in a superconductor, the negative states are
occupied.

Now we introduce the following ``gauge field'' $A_x^{(\pm)}({\bm k})$,
\begin{eqnarray}
A_x^{(\pm)}({\bm k})=i\sum_{n \gtrless 0}\langle u_n({\bm
 k})|\partial_{k_x}u_n({\bm k})\rangle.
\end{eqnarray}
From (\ref{eq:pn}), the gauge field $A_x^{(\pm)}({\bm k})$
satisfies
\begin{eqnarray}
A_x^{(+)}(k_x,k_y)=A_x^{(-)}(-k_x,k_y) .
\label{eq:apm}
\end{eqnarray}
We also find that $A_{x}({\bm k})=A_x^{(+)}({\bm k})+A_x^{(-)}({\bm k})$
is given by a total derivative of a function.
To see this, we rewrite $|u_n({\bm k})\rangle$ in a component,
\begin{eqnarray}
|u_n({\bm k})\rangle=
\left(
\begin{array}{c}
u_n^1({\bm k}) \\
u_n^2({\bm k}) \\
u_n^3({\bm k}) \\
u_n^4({\bm k})
\end{array}
\right), 
\end{eqnarray}
and introduce the following $4\times 4$ unitary matrix $V({\bm k})$ 
\begin{eqnarray}
V_{mn}({\bm k})=u_n^m({\bm k}), 
\quad
(m=1,2,3,4, \,\, n=2,1,-1,-2). 
\end{eqnarray}
Then $A_x({\bm k})$ is written as 
\begin{eqnarray}
A_x({\bm k})=i{\rm tr}(V^{\dagger}({\bm k})\partial_{k_x}V({\bm k}))
=
i\partial_{k_x}\ln {\rm det}V({\bm k}). 
\label{eq:totalgauge}
\end{eqnarray}

In a similar manner as in Ref.\cite{Sato10},
we introduce the ${\bm Z}_2$ topological number 
$
(-1)^{\nu(k_y)}
$
as
\begin{eqnarray}
\nu(k_y)=\frac{1}{\pi}\int_{-\infty}^{\infty} dk_x  A_x^{(-)}(k_x,k_y).
\label{eq:z2def}
\end{eqnarray}
Here we regulate the gap function
as $\hat{\Delta}({\bm k})\rightarrow 0$ far apart from the Fermi surface to
define the topological number.
Eqs. (\ref{eq:apm}) and (\ref{eq:totalgauge}) lead to
\begin{eqnarray}
\nu(k_y)&=&\frac{1}{\pi}\int_{-\infty}^{0}dk_x A_x^{(-)}(k_x,k_y)
+\frac{1}{\pi}\int_{0}^{\infty}dk_x A_x^{(-)}(k_x,k_y) 
\nonumber\\
&=&\frac{1}{\pi}\int_0^{\infty} dk_x 
\left[
A_x^{(+)}(k_x,k_y)+A_x^{(-)}(k_x,k_y)
\right]
\nonumber\\
&=& \frac{i}{\pi}\ln 
\left[\frac{{\rm det}V(\infty,k_y)}{{\rm det}V(0,k_y)}\right],
\end{eqnarray}
thus, we obtain
\begin{eqnarray}
(-1)^{\nu(k_y)}=\frac{{\rm det}V(0,k_y)}{{\rm det}V(\infty,k_y)}. 
\label{eq:z2V}
\end{eqnarray}
We can see that the right hand side of (\ref{eq:z2V}) indeed takes a ${\bm
Z}_2$ value:
From Eq.(\ref{eq:pn}), $V(0,k_y)$ is given by
\begin{eqnarray}
V(0,k_y)=(
|u_2(0, k_y)\rangle, |u_1(0,k_y)\rangle, C'|u_1^*(0,k_y)\rangle,
C'|u_2^*(0,k_y)\rangle),
\end{eqnarray}
thus it satisfies
\begin{eqnarray}
C'V(0,k_y)=(
C'|u_2(0, k_y)\rangle, C'|u_1(0,k_y)\rangle, |u_1^*(0,k_y)\rangle,
|u_2^*(0,k_y)\rangle).
\end{eqnarray}
Calculating the determinants of both sides, we obtain 
${\rm det}V(0,k_y)={\rm det}V^{*}(0,k_y)$.
This implies ${\rm det}V(0,k_y)=\pm 1$ since $V(0,k_y)$ is a unitary matrix. 
In a similar manner, we obtain ${\rm det}V(\infty,k_y)=\pm 1$ from the
regularization in which $k_x=\infty$ and $k_x=-\infty$ are identified.
So 
the right hand side of (\ref{eq:z2V}) takes $\pm 1$. 

Using the explicit form of the quasiparticle wave functions of
(\ref{eq:BdG4x4}),
we obtain
\begin{eqnarray}
{\rm det}V(0,k_y)=-{\rm sgn} [(k_y^2/2m-\mu)^2-(\lambda k_y)^2]
\end{eqnarray}
for the $d_{xy}+p$-wave superconductor.
Furthermore, due to the regularization of $\hat{\Delta}({\bm k})$, we
find that ${\cal H}({\bm k})$ is
almost diagonal at $k_x=\infty$.
Thus we have
$|u_2(\infty,k_y)\rangle\sim {}^{T}(1,0,0,0)$, and 
$|u_1(\infty,k_y)\rangle\sim {}^{T}(0,1,0,0)$,  
which implies ${\rm det}V(\infty,k_y)=-1$.
Therefore, the ${\bm Z}_2$ topological number is evaluated as
\begin{eqnarray}
(-1)^{\nu(k_y)}={\rm sgn} [(k_y^2/2m-\mu)^2-(\lambda k_y)^2]. 
\label{eq:z21}
\end{eqnarray}

Note that there are two differences between the above ${\bm Z}_2$
topological number and that in Ref.\cite{Sato10}. 
First, the ${\bm Z}_2$ topological number (\ref{eq:z2def}) is defined for
any $k_y$, while the ${\bm Z}_2$ topological number in Ref.\cite{Sato10} is
defined only for the time-reversal invariant $k_y$.
This difference comes from the difference in the symmetry
we used.
In contrast to the ${\bm Z}_2$ topological number in Ref.\cite{Sato10},
where the particle-hole symmetry (\ref{eq:particle-hole}) was used, 
in order to define (\ref{eq:z2def}),
we assume the one-dimensional particle-hole symmetry
(\ref{eq:particle-hole2}) which requires a special symmetry of gap
function such as $d_{xy}+p$-wave pairings.
Since the ${\bm Z}_2$ topological number in Ref.\cite{Sato10} is available
only for the discrete values of $k_y$, it cannot explain the existence of
the ABS with flat dispersion, although 
the ${\bm Z}_2$ topological number (\ref{eq:z2def})
can.
Second, the line integral in (\ref{eq:z2def}) performed from $k_x=0$ to
$k_y=\infty$ as we consider the continuum model of the $d_{xy}+p$-wave
superconductor. 

Even in the presence of the Zeeman magnetic field $H_y$ in the
$y$-direction, the symmetry
(\ref{eq:particle-hole2}) persists in the $d_{xy}+p$-wave
superconductor. 
Thus,  
the ${\bm Z}_2$ topological number can be defined in the 
same manner as above.
The resultant ${\bm Z}_2$ number is 
\begin{eqnarray}
(-1)^{\nu(k_y)}={\rm sgn} [(k_y^2/2m-\mu)^2-(\lambda k_y)^2-(\mu_{\rm B}H_y)^2].
\label{eq:z22}
\end{eqnarray}

\section{Relation to the odd-frequency pairing amplitude generated at
 the surface}
\label{sec:appendixd}

In this appendix, we would like to discuss 
the relevance of Andreev bound state and 
odd-frequency pairing function in more detail \cite{TG07,TGKU07,TTG07}. 
It has been clarified that inhomogeneity of superconductivity can 
induce odd-frequency pairing function 
\cite{TG07,TGKU07,TTG07,TAG08,Yokoyama08,Tanuma09,Yokoyama10}. 
For this purpose, it is better to use 
quasi classical Green's function, where 
atomic scale oscillation is removed out 
\cite{Serene,Rammer,Kopnin,Chandrasekhar,Eschrig00,Shelankov00,Schopohl,Ashida,Nagato93}. 
Here, we assume Cooper pair with $S_{z}=0$ for simplicity. \par
First, we consider two-dimensional semi-infinite superconductor 
in $x>0$ where flat surface is located at $x=0$. 
In the following discussion, to express  the symmetry of 
the frequency dependence of Cooper pair explicitly, 
we use Matsubara frequency. 
Quasiclassical Green's function can be written as  
\begin{equation}
\widehat{g}_{+} \ = \ 
\left(
\begin{array}{cc}
g_{+} & f_{+} \\
\bar{f}_{+} & -g_{+}%
\end{array}%
\right)
\ = \
\left(
\begin{array}{cc}
1 + D_{+}F_{+} & 2i F_{+} \\
2 i D_{+}  & -(1 + D_{+}F_{+}) %
\end{array}%
\right) / (1 -D_{+}F_{+}), 
\label{appendix1}
\end{equation}%

\begin{equation}
\widehat{g}_{-} \ = \ 
\left(
\begin{array}{cc}
g_{-} & f_{-} \\
\bar{f}_{-} & -g_{-}%
\end{array}%
\right)
\ = \
\left(
\begin{array}{cc}
1 + D_{-}F_{-} & -2i D_{-} \\
-2 i F_{-}  & -(1 + D_{-}F_{-}) %
\end{array}%
\right) / (1 -D_{-}F_{-}), 
\label{appendix2}
\end{equation}%
by using Riccati parameterization where $g$ and $f (\bar{f})$ are normal
and anomalous Green's functions, respectively.
The subscript $+(-)$ denotes the right and 
left going quasiparticles. 
The above Riccati parameters $D_{\pm}$ and $F_{\pm}$ 
obey the following equations, 
\[
v_{Fx}\partial_{x} D_{\pm}
=-\Delta_{\pm} (1 - D_{\pm}^{2}) + 2\omega_{n} D_{\pm},
\]
\begin{equation}
v_{Fx}\partial_{x} F_{\pm}
=-\Delta_{\pm} (1 - F_{\pm}^{2}) - 2\omega_{n} F_{\pm}. 
\label{appendix3}
\end{equation}
The bulk solution of $D_{\pm}$ for $x>0$ 
can be written as 
\begin{equation}
D_{\pm}=\frac{\Delta_{\pm}}
{\omega_{n} + \sqrt{\Delta^{2}_{\pm} + \omega^{2}_{n}}}.
\label{appendix3a}
\end{equation}
Here, $\Delta_{\pm}$ is a pair potential 
felt by quasiparticle for each trajectory. 
$v_{Fx}$ denotes the $x$-component of 
Fermi velocity with $v_{Fx} \geq 0$. 
For bulk $d_{xy}$-wave superconductor, it is given by 
\[
\Delta_{\pm}=\frac{\pm k_{x}k_{y}}{{\bm k}^{2}}\Delta_{0},  
\]
for $k_{x} \geq 0$. 
On the other hand, the corresponding quantity for $p_{x}$-wave, 
it is given by 
\[
\Delta_{\pm}=\frac{\pm k_{x}}{\sqrt{ {\bm k}^{2}}}\Delta_{0},  
\]
for $k_{x} \geq 0$. 
\par
At the surface $x=0$, 
$F_{+}=-D_{-}$ and $F_{-}=-D_{+}$ are satisfied due to the 
specular reflection. 
Also in the presence of ABS, 
$D_{+}=-D_{-}$ are satisfied. 
Thus, we obtain 
\begin{equation}
f_{+}= f_{-} = \frac{2 i D_{+}}{1 - D_{+}^{2}}.
\label{appendix4}
\end{equation}
Since $D_{+}(-\omega_{n})=1/D_{+}(\omega_{n})$ is satisfied, 
we can show that $f_{+}$ and $f_{-}$ is an odd-function of 
Matsubara frequency $\omega_{n}$. 
This means that a purely odd-frequency pairing state is realized 
at the surface in these two cases. \par

In order to clarify the parity of the pairing, 
we define pairing amplitude which is available both for 
positive and negative $k_{x}$. We write explicitly the 
${\bm k}$ dependence of the pairing amplitude. 
We introduce $f(k_{x},k_{y})$ as 
$f(k_{x},k_{y})=f_{+}(k_{x},k_{y})$ and 
$f(-k_{x},k_{y})=f_{-}(k_{x},k_{y})$ for $k_{x} \geq 0$.   
For even-parity pairing, 
$f(k_{x},k_{y})=f(-k_{x},-k_{y})$ is satisfied. 
On the other hand, for odd-parity case, 
$f(k_{x},k_{y})=-f(-k_{x},-k_{y})$ is satisfied. 
Here, let us discuss the parity of the odd-frequency pairings obtained above. 
For $d_{xy}$-wave pair potential, since
$f_{-}(k_{x},k_{y})=-f_{-}(k_{x},-k_{y})$ is satisfied,  
it is straightforward to derive that 
$f(k_{x},k_{y})=-f(-k_{x},-k_{y})$ is satisfied. 
This means the realization of odd-parity pairing in this case. 
For $p_{x}$-wave pair potential, since
$f_{-}(k_{x},k_{y})=f_{-}(k_{x},-k_{y})$ is satisfied,  
it is easy to see that 
$f(k_{x},k_{y})=f(-k_{x},-k_{y})$ is satisfied. 
This means the generation of even-parity pairing in this case. 
\par
To see concrete form of $f(k_{x},k_{y})$, 
we neglect the spatial dependence of $\Delta_{\pm}$. 
Then, $f_{\pm}$ is given by 
\begin{equation}
f_{+}=f_{-}=
\frac{i \Delta_{0}}{\omega_{n}}\frac{k_{x}k_{y}}{{\bm k}^{2}},
\label{appendix5}
\end{equation}
for $d_{xy}$-wave pairing and 
\begin{equation}
f_{+}=f_{-}=
\frac{i \Delta_{0}}{\omega_{n}}\frac{k_{x}}{\sqrt{{\bm k}^{2}}},
\label{appendix6}
\end{equation}
for $p_{x}$-wave pairing. 
The resulting $f(k_{x},k_{y})$ is given by 
\begin{equation}
f(k_{x},k_{y})=
\frac{i{\rm sgn}(k_{x}) \Delta_{0}}{\omega_{n}}
\frac{k_{x}k_{y}}{{\bm k}^{2}}
=
\frac{i{\rm sgn}(k_{y}) \Delta_{0}}{\omega_{n}}
\frac{\mid k_{x}\mid \mid k_{y} \mid}{{\bm k}^{2}}
,  
\label{appendix7}
\end{equation}
for $d_{xy}$-wave pair potential,  
and 
\begin{equation}
f(k_{x},k_{y})=\frac{i{\rm sgn}(k_{x}) \Delta_{0}}{\omega_{n}}
\frac{k_{x}}{\sqrt{ {\bm k}^{2}}}
=\frac{i \Delta_{0}}{\omega_{n}}
\frac{\mid k_{x} \mid}{\sqrt{ {\bm k}^{2}}}, 
\label{appendix8}
\end{equation}
for $p_{x}$-wave pairing. 
It is evident that odd-frequency odd-parity pairing in the former case 
and odd-frequency even-parity pairing in the latter case are realized. This is consistent with the fact that the spin of Cooper pairing is conserved.
The difference of the parity results in a
remarkable difference when we consider proximity effect 
into DN attached to 
superconductor \cite{TG07,TGKU07,TTG07}. \par
In DN, only $s$-wave even parity 
pairing is possible due to impurity scattering. Thus, odd-parity odd-frequency pairing amplitude
cannot penetrate into DN. 
Thus, for $d_{xy}$-wave superconductor, 
ABS cannot enter into DN since it is expressed by
odd-frequency spin-singlet odd-parity state. 
On the other hand, for $p_{x}$-wave 
superconductor, 
ABS can enter into DN since it is expressed by
odd-frequency spin-singlet even-parity state including $s$-wave channel. 
Thus, it has been shown that 
proximity effect of $d_{xy}$-wave superconductor and 
$p_{x}$-wave superconductor is completely different
\cite{TNS03,TNGK04,TK04,TKY05,ATK06}. \par
 Next, we consider a two-dimensional semi-infinite superconductor 
in $x<0$ where flat surface is located at $x=0$. 
Quasiclassical Green's function can be written as  
\begin{equation}
\widehat{g}_{+} \ = \ 
\left(
\begin{array}{cc}
g_{+} & f_{+} \\
\bar{f}_{+} & -g_{+}%
\end{array}%
\right)
\ = \
\left(
\begin{array}{cc}
1 + D_{+}F_{+} & -2i D_{+} \\
-2 i F_{+}  & -(1 + D_{+}F_{+}) %
\end{array}%
\right) / (1 -D_{+}F_{+}), 
\label{appendix9}
\end{equation}%
\begin{equation}
\widehat{g}_{-} \ = \ 
\left(
\begin{array}{cc}
g_{-} & f_{-} \\
\bar{f}_{-} & -g_{-}%
\end{array}%
\right)
\ = \
\left(
\begin{array}{cc}
1 + D_{-}F_{-} & 2i F_{-} \\
2 i D_{-}  & -(1 + D_{-}F_{-}) %
\end{array}%
\right) / (1 -D_{-}F_{-}), 
\label{appendix10}
\end{equation}%
by using Riccati parameterization. 
The subscript $+(-)$ again denotes the right and 
left going quasiparticles. 
Following the similar discussion in the case 
where two-dimensional semi-infinite superconductor 
is located on $x>0$, 
we can use the relation
$D_{+}=-F_{-}$ and $D_{+}=-F_{+}$. 
In the presence of Andreev bound state (ABS), 
$D_{+}=-D_{-}$ are satisfied. 
Then, we can write 
\begin{equation}
f_{+}=\frac{-2 i D_{+}}{1 - D_{+}^{2}}.
\label{appendix11}
\end{equation}
If we neglect the spatial dependence of $\Delta_{\pm}$, 
$f_{\pm}$ is given by 
\begin{equation}
f_{+}=f_{-}=
\frac{-i \Delta_{0}}{\omega_{n}}\frac{k_{x}k_{y}}{{\bm k}^{2}},
\label{appendix12}
\end{equation}
for spin-singlet $d_{xy}$-wave pairing and 
\begin{equation}
f_{+}=f_{-}=
\frac{-i \Delta_{0}}{\omega_{n}}\frac{k_{x}}{\sqrt{{\bm k}^{2}}},
\label{appendix13}
\end{equation}
for spin-triplet $p_{x}$-wave pairing. 
Following similar discussions below Eq.  (\ref{appendix5}),  
$f(k_{x},k_{y})$ is given by 
\begin{equation}
f(k_{x},k_{y})=
-\frac{i{\rm sgn}(k_{x}) \Delta_{0}}{\omega_{n}}
\frac{k_{x}k_{y}}{{\bm k}^{2}} 
=
-\frac{i{\rm sgn}(k_{y}) \Delta_{0}}{\omega_{n}}
\frac{\mid k_{x} \mid \mid k_{y} \mid}{{\bm k}^{2}}, 
\label{appendix14}
\end{equation}
for spin-singlet $d_{xy}$-wave and 
\begin{equation}
f(k_{x},k_{y})=-\frac{i{\rm sgn}(k_{x}) \Delta_{0}}{\omega_{n}}
\frac{k_{x}}{\sqrt{ {\bm k}^{2}}}
=-\frac{i \Delta_{0}}{\omega_{n}}
\frac{\mid k_{x} \mid}{\sqrt{ {\bm k}^{2}}},
\label{appendix15}
\end{equation}
for spin-triplet $p_{x}$-wave.

\section{diagonalization of $\hat{q}({\bm k},x)$}
\label{sec:diagq}

Let us consider an $N\times N$ matrix $\hat{q}({\bm k},x)$ with eigenvalues
$\lambda_1, \cdots,\lambda_N$.
Then we have two kinds of eigenvectors : The first one
is right eigenvectors, 
\begin{eqnarray}
\hat{q}({\bm k},x)a_n({\bm k},x)=\lambda_n({\bm k},x)a_n({\bm k},x), 
\label{eq:eigena}
\end{eqnarray}
and the second one is left eigenvectors
\begin{eqnarray}
b_n^{\dagger}({\bm k},x)\hat{q}({\bm k},x)
=b_n^{\dagger}({\bm k},x)\lambda_n({\bm k},x),
\label{eq:eigenb}
\end{eqnarray}
with the normalization condition
\begin{eqnarray}
b_n^{\dagger}({\bm k},x)a_m({\bm k},x)=\delta_{mn}. 
\label{eq:ab}
\end{eqnarray}
Note that $a_n({\bm k},x)\neq b_n({\bm k},x)$ in general unless
$\hat{q}({\bm k},x)$ is a hermitian matrix. 
Eqs.(\ref{eq:eigena}) and (\ref{eq:eigenb}) lead to
\begin{eqnarray}
\hat{q}({\bm k},x)A({\bm k},x)=A({\bm k},x)\Lambda({\bm k},x),
\nonumber\\
q^{\dagger}({\bm k},x)B({\bm k},x)=B({\bm k},x)\Lambda^*({\bm k},x),
\end{eqnarray}
with
\begin{eqnarray}
&&A({\bm k},x)=(a_1({\bm k},x), \cdots, a_N({\bm k},x)),
\nonumber\\ 
&&B({\bm k},x)=(b_1({\bm k},x), \cdots, b_N({\bm k},x)), 
\nonumber\\
&&\Lambda({\bm k},x)={\rm diag}(\lambda_1({\bm k},x),\cdots,
\lambda_N({\bm k},x)).
\end{eqnarray} 
Using the normalization condition (\ref{eq:ab}) which reads
\begin{eqnarray}
B^{\dagger}({\bm k},x)A({\bm k},x)= A^{\dagger}({\bm k},x)B({\bm
 k},x)={\bm 1}_{N\times N}, 
\end{eqnarray}
we obtain 
\begin{eqnarray}
\hat{q}({\bm k},x)=A({\bm k},x)\Lambda({\bm k},x)B^{\dagger}({\bm k},x),
\nonumber\\ 
\hat{q}^{\dagger}({\bm k},x)=B({\bm k},x)\Lambda^{*}({\bm
 k},x)A^{\dagger}({\bm k},x). 
\end{eqnarray}

\bibliography{Majorana}

\begin{thebibliography}{143}
\expandafter\ifx\csname natexlab\endcsname\relax\def\natexlab#1{#1}\fi
\expandafter\ifx\csname bibnamefont\endcsname\relax
  \def\bibnamefont#1{#1}\fi
\expandafter\ifx\csname bibfnamefont\endcsname\relax
  \def\bibfnamefont#1{#1}\fi
\expandafter\ifx\csname citenamefont\endcsname\relax
  \def\citenamefont#1{#1}\fi
\expandafter\ifx\csname url\endcsname\relax
  \def\url#1{\texttt{#1}}\fi
\expandafter\ifx\csname urlprefix\endcsname\relax\def\urlprefix{URL }\fi
\providecommand{\bibinfo}[2]{#2}
\providecommand{\eprint}[2][]{\url{#2}}

\bibitem[{\citenamefont{Andreev}(1964)}]{Andreev64}
\bibinfo{author}{\bibfnamefont{A.~F.} \bibnamefont{Andreev}},
  \bibinfo{journal}{Sov. Phys. JETP} \textbf{\bibinfo{volume}{19}},
  \bibinfo{pages}{1228} (\bibinfo{year}{1964}).

\bibitem[{\citenamefont{Blonder et~al.}(1982)\citenamefont{Blonder, Tinkham,
  and Klapwijk}}]{BTK82}
\bibinfo{author}{\bibfnamefont{G.~E.} \bibnamefont{Blonder}},
  \bibinfo{author}{\bibfnamefont{M.}~\bibnamefont{Tinkham}}, \bibnamefont{and}
  \bibinfo{author}{\bibfnamefont{T.}~\bibnamefont{Klapwijk}},
  \bibinfo{journal}{Phys. Rev. B} \textbf{\bibinfo{volume}{25}},
  \bibinfo{pages}{4515} (\bibinfo{year}{1982}).

\bibitem[{\citenamefont{Bruder}(1990)}]{Bruder90}
\bibinfo{author}{\bibfnamefont{C.}~\bibnamefont{Bruder}},
  \bibinfo{journal}{Phys. Rev. B} \textbf{\bibinfo{volume}{41}},
  \bibinfo{pages}{4017} (\bibinfo{year}{1990}).

\bibitem[{\citenamefont{Buchholtz and Zwicknagl}(1981)}]{BZ81}
\bibinfo{author}{\bibfnamefont{L.~J.} \bibnamefont{Buchholtz}}
  \bibnamefont{and}
  \bibinfo{author}{\bibfnamefont{G.}~\bibnamefont{Zwicknagl}},
  \bibinfo{journal}{Phys. Rev. B} \textbf{\bibinfo{volume}{23}},
  \bibinfo{pages}{5788} (\bibinfo{year}{1981}).

\bibitem[{\citenamefont{Hara and Nagai}(1986)}]{HN86}
\bibinfo{author}{\bibfnamefont{J.}~\bibnamefont{Hara}} \bibnamefont{and}
  \bibinfo{author}{\bibfnamefont{K.}~\bibnamefont{Nagai}},
  \bibinfo{journal}{Prog. Theor. Phys.} \textbf{\bibinfo{volume}{74}},
  \bibinfo{pages}{1237} (\bibinfo{year}{1986}).

\bibitem[{\citenamefont{Hu}(1994)}]{Hu94}
\bibinfo{author}{\bibfnamefont{C.~R.} \bibnamefont{Hu}},
  \bibinfo{journal}{Phys. Rev. Lett.} \textbf{\bibinfo{volume}{72}},
  \bibinfo{pages}{1526} (\bibinfo{year}{1994}).

\bibitem[{\citenamefont{Kashiwaya and Tanaka}(2000)}]{KT00}
\bibinfo{author}{\bibfnamefont{S.}~\bibnamefont{Kashiwaya}} \bibnamefont{and}
  \bibinfo{author}{\bibfnamefont{Y.}~\bibnamefont{Tanaka}},
  \bibinfo{journal}{Rep. Prog. Phys.} \textbf{\bibinfo{volume}{63}},
  \bibinfo{pages}{1641} (\bibinfo{year}{2000}).

\bibitem[{\citenamefont{L{\"o}fwander et~al.}(2001)\citenamefont{L{\"o}fwander,
  Shumeiko, and Wendin}}]{LSW01}
\bibinfo{author}{\bibfnamefont{T.}~\bibnamefont{L{\"o}fwander}},
  \bibinfo{author}{\bibfnamefont{V.~S.} \bibnamefont{Shumeiko}},
  \bibnamefont{and} \bibinfo{author}{\bibfnamefont{G.}~\bibnamefont{Wendin}},
  \bibinfo{journal}{Supercond. Sci. Technol.} \textbf{\bibinfo{volume}{14}},
  \bibinfo{pages}{R53} (\bibinfo{year}{2001}).

\bibitem[{\citenamefont{Asano et~al.}(2004)\citenamefont{Asano, Tanaka, and
  Kashiwaya}}]{ATK04}
\bibinfo{author}{\bibfnamefont{Y.}~\bibnamefont{Asano}},
  \bibinfo{author}{\bibfnamefont{Y.}~\bibnamefont{Tanaka}}, \bibnamefont{and}
  \bibinfo{author}{\bibfnamefont{S.}~\bibnamefont{Kashiwaya}},
  \bibinfo{journal}{Phys. Rev. B} \textbf{\bibinfo{volume}{69}},
  \bibinfo{pages}{134501} (\bibinfo{year}{2004}).

\bibitem[{\citenamefont{Tanaka and Kashiwaya}(1995)}]{TK95}
\bibinfo{author}{\bibfnamefont{Y.}~\bibnamefont{Tanaka}} \bibnamefont{and}
  \bibinfo{author}{\bibfnamefont{S.}~\bibnamefont{Kashiwaya}},
  \bibinfo{journal}{Phys. Rev. Lett.} \textbf{\bibinfo{volume}{74}},
  \bibinfo{pages}{3451} (\bibinfo{year}{1995}).

\bibitem[{\citenamefont{Kashiwaya et~al.}(1995)\citenamefont{Kashiwaya, Tanaka,
  Koyanagi, Takashima, and Kajimura}}]{KTKTK95}
\bibinfo{author}{\bibfnamefont{S.}~\bibnamefont{Kashiwaya}},
  \bibinfo{author}{\bibfnamefont{Y.}~\bibnamefont{Tanaka}},
  \bibinfo{author}{\bibfnamefont{M.}~\bibnamefont{Koyanagi}},
  \bibinfo{author}{\bibfnamefont{H.}~\bibnamefont{Takashima}},
  \bibnamefont{and} \bibinfo{author}{\bibfnamefont{K.}~\bibnamefont{Kajimura}},
  \bibinfo{journal}{Phys. Rev. B} \textbf{\bibinfo{volume}{51}},
  \bibinfo{pages}{1350} (\bibinfo{year}{1995}).

\bibitem[{\citenamefont{Kashiwaya et~al.}(1998)\citenamefont{Kashiwaya, Tanaka,
  Terada, Koyanagi, Ueno, Alff, Takashima, Tanuma, and Kajimura}}]{KTTKUATTK98}
\bibinfo{author}{\bibfnamefont{S.}~\bibnamefont{Kashiwaya}},
  \bibinfo{author}{\bibfnamefont{Y.}~\bibnamefont{Tanaka}},
  \bibinfo{author}{\bibfnamefont{N.}~\bibnamefont{Terada}},
  \bibinfo{author}{\bibfnamefont{M.}~\bibnamefont{Koyanagi}},
  \bibinfo{author}{\bibfnamefont{S.}~\bibnamefont{Ueno}},
  \bibinfo{author}{\bibfnamefont{L.}~\bibnamefont{Alff}},
  \bibinfo{author}{\bibfnamefont{H.}~\bibnamefont{Takashima}},
  \bibinfo{author}{\bibfnamefont{Y.}~\bibnamefont{Tanuma}}, \bibnamefont{and}
  \bibinfo{author}{\bibfnamefont{K.}~\bibnamefont{Kajimura}},
  \bibinfo{journal}{J. Phys. Chem. Solids} \textbf{\bibinfo{volume}{59}},
  \bibinfo{pages}{2034} (\bibinfo{year}{1998}).

\bibitem[{\citenamefont{Covington et~al.}(1997)\citenamefont{Covington, Aprili,
  Paraoanu, Greene, Xu, Zhu, and Mirkin}}]{CAPGXZM97}
\bibinfo{author}{\bibfnamefont{M.}~\bibnamefont{Covington}},
  \bibinfo{author}{\bibfnamefont{M.}~\bibnamefont{Aprili}},
  \bibinfo{author}{\bibfnamefont{E.}~\bibnamefont{Paraoanu}},
  \bibinfo{author}{\bibfnamefont{L.~H.} \bibnamefont{Greene}},
  \bibinfo{author}{\bibfnamefont{F.}~\bibnamefont{Xu}},
  \bibinfo{author}{\bibfnamefont{J.}~\bibnamefont{Zhu}}, \bibnamefont{and}
  \bibinfo{author}{\bibfnamefont{C.~A.} \bibnamefont{Mirkin}},
  \bibinfo{journal}{Phys. Rev. Lett.} \textbf{\bibinfo{volume}{79}},
  \bibinfo{pages}{277} (\bibinfo{year}{1997}).

\bibitem[{\citenamefont{Alff et~al.}(1997)\citenamefont{Alff, Takashima,
  Kashiwaya, Terada, Ihara, Tanaka, Koyanagi, and Kajimura}}]{ATKTITKK97}
\bibinfo{author}{\bibfnamefont{L.}~\bibnamefont{Alff}},
  \bibinfo{author}{\bibfnamefont{H.}~\bibnamefont{Takashima}},
  \bibinfo{author}{\bibfnamefont{S.}~\bibnamefont{Kashiwaya}},
  \bibinfo{author}{\bibfnamefont{N.}~\bibnamefont{Terada}},
  \bibinfo{author}{\bibfnamefont{H.}~\bibnamefont{Ihara}},
  \bibinfo{author}{\bibfnamefont{Y.}~\bibnamefont{Tanaka}},
  \bibinfo{author}{\bibfnamefont{M.}~\bibnamefont{Koyanagi}}, \bibnamefont{and}
  \bibinfo{author}{\bibfnamefont{K.}~\bibnamefont{Kajimura}},
  \bibinfo{journal}{Phys. Rev. B} \textbf{\bibinfo{volume}{55}},
  \bibinfo{pages}{R14757} (\bibinfo{year}{1997}).

\bibitem[{\citenamefont{Wei et~al.}(1998)\citenamefont{Wei, Yeh, Garrigus, and
  Strasik}}]{WYGS98}
\bibinfo{author}{\bibfnamefont{J.~Y.~T.} \bibnamefont{Wei}},
  \bibinfo{author}{\bibfnamefont{N.-C.} \bibnamefont{Yeh}},
  \bibinfo{author}{\bibfnamefont{D.~F.} \bibnamefont{Garrigus}},
  \bibnamefont{and} \bibinfo{author}{\bibfnamefont{M.}~\bibnamefont{Strasik}},
  \bibinfo{journal}{Phys. Rev. Lett.} \textbf{\bibinfo{volume}{81}},
  \bibinfo{pages}{2542} (\bibinfo{year}{1998}).

\bibitem[{\citenamefont{Iguchi et~al.}(2000)\citenamefont{Iguchi, Wang,
  Yamazaki, Tanaka, and Kashiwaya}}]{IWYTK00}
\bibinfo{author}{\bibfnamefont{I.}~\bibnamefont{Iguchi}},
  \bibinfo{author}{\bibfnamefont{W.}~\bibnamefont{Wang}},
  \bibinfo{author}{\bibfnamefont{M.}~\bibnamefont{Yamazaki}},
  \bibinfo{author}{\bibfnamefont{Y.}~\bibnamefont{Tanaka}}, \bibnamefont{and}
  \bibinfo{author}{\bibfnamefont{S.}~\bibnamefont{Kashiwaya}},
  \bibinfo{journal}{Phys. Rev. B} \textbf{\bibinfo{volume}{62}},
  \bibinfo{pages}{R6131} (\bibinfo{year}{2000}).

\bibitem[{\citenamefont{Biswas et~al.}(2002)\citenamefont{Biswas, Fournier,
  Qazilbash, Smolyaninova, Balci, and Greene}}]{BFQSBG02}
\bibinfo{author}{\bibfnamefont{A.}~\bibnamefont{Biswas}},
  \bibinfo{author}{\bibfnamefont{P.}~\bibnamefont{Fournier}},
  \bibinfo{author}{\bibfnamefont{M.~M.} \bibnamefont{Qazilbash}},
  \bibinfo{author}{\bibfnamefont{V.~N.} \bibnamefont{Smolyaninova}},
  \bibinfo{author}{\bibfnamefont{H.}~\bibnamefont{Balci}}, \bibnamefont{and}
  \bibinfo{author}{\bibfnamefont{R.~L.} \bibnamefont{Greene}},
  \bibinfo{journal}{Phys. Rev. Lett.} \textbf{\bibinfo{volume}{88}},
  \bibinfo{pages}{207004} (\bibinfo{year}{2002}).

\bibitem[{\citenamefont{Chesca et~al.}(2005)\citenamefont{Chesca, Seifried,
  Dahm, Schopohl, Koelle, Kleiner, and Tsukada}}]{CSDSKKT05}
\bibinfo{author}{\bibfnamefont{B.}~\bibnamefont{Chesca}},
  \bibinfo{author}{\bibfnamefont{M.}~\bibnamefont{Seifried}},
  \bibinfo{author}{\bibfnamefont{T.}~\bibnamefont{Dahm}},
  \bibinfo{author}{\bibfnamefont{N.}~\bibnamefont{Schopohl}},
  \bibinfo{author}{\bibfnamefont{D.}~\bibnamefont{Koelle}},
  \bibinfo{author}{\bibfnamefont{R.}~\bibnamefont{Kleiner}}, \bibnamefont{and}
  \bibinfo{author}{\bibfnamefont{A.}~\bibnamefont{Tsukada}},
  \bibinfo{journal}{Phys. Rev. B} \textbf{\bibinfo{volume}{71}},
  \bibinfo{pages}{104504} (\bibinfo{year}{2005}).

\bibitem[{\citenamefont{Chesca et~al.}(2006)\citenamefont{Chesca, Doenitz,
  Dahm, Huebener, Koelle, Kleiner, Ariando, Smilde, and
  Hilgenkamp}}]{CDDHKKASH06}
\bibinfo{author}{\bibfnamefont{B.}~\bibnamefont{Chesca}},
  \bibinfo{author}{\bibfnamefont{D.}~\bibnamefont{Doenitz}},
  \bibinfo{author}{\bibfnamefont{T.}~\bibnamefont{Dahm}},
  \bibinfo{author}{\bibfnamefont{R.~P.} \bibnamefont{Huebener}},
  \bibinfo{author}{\bibfnamefont{D.}~\bibnamefont{Koelle}},
  \bibinfo{author}{\bibfnamefont{R.}~\bibnamefont{Kleiner}},
  \bibinfo{author}{\bibnamefont{Ariando}},
  \bibinfo{author}{\bibfnamefont{H.~J.~H.} \bibnamefont{Smilde}},
  \bibnamefont{and}
  \bibinfo{author}{\bibfnamefont{H.}~\bibnamefont{Hilgenkamp}},
  \bibinfo{journal}{Phys. Rev. B} \textbf{\bibinfo{volume}{73}},
  \bibinfo{pages}{014529} (\bibinfo{year}{2006}).

\bibitem[{\citenamefont{Chesca et~al.}(2008)\citenamefont{Chesca, Smilde, and
  Hilgenkamp}}]{CSH08}
\bibinfo{author}{\bibfnamefont{B.}~\bibnamefont{Chesca}},
  \bibinfo{author}{\bibfnamefont{H.~J.~H.} \bibnamefont{Smilde}},
  \bibnamefont{and}
  \bibinfo{author}{\bibfnamefont{H.}~\bibnamefont{Hilgenkamp}},
  \bibinfo{journal}{Phys. Rev. B} \textbf{\bibinfo{volume}{77}},
  \bibinfo{pages}{184510} (\bibinfo{year}{2008}).

\bibitem[{\citenamefont{Matsumoto and Shiba}(1995)}]{MS95}
\bibinfo{author}{\bibfnamefont{M.}~\bibnamefont{Matsumoto}} \bibnamefont{and}
  \bibinfo{author}{\bibfnamefont{H.}~\bibnamefont{Shiba}}, \bibinfo{journal}{J.
  Phys. Soc. Jpn} \textbf{\bibinfo{volume}{64}}, \bibinfo{pages}{1703}
  (\bibinfo{year}{1995}).

\bibitem[{\citenamefont{Nagato and Nagai}(1995)}]{Nagato95}
\bibinfo{author}{\bibfnamefont{Y.}~\bibnamefont{Nagato}} \bibnamefont{and}
  \bibinfo{author}{\bibfnamefont{K.}~\bibnamefont{Nagai}},
  \bibinfo{journal}{Phys. Rev. B} \textbf{\bibinfo{volume}{51}},
  \bibinfo{pages}{16254} (\bibinfo{year}{1995}).

\bibitem[{\citenamefont{Tanaka and Kashiwaya}(1996{\natexlab{a}})}]{TKPRB96}
\bibinfo{author}{\bibfnamefont{Y.}~\bibnamefont{Tanaka}} \bibnamefont{and}
  \bibinfo{author}{\bibfnamefont{S.}~\bibnamefont{Kashiwaya}},
  \bibinfo{journal}{Phys. Rev. B} \textbf{\bibinfo{volume}{53}},
  \bibinfo{pages}{9371} (\bibinfo{year}{1996}{\natexlab{a}}).

\bibitem[{\citenamefont{Tanuma et~al.}(1998{\natexlab{a}})\citenamefont{Tanuma,
  Tanaka, Yamashiro, and Kashiwaya}}]{Tanuma98}
\bibinfo{author}{\bibfnamefont{Y.}~\bibnamefont{Tanuma}},
  \bibinfo{author}{\bibfnamefont{Y.}~\bibnamefont{Tanaka}},
  \bibinfo{author}{\bibfnamefont{M.}~\bibnamefont{Yamashiro}},
  \bibnamefont{and}
  \bibinfo{author}{\bibfnamefont{S.}~\bibnamefont{Kashiwaya}},
  \bibinfo{journal}{Phys. Rev. B} \textbf{\bibinfo{volume}{57}},
  \bibinfo{pages}{7997} (\bibinfo{year}{1998}{\natexlab{a}}).

\bibitem[{\citenamefont{Tanuma et~al.}(1998{\natexlab{b}})\citenamefont{Tanuma,
  Tanaka, Yamashiro, and Kashiwaya}}]{TanumaJ99}
\bibinfo{author}{\bibfnamefont{Y.}~\bibnamefont{Tanuma}},
  \bibinfo{author}{\bibfnamefont{Y.}~\bibnamefont{Tanaka}},
  \bibinfo{author}{\bibfnamefont{M.}~\bibnamefont{Yamashiro}},
  \bibnamefont{and}
  \bibinfo{author}{\bibfnamefont{S.}~\bibnamefont{Kashiwaya}},
  \bibinfo{journal}{J. Phys. Soc. Jpn.} \textbf{\bibinfo{volume}{67}},
  \bibinfo{pages}{1118} (\bibinfo{year}{1998}{\natexlab{b}}).

\bibitem[{\citenamefont{Tanuma et~al.}(1999)\citenamefont{Tanuma, Tanaka,
  Ogata, and Kashiwaya}}]{Tanuma99}
\bibinfo{author}{\bibfnamefont{Y.}~\bibnamefont{Tanuma}},
  \bibinfo{author}{\bibfnamefont{Y.}~\bibnamefont{Tanaka}},
  \bibinfo{author}{\bibfnamefont{M.}~\bibnamefont{Ogata}}, \bibnamefont{and}
  \bibinfo{author}{\bibfnamefont{S.}~\bibnamefont{Kashiwaya}},
  \bibinfo{journal}{Phys. Rev. B} \textbf{\bibinfo{volume}{60}},
  \bibinfo{pages}{9817} (\bibinfo{year}{1999}).

\bibitem[{\citenamefont{Barash et~al.}(1997)\citenamefont{Barash, Svidzinsky,
  and Burkhardt}}]{Barash97}
\bibinfo{author}{\bibfnamefont{Y.}~\bibnamefont{Barash}},
  \bibinfo{author}{\bibfnamefont{A.}~\bibnamefont{Svidzinsky}},
  \bibnamefont{and}
  \bibinfo{author}{\bibfnamefont{H.}~\bibnamefont{Burkhardt}},
  \bibinfo{journal}{Phys. Rev. B} \textbf{\bibinfo{volume}{55}},
  \bibinfo{pages}{15282} (\bibinfo{year}{1997}).

\bibitem[{\citenamefont{Tanuma et~al.}(2001)\citenamefont{Tanuma, Tanaka, and
  Kashiwaya}}]{Tanuma01}
\bibinfo{author}{\bibfnamefont{Y.}~\bibnamefont{Tanuma}},
  \bibinfo{author}{\bibfnamefont{Y.}~\bibnamefont{Tanaka}}, \bibnamefont{and}
  \bibinfo{author}{\bibfnamefont{S.}~\bibnamefont{Kashiwaya}},
  \bibinfo{journal}{Phys. Rev. B} \textbf{\bibinfo{volume}{64}},
  \bibinfo{pages}{214519} (\bibinfo{year}{2001}).

\bibitem[{\citenamefont{Tanaka and Kashiwaya}(1996{\natexlab{b}})}]{TK96}
\bibinfo{author}{\bibfnamefont{Y.}~\bibnamefont{Tanaka}} \bibnamefont{and}
  \bibinfo{author}{\bibfnamefont{S.}~\bibnamefont{Kashiwaya}},
  \bibinfo{journal}{Phys. Rev. B} \textbf{\bibinfo{volume}{53}},
  \bibinfo{pages}{R11957} (\bibinfo{year}{1996}{\natexlab{b}}).

\bibitem[{\citenamefont{Tanaka and Kashiwaya}(1997)}]{TK97}
\bibinfo{author}{\bibfnamefont{Y.}~\bibnamefont{Tanaka}} \bibnamefont{and}
  \bibinfo{author}{\bibfnamefont{S.}~\bibnamefont{Kashiwaya}},
  \bibinfo{journal}{Phys. Rev. B} \textbf{\bibinfo{volume}{56}},
  \bibinfo{pages}{892} (\bibinfo{year}{1997}).

\bibitem[{\citenamefont{Barash et~al.}(1996)\citenamefont{Barash, Burkhardt,
  and Rainer}}]{BBR96}
\bibinfo{author}{\bibfnamefont{Y.~S.} \bibnamefont{Barash}},
  \bibinfo{author}{\bibfnamefont{H.}~\bibnamefont{Burkhardt}},
  \bibnamefont{and} \bibinfo{author}{\bibfnamefont{D.}~\bibnamefont{Rainer}},
  \bibinfo{journal}{Phys. Rev. Lett.} \textbf{\bibinfo{volume}{77}},
  \bibinfo{pages}{4070} (\bibinfo{year}{1996}).

\bibitem[{\citenamefont{Kwon et~al.}(2004)\citenamefont{Kwon, Sengupta, and
  Yakovenko}}]{Yakovenko}
\bibinfo{author}{\bibfnamefont{H.}~\bibnamefont{Kwon}},
  \bibinfo{author}{\bibfnamefont{K.}~\bibnamefont{Sengupta}}, \bibnamefont{and}
  \bibinfo{author}{\bibfnamefont{V.}~\bibnamefont{Yakovenko}},
  \bibinfo{journal}{Eur. Phys. J. B} \textbf{\bibinfo{volume}{37}},
  \bibinfo{pages}{349} (\bibinfo{year}{2004}).

\bibitem[{\citenamefont{Tanaka and Kashiwaya}(1999)}]{TK99}
\bibinfo{author}{\bibfnamefont{Y.}~\bibnamefont{Tanaka}} \bibnamefont{and}
  \bibinfo{author}{\bibfnamefont{S.}~\bibnamefont{Kashiwaya}},
  \bibinfo{journal}{J. Phys. Soc. Jpn.} \textbf{\bibinfo{volume}{68}},
  \bibinfo{pages}{3485} (\bibinfo{year}{1999}).

\bibitem[{\citenamefont{Tanaka and Kashiwaya}(2000)}]{TK00}
\bibinfo{author}{\bibfnamefont{Y.}~\bibnamefont{Tanaka}} \bibnamefont{and}
  \bibinfo{author}{\bibfnamefont{S.}~\bibnamefont{Kashiwaya}},
  \bibinfo{journal}{J. Phys. Soc. Jpn.} \textbf{\bibinfo{volume}{69}},
  \bibinfo{pages}{1152} (\bibinfo{year}{2000}).

\bibitem[{\citenamefont{Kashiwaya et~al.}(1999)\citenamefont{Kashiwaya, Tanaka,
  Yoshida, and Beasley}}]{KTYB99}
\bibinfo{author}{\bibfnamefont{S.}~\bibnamefont{Kashiwaya}},
  \bibinfo{author}{\bibfnamefont{Y.}~\bibnamefont{Tanaka}},
  \bibinfo{author}{\bibfnamefont{N.}~\bibnamefont{Yoshida}}, \bibnamefont{and}
  \bibinfo{author}{\bibfnamefont{M.~R.} \bibnamefont{Beasley}},
  \bibinfo{journal}{Phys. Rev. B} \textbf{\bibinfo{volume}{60}},
  \bibinfo{pages}{3572} (\bibinfo{year}{1999}).

\bibitem[{\citenamefont{Hirai et~al.}(2003)\citenamefont{Hirai, Tanaka,
  Yoshida, Asano, Inoue, and Kashiwaya}}]{HTYAIK03}
\bibinfo{author}{\bibfnamefont{T.}~\bibnamefont{Hirai}},
  \bibinfo{author}{\bibfnamefont{Y.}~\bibnamefont{Tanaka}},
  \bibinfo{author}{\bibfnamefont{N.}~\bibnamefont{Yoshida}},
  \bibinfo{author}{\bibfnamefont{Y.}~\bibnamefont{Asano}},
  \bibinfo{author}{\bibfnamefont{J.}~\bibnamefont{Inoue}}, \bibnamefont{and}
  \bibinfo{author}{\bibfnamefont{S.}~\bibnamefont{Kashiwaya}},
  \bibinfo{journal}{Phys. Rev. B} \textbf{\bibinfo{volume}{67}},
  \bibinfo{pages}{174501} (\bibinfo{year}{2003}).

\bibitem[{\citenamefont{Higashitani}(1997)}]{Higashitani97}
\bibinfo{author}{\bibfnamefont{S.}~\bibnamefont{Higashitani}},
  \bibinfo{journal}{J. Phys. Soc. Jpn.} \textbf{\bibinfo{volume}{66}},
  \bibinfo{pages}{2556} (\bibinfo{year}{1997}).

\bibitem[{\citenamefont{Barash et~al.}(2000)\citenamefont{Barash, Kalenkov, and
  Kurkij{\"a}rvi}}]{BKK00}
\bibinfo{author}{\bibfnamefont{Y.~S.} \bibnamefont{Barash}},
  \bibinfo{author}{\bibfnamefont{M.~S.} \bibnamefont{Kalenkov}},
  \bibnamefont{and}
  \bibinfo{author}{\bibfnamefont{J.}~\bibnamefont{Kurkij{\"a}rvi}},
  \bibinfo{journal}{Phys. Rev. B} \textbf{\bibinfo{volume}{62}},
  \bibinfo{pages}{6665} (\bibinfo{year}{2000}).

\bibitem[{\citenamefont{Walter et~al.}(1998)\citenamefont{Walter, Prusseit,
  Semerad, Kinder, Assmann, Huber, Burkhardt, Rainer, and Sauls}}]{WPSKAHBRS98}
\bibinfo{author}{\bibfnamefont{H.}~\bibnamefont{Walter}},
  \bibinfo{author}{\bibfnamefont{W.}~\bibnamefont{Prusseit}},
  \bibinfo{author}{\bibfnamefont{R.}~\bibnamefont{Semerad}},
  \bibinfo{author}{\bibfnamefont{H.}~\bibnamefont{Kinder}},
  \bibinfo{author}{\bibfnamefont{W.}~\bibnamefont{Assmann}},
  \bibinfo{author}{\bibfnamefont{H.}~\bibnamefont{Huber}},
  \bibinfo{author}{\bibfnamefont{H.}~\bibnamefont{Burkhardt}},
  \bibinfo{author}{\bibfnamefont{D.}~\bibnamefont{Rainer}}, \bibnamefont{and}
  \bibinfo{author}{\bibfnamefont{J.}~\bibnamefont{Sauls}},
  \bibinfo{journal}{Phys. Rev. Lett.} \textbf{\bibinfo{volume}{80}},
  \bibinfo{pages}{3598} (\bibinfo{year}{1998}).

\bibitem[{\citenamefont{Tanaka et~al.}(2005{\natexlab{a}})\citenamefont{Tanaka,
  Asano, Golubov, and Kashiwaya}}]{TAGK05}
\bibinfo{author}{\bibfnamefont{Y.}~\bibnamefont{Tanaka}},
  \bibinfo{author}{\bibfnamefont{Y.}~\bibnamefont{Asano}},
  \bibinfo{author}{\bibfnamefont{A.~A.} \bibnamefont{Golubov}},
  \bibnamefont{and}
  \bibinfo{author}{\bibfnamefont{S.}~\bibnamefont{Kashiwaya}},
  \bibinfo{journal}{Phys. Rev. B} \textbf{\bibinfo{volume}{72}},
  \bibinfo{pages}{140503(R)} (\bibinfo{year}{2005}{\natexlab{a}}).

\bibitem[{\citenamefont{Kawabata et~al.}(2004)\citenamefont{Kawabata,
  Kashiwaya, Asano, and Tanaka}}]{Kawabata04}
\bibinfo{author}{\bibfnamefont{S.}~\bibnamefont{Kawabata}},
  \bibinfo{author}{\bibfnamefont{S.}~\bibnamefont{Kashiwaya}},
  \bibinfo{author}{\bibfnamefont{Y.}~\bibnamefont{Asano}}, \bibnamefont{and}
  \bibinfo{author}{\bibfnamefont{Y.}~\bibnamefont{Tanaka}},
  \bibinfo{journal}{Phys. Rev. B} \textbf{\bibinfo{volume}{70}},
  \bibinfo{pages}{132505} (\bibinfo{year}{2004}).

\bibitem[{\citenamefont{Kawabata et~al.}(2005)\citenamefont{Kawabata,
  Kashiwaya, Asano, and Tanaka}}]{Kawabata05}
\bibinfo{author}{\bibfnamefont{S.}~\bibnamefont{Kawabata}},
  \bibinfo{author}{\bibfnamefont{S.}~\bibnamefont{Kashiwaya}},
  \bibinfo{author}{\bibfnamefont{Y.}~\bibnamefont{Asano}}, \bibnamefont{and}
  \bibinfo{author}{\bibfnamefont{Y.}~\bibnamefont{Tanaka}},
  \bibinfo{journal}{Phys. Rev. B} \textbf{\bibinfo{volume}{72}},
  \bibinfo{pages}{052506} (\bibinfo{year}{2005}).

\bibitem[{\citenamefont{Yokoyama
  et~al.}(2007{\natexlab{a}})\citenamefont{Yokoyama, Kawabata, Kato, and
  Tanaka}}]{Yokoyama07}
\bibinfo{author}{\bibfnamefont{T.}~\bibnamefont{Yokoyama}},
  \bibinfo{author}{\bibfnamefont{S.}~\bibnamefont{Kawabata}},
  \bibinfo{author}{\bibfnamefont{T.}~\bibnamefont{Kato}}, \bibnamefont{and}
  \bibinfo{author}{\bibfnamefont{Y.}~\bibnamefont{Tanaka}},
  \bibinfo{journal}{Phys. Rev. B} \textbf{\bibinfo{volume}{76}},
  \bibinfo{pages}{134501} (\bibinfo{year}{2007}{\natexlab{a}}).

\bibitem[{\citenamefont{Tanaka et~al.}(2003)\citenamefont{Tanaka, Nazarov, and
  Kashiwaya}}]{TNS03}
\bibinfo{author}{\bibfnamefont{Y.}~\bibnamefont{Tanaka}},
  \bibinfo{author}{\bibfnamefont{Y.~V.} \bibnamefont{Nazarov}},
  \bibnamefont{and}
  \bibinfo{author}{\bibfnamefont{S.}~\bibnamefont{Kashiwaya}},
  \bibinfo{journal}{Phys. Rev. Lett.} \textbf{\bibinfo{volume}{90}},
  \bibinfo{pages}{167003} (\bibinfo{year}{2003}).

\bibitem[{\citenamefont{Tanaka et~al.}(2004)\citenamefont{Tanaka, Nazarov,
  Golubov, and Kashiwaya}}]{TNGK04}
\bibinfo{author}{\bibfnamefont{Y.}~\bibnamefont{Tanaka}},
  \bibinfo{author}{\bibfnamefont{Y.}~\bibnamefont{Nazarov}},
  \bibinfo{author}{\bibfnamefont{A.}~\bibnamefont{Golubov}}, \bibnamefont{and}
  \bibinfo{author}{\bibfnamefont{S.}~\bibnamefont{Kashiwaya}},
  \bibinfo{journal}{Phys. Rev. B} \textbf{\bibinfo{volume}{69}},
  \bibinfo{pages}{144519} (\bibinfo{year}{2004}).

\bibitem[{\citenamefont{Tanaka and Kashiwaya}(2004)}]{TK04}
\bibinfo{author}{\bibfnamefont{Y.}~\bibnamefont{Tanaka}} \bibnamefont{and}
  \bibinfo{author}{\bibfnamefont{S.}~\bibnamefont{Kashiwaya}},
  \bibinfo{journal}{Phys. Rev. B} \textbf{\bibinfo{volume}{70}},
  \bibinfo{pages}{012507} (\bibinfo{year}{2004}).

\bibitem[{\citenamefont{Tanaka et~al.}(2005{\natexlab{b}})\citenamefont{Tanaka,
  Kashiwaya, and Yokoyama}}]{TKY05}
\bibinfo{author}{\bibfnamefont{Y.}~\bibnamefont{Tanaka}},
  \bibinfo{author}{\bibfnamefont{S.}~\bibnamefont{Kashiwaya}},
  \bibnamefont{and} \bibinfo{author}{\bibfnamefont{T.}~\bibnamefont{Yokoyama}},
  \bibinfo{journal}{Phys. Rev. B} \textbf{\bibinfo{volume}{71}},
  \bibinfo{pages}{094513} (\bibinfo{year}{2005}{\natexlab{b}}).

\bibitem[{\citenamefont{Asano et~al.}(2006)\citenamefont{Asano, Tanaka, and
  Kashiwaya}}]{ATK06}
\bibinfo{author}{\bibfnamefont{Y.}~\bibnamefont{Asano}},
  \bibinfo{author}{\bibfnamefont{Y.}~\bibnamefont{Tanaka}}, \bibnamefont{and}
  \bibinfo{author}{\bibfnamefont{S.}~\bibnamefont{Kashiwaya}},
  \bibinfo{journal}{Phys. Rev. Lett.} \textbf{\bibinfo{volume}{96}},
  \bibinfo{pages}{097007} (\bibinfo{year}{2006}).

\bibitem[{\citenamefont{Tanaka and Golubov}(2007)}]{TG07}
\bibinfo{author}{\bibfnamefont{Y.}~\bibnamefont{Tanaka}} \bibnamefont{and}
  \bibinfo{author}{\bibfnamefont{A.~A.} \bibnamefont{Golubov}},
  \bibinfo{journal}{Phys. Rev. Lett.} \textbf{\bibinfo{volume}{98}},
  \bibinfo{pages}{037003} (\bibinfo{year}{2007}).

\bibitem[{\citenamefont{Eschrig et~al.}(2007)\citenamefont{Eschrig, Lofwander,
  Champel, Cuevas, and Schon}}]{ELCCS07}
\bibinfo{author}{\bibfnamefont{M.}~\bibnamefont{Eschrig}},
  \bibinfo{author}{\bibfnamefont{T.}~\bibnamefont{Lofwander}},
  \bibinfo{author}{\bibfnamefont{T.}~\bibnamefont{Champel}},
  \bibinfo{author}{\bibfnamefont{J.~C.} \bibnamefont{Cuevas}},
  \bibnamefont{and} \bibinfo{author}{\bibfnamefont{G.}~\bibnamefont{Schon}},
  \bibinfo{journal}{J. Low Temp. Phys.} \textbf{\bibinfo{volume}{147}},
  \bibinfo{pages}{457} (\bibinfo{year}{2007}).

\bibitem[{\citenamefont{Tanaka et~al.}(2007{\natexlab{a}})\citenamefont{Tanaka,
  Golubov, Kashiwaya, and Ueda}}]{TGKU07}
\bibinfo{author}{\bibfnamefont{Y.}~\bibnamefont{Tanaka}},
  \bibinfo{author}{\bibfnamefont{A.~A.} \bibnamefont{Golubov}},
  \bibinfo{author}{\bibfnamefont{S.}~\bibnamefont{Kashiwaya}},
  \bibnamefont{and} \bibinfo{author}{\bibfnamefont{M.}~\bibnamefont{Ueda}},
  \bibinfo{journal}{Phys. Rev. Lett.} \textbf{\bibinfo{volume}{99}},
  \bibinfo{pages}{037005} (\bibinfo{year}{2007}{\natexlab{a}}).

\bibitem[{\citenamefont{Tanaka et~al.}(2007{\natexlab{b}})\citenamefont{Tanaka,
  Tanuma, and Golubov}}]{TTG07}
\bibinfo{author}{\bibfnamefont{Y.}~\bibnamefont{Tanaka}},
  \bibinfo{author}{\bibfnamefont{Y.}~\bibnamefont{Tanuma}}, \bibnamefont{and}
  \bibinfo{author}{\bibfnamefont{A.~A.} \bibnamefont{Golubov}},
  \bibinfo{journal}{Phys. Rev. B} \textbf{\bibinfo{volume}{76}},
  \bibinfo{pages}{054522} (\bibinfo{year}{2007}{\natexlab{b}}).

\bibitem[{\citenamefont{Asano et~al.}(2007)\citenamefont{Asano, Tanaka,
  Golubov, and Kashiwaya}}]{ATGK07}
\bibinfo{author}{\bibfnamefont{Y.}~\bibnamefont{Asano}},
  \bibinfo{author}{\bibfnamefont{Y.}~\bibnamefont{Tanaka}},
  \bibinfo{author}{\bibfnamefont{A.}~\bibnamefont{Golubov}}, \bibnamefont{and}
  \bibinfo{author}{\bibfnamefont{S.}~\bibnamefont{Kashiwaya}},
  \bibinfo{journal}{Phys. Rev. Lett} \textbf{\bibinfo{volume}{99}},
  \bibinfo{pages}{067005} (\bibinfo{year}{2007}).

\bibitem[{\citenamefont{Higashitani et~al.}(2009)\citenamefont{Higashitani,
  Nagato, and Nagai}}]{Higashitani09}
\bibinfo{author}{\bibfnamefont{S.}~\bibnamefont{Higashitani}},
  \bibinfo{author}{\bibfnamefont{Y.}~\bibnamefont{Nagato}}, \bibnamefont{and}
  \bibinfo{author}{\bibfnamefont{K.}~\bibnamefont{Nagai}}, \bibinfo{journal}{J.
  Low. Temp. Phys.} \textbf{\bibinfo{volume}{155}}, \bibinfo{pages}{83}
  (\bibinfo{year}{2009}).

\bibitem[{\citenamefont{Maeno et~al.}(1994)\citenamefont{Maeno, Hashimoto,
  Yoshida, Nishizaki, Fujita, Bednorz, and Lichtenberg}}]{MHYNFBL94}
\bibinfo{author}{\bibfnamefont{Y.}~\bibnamefont{Maeno}},
  \bibinfo{author}{\bibfnamefont{H.}~\bibnamefont{Hashimoto}},
  \bibinfo{author}{\bibfnamefont{K.}~\bibnamefont{Yoshida}},
  \bibinfo{author}{\bibfnamefont{S.}~\bibnamefont{Nishizaki}},
  \bibinfo{author}{\bibfnamefont{T.}~\bibnamefont{Fujita}},
  \bibinfo{author}{\bibfnamefont{J.~G.} \bibnamefont{Bednorz}},
  \bibnamefont{and}
  \bibinfo{author}{\bibfnamefont{F.}~\bibnamefont{Lichtenberg}},
  \bibinfo{journal}{Nature} \textbf{\bibinfo{volume}{372}},
  \bibinfo{pages}{532} (\bibinfo{year}{1994}).

\bibitem[{\citenamefont{Matsumoto and Sigrist}(1999)}]{MS99}
\bibinfo{author}{\bibfnamefont{M.}~\bibnamefont{Matsumoto}} \bibnamefont{and}
  \bibinfo{author}{\bibfnamefont{M.}~\bibnamefont{Sigrist}},
  \bibinfo{journal}{J. Phys. Soc. Jpn.} \textbf{\bibinfo{volume}{68}},
  \bibinfo{pages}{994} (\bibinfo{year}{1999}).

\bibitem[{\citenamefont{Honerkamp and Sigrist}(1998)}]{HS98}
\bibinfo{author}{\bibfnamefont{C.}~\bibnamefont{Honerkamp}} \bibnamefont{and}
  \bibinfo{author}{\bibfnamefont{M.}~\bibnamefont{Sigrist}},
  \bibinfo{journal}{J. Low Temp. Phys.} \textbf{\bibinfo{volume}{111}},
  \bibinfo{pages}{895} (\bibinfo{year}{1998}).

\bibitem[{\citenamefont{Yamashiro et~al.}(1997)\citenamefont{Yamashiro, Tanaka,
  and Kashiwaya}}]{YTK97}
\bibinfo{author}{\bibfnamefont{M.}~\bibnamefont{Yamashiro}},
  \bibinfo{author}{\bibfnamefont{Y.}~\bibnamefont{Tanaka}}, \bibnamefont{and}
  \bibinfo{author}{\bibfnamefont{S.}~\bibnamefont{Kashiwaya}},
  \bibinfo{journal}{Phys. Rev. B} \textbf{\bibinfo{volume}{56}},
  \bibinfo{pages}{7847} (\bibinfo{year}{1997}).

\bibitem[{\citenamefont{G.E.Volovik}(1997)}]{Volovik97}
\bibinfo{author}{\bibnamefont{G.E.Volovik}}, \bibinfo{journal}{JETP Lett.}
  \textbf{\bibinfo{volume}{66}}, \bibinfo{pages}{522} (\bibinfo{year}{1997}).

\bibitem[{\citenamefont{Read and Green}(2000)}]{RG00}
\bibinfo{author}{\bibfnamefont{N.}~\bibnamefont{Read}} \bibnamefont{and}
  \bibinfo{author}{\bibfnamefont{D.}~\bibnamefont{Green}},
  \bibinfo{journal}{Phys. Rev. B} \textbf{\bibinfo{volume}{61}},
  \bibinfo{pages}{10267} (\bibinfo{year}{2000}).

\bibitem[{\citenamefont{Goryo and Ishikawa}(1998)}]{GI98}
\bibinfo{author}{\bibfnamefont{J.}~\bibnamefont{Goryo}} \bibnamefont{and}
  \bibinfo{author}{\bibfnamefont{K.}~\bibnamefont{Ishikawa}},
  \bibinfo{journal}{J. Phys. Soc. Jpn.} \textbf{\bibinfo{volume}{67}},
  \bibinfo{pages}{3006} (\bibinfo{year}{1998}).

\bibitem[{\citenamefont{Furusaki et~al.}(2001)\citenamefont{Furusaki,
  Matsumoto, and Sigrist}}]{FMS01}
\bibinfo{author}{\bibfnamefont{A.}~\bibnamefont{Furusaki}},
  \bibinfo{author}{\bibfnamefont{M.}~\bibnamefont{Matsumoto}},
  \bibnamefont{and} \bibinfo{author}{\bibfnamefont{M.}~\bibnamefont{Sigrist}},
  \bibinfo{journal}{Phys. Rev. B} \textbf{\bibinfo{volume}{64}},
  \bibinfo{pages}{054514} (\bibinfo{year}{2001}).

\bibitem[{\citenamefont{Kitaev}(2003)}]{Kitaev03}
\bibinfo{author}{\bibfnamefont{A.}~\bibnamefont{Kitaev}},
  \bibinfo{journal}{Ann. Phys.} \textbf{\bibinfo{volume}{302}},
  \bibinfo{pages}{2} (\bibinfo{year}{2003}).

\bibitem[{\citenamefont{Freedman et~al.}(2003)\citenamefont{Freedman, Kitaev,
  Larsen, and Wang}}]{FKLW03}
\bibinfo{author}{\bibfnamefont{M.}~\bibnamefont{Freedman}},
  \bibinfo{author}{\bibfnamefont{A.}~\bibnamefont{Kitaev}},
  \bibinfo{author}{\bibfnamefont{M.}~\bibnamefont{Larsen}}, \bibnamefont{and}
  \bibinfo{author}{\bibfnamefont{Z.}~\bibnamefont{Wang}},
  \bibinfo{journal}{Bull. Am. Math. Soc.} \textbf{\bibinfo{volume}{40}},
  \bibinfo{pages}{31} (\bibinfo{year}{2003}).

\bibitem[{\citenamefont{Ivanov}(2001)}]{Ivanov01}
\bibinfo{author}{\bibfnamefont{D.~A.} \bibnamefont{Ivanov}},
  \bibinfo{journal}{Phys. Rev. Lett.} \textbf{\bibinfo{volume}{86}},
  \bibinfo{pages}{268} (\bibinfo{year}{2001}).

\bibitem[{\citenamefont{Sato and Fujimoto}(2009)}]{SF09}
\bibinfo{author}{\bibfnamefont{M.}~\bibnamefont{Sato}} \bibnamefont{and}
  \bibinfo{author}{\bibfnamefont{S.}~\bibnamefont{Fujimoto}},
  \bibinfo{journal}{Phys. Rev. B} \textbf{\bibinfo{volume}{79}},
  \bibinfo{pages}{094504} (\bibinfo{year}{2009}).

\bibitem[{\citenamefont{Sato}(2010{\natexlab{a}})}]{Sato10}
\bibinfo{author}{\bibfnamefont{M.}~\bibnamefont{Sato}}, \bibinfo{journal}{Phys.
  Rev. B} \textbf{\bibinfo{volume}{81}}, \bibinfo{pages}{220504(R)}
  (\bibinfo{year}{2010}{\natexlab{a}}).

\bibitem[{\citenamefont{M.Sato}(2003)}]{Sato03}
\bibinfo{author}{\bibnamefont{M.Sato}}, \bibinfo{journal}{Phys. Lett. B}
  \textbf{\bibinfo{volume}{575}}, \bibinfo{pages}{126} (\bibinfo{year}{2003}).

\bibitem[{\citenamefont{L.Fu and Kane}(2008)}]{FK08}
\bibinfo{author}{\bibnamefont{L.Fu}} \bibnamefont{and}
  \bibinfo{author}{\bibfnamefont{C.~L.} \bibnamefont{Kane}},
  \bibinfo{journal}{Phys. Rev. Lett.} \textbf{\bibinfo{volume}{100}},
  \bibinfo{pages}{096407} (\bibinfo{year}{2008}).

\bibitem[{\citenamefont{Sato et~al.}(2009)\citenamefont{Sato, Takahashi, and
  Fujimoto}}]{STF09}
\bibinfo{author}{\bibfnamefont{M.}~\bibnamefont{Sato}},
  \bibinfo{author}{\bibfnamefont{Y.}~\bibnamefont{Takahashi}},
  \bibnamefont{and} \bibinfo{author}{\bibfnamefont{S.}~\bibnamefont{Fujimoto}},
  \bibinfo{journal}{Phys. Rev. Lett.} \textbf{\bibinfo{volume}{103}},
  \bibinfo{pages}{020401} (\bibinfo{year}{2009}).

\bibitem[{\citenamefont{Sato et~al.}(2010)\citenamefont{Sato, Takahashi, and
  Fujimoto}}]{STF10}
\bibinfo{author}{\bibfnamefont{M.}~\bibnamefont{Sato}},
  \bibinfo{author}{\bibfnamefont{Y.}~\bibnamefont{Takahashi}},
  \bibnamefont{and} \bibinfo{author}{\bibfnamefont{S.}~\bibnamefont{Fujimoto}},
  \bibinfo{journal}{Phys. Rev. B} \textbf{\bibinfo{volume}{82}},
  \bibinfo{pages}{134521} (\bibinfo{year}{2010}).

\bibitem[{\citenamefont{Sau et~al.}(2010)\citenamefont{Sau, Lutchyn, Tewari,
  and Sarma}}]{SLTD10}
\bibinfo{author}{\bibfnamefont{J.~D.} \bibnamefont{Sau}},
  \bibinfo{author}{\bibfnamefont{R.~M.} \bibnamefont{Lutchyn}},
  \bibinfo{author}{\bibfnamefont{S.}~\bibnamefont{Tewari}}, \bibnamefont{and}
  \bibinfo{author}{\bibfnamefont{S.~D.} \bibnamefont{Sarma}},
  \bibinfo{journal}{Phys. Rev. Lett.} \textbf{\bibinfo{volume}{104}},
  \bibinfo{pages}{040502} (\bibinfo{year}{2010}).

\bibitem[{\citenamefont{Alicea}(2010)}]{Alicea10}
\bibinfo{author}{\bibfnamefont{J.}~\bibnamefont{Alicea}},
  \bibinfo{journal}{Phys. Rev. B} \textbf{\bibinfo{volume}{81}},
  \bibinfo{pages}{125318} (\bibinfo{year}{2010}).

\bibitem[{\citenamefont{Sato and Fujimoto}(2010)}]{SF10}
\bibinfo{author}{\bibfnamefont{M.}~\bibnamefont{Sato}} \bibnamefont{and}
  \bibinfo{author}{\bibfnamefont{S.}~\bibnamefont{Fujimoto}},
  \bibinfo{journal}{Phys. Rev. Lett.} \textbf{\bibinfo{volume}{105}},
  \bibinfo{pages}{217001} (\bibinfo{year}{2010}).

\bibitem[{\citenamefont{Fu and Kane}(2009)}]{FK09}
\bibinfo{author}{\bibfnamefont{L.}~\bibnamefont{Fu}} \bibnamefont{and}
  \bibinfo{author}{\bibfnamefont{C.~L.} \bibnamefont{Kane}},
  \bibinfo{journal}{Phys. Rev. Lett.} \textbf{\bibinfo{volume}{102}},
  \bibinfo{pages}{216403} (\bibinfo{year}{2009}).

\bibitem[{\citenamefont{Nilsson et~al.}(2008)\citenamefont{Nilsson, Akhmerov,
  and Beenakker}}]{NAB08}
\bibinfo{author}{\bibfnamefont{J.}~\bibnamefont{Nilsson}},
  \bibinfo{author}{\bibfnamefont{A.~R.} \bibnamefont{Akhmerov}},
  \bibnamefont{and} \bibinfo{author}{\bibfnamefont{C.~W.~J.}
  \bibnamefont{Beenakker}}, \bibinfo{journal}{Phys. Rev. Lett.}
  \textbf{\bibinfo{volume}{101}}, \bibinfo{pages}{120403}
  (\bibinfo{year}{2008}).

\bibitem[{\citenamefont{Akhmerov et~al.}(2009)\citenamefont{Akhmerov, Nilsson,
  and Beenakker}}]{ANB09}
\bibinfo{author}{\bibfnamefont{A.~R.} \bibnamefont{Akhmerov}},
  \bibinfo{author}{\bibfnamefont{J.}~\bibnamefont{Nilsson}}, \bibnamefont{and}
  \bibinfo{author}{\bibfnamefont{C.~W.~J.} \bibnamefont{Beenakker}},
  \bibinfo{journal}{Phys. Rev. Lett.} \textbf{\bibinfo{volume}{102}},
  \bibinfo{pages}{216404} (\bibinfo{year}{2009}).

\bibitem[{\citenamefont{Tanaka et~al.}(2009{\natexlab{a}})\citenamefont{Tanaka,
  Yokoyama, and Nagaosa}}]{TYN09}
\bibinfo{author}{\bibfnamefont{Y.}~\bibnamefont{Tanaka}},
  \bibinfo{author}{\bibfnamefont{T.}~\bibnamefont{Yokoyama}}, \bibnamefont{and}
  \bibinfo{author}{\bibfnamefont{N.}~\bibnamefont{Nagaosa}},
  \bibinfo{journal}{Phys. Rev. Lett.} \textbf{\bibinfo{volume}{103}},
  \bibinfo{pages}{107002} (\bibinfo{year}{2009}{\natexlab{a}}).

\bibitem[{\citenamefont{Law et~al.}(2009)\citenamefont{Law, Lee, and
  Ng}}]{LLN09}
\bibinfo{author}{\bibfnamefont{K.~T.} \bibnamefont{Law}},
  \bibinfo{author}{\bibfnamefont{P.~A.} \bibnamefont{Lee}}, \bibnamefont{and}
  \bibinfo{author}{\bibfnamefont{T.~K.} \bibnamefont{Ng}},
  \bibinfo{journal}{Phys. Rev. Lett.} \textbf{\bibinfo{volume}{103}},
  \bibinfo{pages}{237001} (\bibinfo{year}{2009}).

\bibitem[{\citenamefont{Linder et~al.}(2010{\natexlab{a}})\citenamefont{Linder,
  Tanaka, Yokoyama, Sudbo, and Nagaosa}}]{LTYSN10a}
\bibinfo{author}{\bibfnamefont{J.}~\bibnamefont{Linder}},
  \bibinfo{author}{\bibfnamefont{Y.}~\bibnamefont{Tanaka}},
  \bibinfo{author}{\bibfnamefont{T.}~\bibnamefont{Yokoyama}},
  \bibinfo{author}{\bibfnamefont{A.}~\bibnamefont{Sudbo}}, \bibnamefont{and}
  \bibinfo{author}{\bibfnamefont{N.}~\bibnamefont{Nagaosa}},
  \bibinfo{journal}{Phys. Rev. Lett.} \textbf{\bibinfo{volume}{104}},
  \bibinfo{pages}{067001} (\bibinfo{year}{2010}{\natexlab{a}}).

\bibitem[{\citenamefont{Linder et~al.}(2010{\natexlab{b}})\citenamefont{Linder,
  Tanaka, Yokoyama, Sudbo, and Nagaosa}}]{LTYSN10b}
\bibinfo{author}{\bibfnamefont{J.}~\bibnamefont{Linder}},
  \bibinfo{author}{\bibfnamefont{Y.}~\bibnamefont{Tanaka}},
  \bibinfo{author}{\bibfnamefont{T.}~\bibnamefont{Yokoyama}},
  \bibinfo{author}{\bibfnamefont{A.}~\bibnamefont{Sudbo}}, \bibnamefont{and}
  \bibinfo{author}{\bibfnamefont{N.}~\bibnamefont{Nagaosa}},
  \bibinfo{journal}{Phys. Rev. B} \textbf{\bibinfo{volume}{81}},
  \bibinfo{pages}{184525} (\bibinfo{year}{2010}{\natexlab{b}}).

\bibitem[{\citenamefont{Shivamoggi et~al.}(2010)\citenamefont{Shivamoggi,
  Refael, and Moore}}]{SRM10}
\bibinfo{author}{\bibfnamefont{V.}~\bibnamefont{Shivamoggi}},
  \bibinfo{author}{\bibfnamefont{G.}~\bibnamefont{Refael}}, \bibnamefont{and}
  \bibinfo{author}{\bibfnamefont{J.~E.} \bibnamefont{Moore}},
  \bibinfo{journal}{Phys. Rev. B} \textbf{\bibinfo{volume}{82}},
  \bibinfo{pages}{041405(R)} (\bibinfo{year}{2010}).

\bibitem[{\citenamefont{Flensberg}(2010)}]{Flensberg10}
\bibinfo{author}{\bibfnamefont{K.}~\bibnamefont{Flensberg}},
  \bibinfo{journal}{Phys. Rev. B} \textbf{\bibinfo{volume}{82}},
  \bibinfo{pages}{180516(R)} (\bibinfo{year}{2010}).

\bibitem[{\citenamefont{Shindou et~al.}(2010)\citenamefont{Shindou, Furusaki,
  and Nagaosa}}]{SFN10}
\bibinfo{author}{\bibfnamefont{R.}~\bibnamefont{Shindou}},
  \bibinfo{author}{\bibfnamefont{A.}~\bibnamefont{Furusaki}}, \bibnamefont{and}
  \bibinfo{author}{\bibfnamefont{N.}~\bibnamefont{Nagaosa}},
  \bibinfo{journal}{Phys. Rev. B} \textbf{\bibinfo{volume}{82}},
  \bibinfo{pages}{180505} (\bibinfo{year}{2010}).

\bibitem[{\citenamefont{Mao and Zhang}(2010)}]{MZ10}
\bibinfo{author}{\bibfnamefont{L.}~\bibnamefont{Mao}} \bibnamefont{and}
  \bibinfo{author}{\bibfnamefont{C.}~\bibnamefont{Zhang}},
  \bibinfo{journal}{Phys. Rev. B} \textbf{\bibinfo{volume}{82}},
  \bibinfo{pages}{174506} (\bibinfo{year}{2010}).

\bibitem[{\citenamefont{Neupert et~al.}(2010)\citenamefont{Neupert, Onoda, and
  Furusaki}}]{NOF10}
\bibinfo{author}{\bibfnamefont{T.}~\bibnamefont{Neupert}},
  \bibinfo{author}{\bibfnamefont{S.}~\bibnamefont{Onoda}}, \bibnamefont{and}
  \bibinfo{author}{\bibfnamefont{A.}~\bibnamefont{Furusaki}},
  \bibinfo{journal}{Phys. Rev. Lett.} \textbf{\bibinfo{volume}{105}},
  \bibinfo{pages}{206404} (\bibinfo{year}{2010}).

\bibitem[{\citenamefont{Bauer et~al.}(2004)\citenamefont{Bauer, Hilscher,
  Michor, Paul, Scheidt, Gribanov, Seropegin, No{\"e}l, Sigrist, and
  Rogl}}]{BHMPSGSNSR04}
\bibinfo{author}{\bibfnamefont{E.}~\bibnamefont{Bauer}},
  \bibinfo{author}{\bibfnamefont{G.}~\bibnamefont{Hilscher}},
  \bibinfo{author}{\bibfnamefont{H.}~\bibnamefont{Michor}},
  \bibinfo{author}{\bibfnamefont{C.}~\bibnamefont{Paul}},
  \bibinfo{author}{\bibfnamefont{E.}~\bibnamefont{Scheidt}},
  \bibinfo{author}{\bibfnamefont{A.}~\bibnamefont{Gribanov}},
  \bibinfo{author}{\bibfnamefont{Y.}~\bibnamefont{Seropegin}},
  \bibinfo{author}{\bibfnamefont{H.}~\bibnamefont{No{\"e}l}},
  \bibinfo{author}{\bibfnamefont{M.}~\bibnamefont{Sigrist}}, \bibnamefont{and}
  \bibinfo{author}{\bibfnamefont{P.}~\bibnamefont{Rogl}},
  \bibinfo{journal}{Phys. Rev. Lett.} \textbf{\bibinfo{volume}{92}},
  \bibinfo{pages}{027003} (\bibinfo{year}{2004}).

\bibitem[{\citenamefont{Togano et~al.}(2004)\citenamefont{Togano, Badica,
  Nakamori, Orimo, Takeya, and Hirata}}]{TBNOTH04}
\bibinfo{author}{\bibfnamefont{K.}~\bibnamefont{Togano}},
  \bibinfo{author}{\bibfnamefont{P.}~\bibnamefont{Badica}},
  \bibinfo{author}{\bibfnamefont{Y.}~\bibnamefont{Nakamori}},
  \bibinfo{author}{\bibfnamefont{S.}~\bibnamefont{Orimo}},
  \bibinfo{author}{\bibfnamefont{H.}~\bibnamefont{Takeya}}, \bibnamefont{and}
  \bibinfo{author}{\bibfnamefont{K.}~\bibnamefont{Hirata}},
  \bibinfo{journal}{Phys. Rev. Lett.} \textbf{\bibinfo{volume}{93}},
  \bibinfo{pages}{247004} (\bibinfo{year}{2004}).

\bibitem[{\citenamefont{Nishiyama et~al.}(2005)\citenamefont{Nishiyama, Inada,
  and Zheng}}]{NIZ05}
\bibinfo{author}{\bibfnamefont{M.}~\bibnamefont{Nishiyama}},
  \bibinfo{author}{\bibfnamefont{Y.}~\bibnamefont{Inada}}, \bibnamefont{and}
  \bibinfo{author}{\bibfnamefont{G.~Q.} \bibnamefont{Zheng}},
  \bibinfo{journal}{Phys. Rev. B} \textbf{\bibinfo{volume}{71}},
  \bibinfo{pages}{220505(R)} (\bibinfo{year}{2005}).

\bibitem[{\citenamefont{Hillier et~al.}(2009)\citenamefont{Hillier,
  Quintanilla, and Cywinski}}]{HQC09}
\bibinfo{author}{\bibfnamefont{A.~D.} \bibnamefont{Hillier}},
  \bibinfo{author}{\bibfnamefont{J.}~\bibnamefont{Quintanilla}},
  \bibnamefont{and} \bibinfo{author}{\bibfnamefont{R.}~\bibnamefont{Cywinski}},
  \bibinfo{journal}{Phys. Rev. Lett.} \textbf{\bibinfo{volume}{102}},
  \bibinfo{pages}{117007} (\bibinfo{year}{2009}).

\bibitem[{\citenamefont{Reyren et~al.}(2007)\citenamefont{Reyren, Thiel, S,
  Kourkoutis, Hammerl, Richter, Schneider, Kopp, Ruetschi, Jaccard
  et~al.}}]{RTCKHRSKRJGMTM07}
\bibinfo{author}{\bibfnamefont{N.}~\bibnamefont{Reyren}},
  \bibinfo{author}{\bibfnamefont{S.}~\bibnamefont{Thiel}},
  \bibinfo{author}{\bibfnamefont{A.~D.~C.} \bibnamefont{S}},
  \bibinfo{author}{\bibfnamefont{L.~F.} \bibnamefont{Kourkoutis}},
  \bibinfo{author}{\bibfnamefont{G.}~\bibnamefont{Hammerl}},
  \bibinfo{author}{\bibfnamefont{C.}~\bibnamefont{Richter}},
  \bibinfo{author}{\bibfnamefont{C.~W.} \bibnamefont{Schneider}},
  \bibinfo{author}{\bibfnamefont{T.}~\bibnamefont{Kopp}},
  \bibinfo{author}{\bibfnamefont{A.~S.} \bibnamefont{Ruetschi}},
  \bibinfo{author}{\bibfnamefont{D.}~\bibnamefont{Jaccard}},
  \bibnamefont{et~al.}, \bibinfo{journal}{Science}
  \textbf{\bibinfo{volume}{317}}, \bibinfo{pages}{1196} (\bibinfo{year}{2007}).

\bibitem[{\citenamefont{Gor'kov and Rashba}(2001)}]{GR01}
\bibinfo{author}{\bibfnamefont{L.~P.} \bibnamefont{Gor'kov}} \bibnamefont{and}
  \bibinfo{author}{\bibfnamefont{E.~I.} \bibnamefont{Rashba}},
  \bibinfo{journal}{Phys. Rev. Lett.} \textbf{\bibinfo{volume}{87}},
  \bibinfo{pages}{037004} (\bibinfo{year}{2001}).

\bibitem[{\citenamefont{Frigeri et~al.}(2004)\citenamefont{Frigeri, Agterberg,
  Koga, and Sigrist}}]{FAKS04}
\bibinfo{author}{\bibfnamefont{P.~A.} \bibnamefont{Frigeri}},
  \bibinfo{author}{\bibfnamefont{D.~F.} \bibnamefont{Agterberg}},
  \bibinfo{author}{\bibfnamefont{A.}~\bibnamefont{Koga}}, \bibnamefont{and}
  \bibinfo{author}{\bibfnamefont{M.}~\bibnamefont{Sigrist}},
  \bibinfo{journal}{Phys. Rev. Lett.} \textbf{\bibinfo{volume}{92}},
  \bibinfo{pages}{097001} (\bibinfo{year}{2004}).

\bibitem[{\citenamefont{Yokoyama et~al.}(2005)\citenamefont{Yokoyama, Tanaka,
  and Inoue}}]{YTJ05}
\bibinfo{author}{\bibfnamefont{T.}~\bibnamefont{Yokoyama}},
  \bibinfo{author}{\bibfnamefont{Y.}~\bibnamefont{Tanaka}}, \bibnamefont{and}
  \bibinfo{author}{\bibfnamefont{J.}~\bibnamefont{Inoue}},
  \bibinfo{journal}{Phys. Rev. B} \textbf{\bibinfo{volume}{72}},
  \bibinfo{pages}{220504(R)} (\bibinfo{year}{2005}).

\bibitem[{\citenamefont{Iniotakis et~al.}(2007)\citenamefont{Iniotakis,
  Hayashi, Sawa, Yokoyama, May, Tanaka, and Sigrist}}]{IHSYMTS07}
\bibinfo{author}{\bibfnamefont{C.}~\bibnamefont{Iniotakis}},
  \bibinfo{author}{\bibfnamefont{N.}~\bibnamefont{Hayashi}},
  \bibinfo{author}{\bibfnamefont{Y.}~\bibnamefont{Sawa}},
  \bibinfo{author}{\bibfnamefont{T.}~\bibnamefont{Yokoyama}},
  \bibinfo{author}{\bibfnamefont{U.}~\bibnamefont{May}},
  \bibinfo{author}{\bibfnamefont{Y.}~\bibnamefont{Tanaka}}, \bibnamefont{and}
  \bibinfo{author}{\bibfnamefont{M.}~\bibnamefont{Sigrist}},
  \bibinfo{journal}{Phys. Rev. B} \textbf{\bibinfo{volume}{76}},
  \bibinfo{pages}{012501} (\bibinfo{year}{2007}).

\bibitem[{\citenamefont{Eschrig et~al.}()\citenamefont{Eschrig, Inotakis, and
  Tanaka}}]{EIT10}
\bibinfo{author}{\bibfnamefont{M.}~\bibnamefont{Eschrig}},
  \bibinfo{author}{\bibfnamefont{C.}~\bibnamefont{Inotakis}}, \bibnamefont{and}
  \bibinfo{author}{\bibfnamefont{Y.}~\bibnamefont{Tanaka}},
  \eprint{arXiv:1001.2486}.

\bibitem[{\citenamefont{Vorontsov et~al.}(2008)\citenamefont{Vorontsov,
  Vekhter, and Eschrig}}]{VVE08}
\bibinfo{author}{\bibfnamefont{A.}~\bibnamefont{Vorontsov}},
  \bibinfo{author}{\bibfnamefont{I.}~\bibnamefont{Vekhter}}, \bibnamefont{and}
  \bibinfo{author}{\bibfnamefont{M.}~\bibnamefont{Eschrig}},
  \bibinfo{journal}{Phys. Rev. Lett.} \textbf{\bibinfo{volume}{101}},
  \bibinfo{pages}{127003} (\bibinfo{year}{2008}).

\bibitem[{\citenamefont{Tanaka et~al.}(2009{\natexlab{b}})\citenamefont{Tanaka,
  Yokoyama, Balatsky, and Nagaosa}}]{TYBN09}
\bibinfo{author}{\bibfnamefont{Y.}~\bibnamefont{Tanaka}},
  \bibinfo{author}{\bibfnamefont{T.}~\bibnamefont{Yokoyama}},
  \bibinfo{author}{\bibfnamefont{A.~V.} \bibnamefont{Balatsky}},
  \bibnamefont{and} \bibinfo{author}{\bibfnamefont{N.}~\bibnamefont{Nagaosa}},
  \bibinfo{journal}{Phys. Rev. B} \textbf{\bibinfo{volume}{79}},
  \bibinfo{pages}{060505(R)} (\bibinfo{year}{2009}{\natexlab{b}}).

\bibitem[{\citenamefont{Sato}(2006)}]{Sato06}
\bibinfo{author}{\bibfnamefont{M.}~\bibnamefont{Sato}}, \bibinfo{journal}{Phys.
  Rev. B} \textbf{\bibinfo{volume}{73}}, \bibinfo{pages}{214502}
  (\bibinfo{year}{2006}).

\bibitem[{\citenamefont{Lu and Yip}(2009)}]{LY09}
\bibinfo{author}{\bibfnamefont{C.}~\bibnamefont{Lu}} \bibnamefont{and}
  \bibinfo{author}{\bibfnamefont{S.}~\bibnamefont{Yip}},
  \bibinfo{journal}{Phys. Rev. B} \textbf{\bibinfo{volume}{80}},
  \bibinfo{pages}{024504} (\bibinfo{year}{2009}).

\bibitem[{\citenamefont{Schnyder et~al.}(2010)\citenamefont{Schnyder, Brydon,
  Manske, and Timm}}]{SBMT10}
\bibinfo{author}{\bibfnamefont{A.~P.} \bibnamefont{Schnyder}},
  \bibinfo{author}{\bibfnamefont{P.~M.~R.} \bibnamefont{Brydon}},
  \bibinfo{author}{\bibfnamefont{D.}~\bibnamefont{Manske}}, \bibnamefont{and}
  \bibinfo{author}{\bibfnamefont{C.}~\bibnamefont{Timm}},
  \bibinfo{journal}{Phys. Rev. B} \textbf{\bibinfo{volume}{82}},
  \bibinfo{pages}{184508} (\bibinfo{year}{2010}).

\bibitem[{\citenamefont{Kane and Mele}(2005{\natexlab{a}})}]{KM05a}
\bibinfo{author}{\bibfnamefont{C.~L.} \bibnamefont{Kane}} \bibnamefont{and}
  \bibinfo{author}{\bibfnamefont{E.~J.} \bibnamefont{Mele}},
  \bibinfo{journal}{Phys. Rev. Lett.} \textbf{\bibinfo{volume}{95}},
  \bibinfo{pages}{146802} (\bibinfo{year}{2005}{\natexlab{a}}).

\bibitem[{\citenamefont{Kane and Mele}(2005{\natexlab{b}})}]{KM05b}
\bibinfo{author}{\bibfnamefont{C.~L.} \bibnamefont{Kane}} \bibnamefont{and}
  \bibinfo{author}{\bibfnamefont{E.~J.} \bibnamefont{Mele}},
  \bibinfo{journal}{Phys. Rev. Lett.} \textbf{\bibinfo{volume}{95}},
  \bibinfo{pages}{226801} (\bibinfo{year}{2005}{\natexlab{b}}).

\bibitem[{\citenamefont{Bernevig and Zhang}(2006)}]{BZ06}
\bibinfo{author}{\bibfnamefont{B.~A.} \bibnamefont{Bernevig}} \bibnamefont{and}
  \bibinfo{author}{\bibfnamefont{S.~C.} \bibnamefont{Zhang}},
  \bibinfo{journal}{Phys. Rev. Lett.} \textbf{\bibinfo{volume}{96}},
  \bibinfo{pages}{106802} (\bibinfo{year}{2006}).

\bibitem[{\citenamefont{Bernevig et~al.}(2006)\citenamefont{Bernevig, Hughes,
  and Zhang}}]{BHZ06}
\bibinfo{author}{\bibfnamefont{B.~A.} \bibnamefont{Bernevig}},
  \bibinfo{author}{\bibfnamefont{T.~L.} \bibnamefont{Hughes}},
  \bibnamefont{and} \bibinfo{author}{\bibfnamefont{S.~C.} \bibnamefont{Zhang}},
  \bibinfo{journal}{Science} \textbf{\bibinfo{volume}{314}},
  \bibinfo{pages}{1757} (\bibinfo{year}{2006}).

\bibitem[{\citenamefont{Asano et~al.}(2010)\citenamefont{Asano, Tanaka, and
  Nagaosa}}]{ATN10}
\bibinfo{author}{\bibfnamefont{Y.}~\bibnamefont{Asano}},
  \bibinfo{author}{\bibfnamefont{Y.}~\bibnamefont{Tanaka}}, \bibnamefont{and}
  \bibinfo{author}{\bibfnamefont{N.}~\bibnamefont{Nagaosa}},
  \bibinfo{journal}{Phys. Rev. Lett.} \textbf{\bibinfo{volume}{105}},
  \bibinfo{pages}{056402} (\bibinfo{year}{2010}).

\bibitem[{\citenamefont{Tanaka et~al.}(2010)\citenamefont{Tanaka, Mizuno,
  Yokoyama, Yada, and Sato}}]{TMYYS10}
\bibinfo{author}{\bibfnamefont{Y.}~\bibnamefont{Tanaka}},
  \bibinfo{author}{\bibfnamefont{Y.}~\bibnamefont{Mizuno}},
  \bibinfo{author}{\bibfnamefont{T.}~\bibnamefont{Yokoyama}},
  \bibinfo{author}{\bibfnamefont{K.}~\bibnamefont{Yada}}, \bibnamefont{and}
  \bibinfo{author}{\bibfnamefont{M.}~\bibnamefont{Sato}},
  \bibinfo{journal}{Phys. Rev. Lett.} \textbf{\bibinfo{volume}{105}},
  \bibinfo{pages}{097002} (\bibinfo{year}{2010}).

\bibitem[{\citenamefont{Yada et~al.}()\citenamefont{Yada, Sato, Tanak, and
  Yokoyama}}]{YSTY10}
\bibinfo{author}{\bibfnamefont{K.}~\bibnamefont{Yada}},
  \bibinfo{author}{\bibfnamefont{M.}~\bibnamefont{Sato}},
  \bibinfo{author}{\bibfnamefont{Y.}~\bibnamefont{Tanak}}, \bibnamefont{and}
  \bibinfo{author}{\bibfnamefont{T.}~\bibnamefont{Yokoyama}},
  \eprint{arXiv:1011.2002, to be published in Phys. Rev. B.}

\bibitem[{\citenamefont{Yokoyama
  et~al.}(2007{\natexlab{b}})\citenamefont{Yokoyama, Onari, and
  Tanaka}}]{YOT07a}
\bibinfo{author}{\bibfnamefont{T.}~\bibnamefont{Yokoyama}},
  \bibinfo{author}{\bibfnamefont{S.}~\bibnamefont{Onari}}, \bibnamefont{and}
  \bibinfo{author}{\bibfnamefont{Y.}~\bibnamefont{Tanaka}},
  \bibinfo{journal}{Phys. Rev. B} \textbf{\bibinfo{volume}{75}},
  \bibinfo{pages}{172511} (\bibinfo{year}{2007}{\natexlab{b}}).

\bibitem[{\citenamefont{Yokoyama
  et~al.}(2008{\natexlab{a}})\citenamefont{Yokoyama, Onari, and
  Tanaka}}]{YOT07b}
\bibinfo{author}{\bibfnamefont{T.}~\bibnamefont{Yokoyama}},
  \bibinfo{author}{\bibfnamefont{S.}~\bibnamefont{Onari}}, \bibnamefont{and}
  \bibinfo{author}{\bibfnamefont{Y.}~\bibnamefont{Tanaka}},
  \bibinfo{journal}{Phys. Rev. B} \textbf{\bibinfo{volume}{78}},
  \bibinfo{pages}{029902(E)} (\bibinfo{year}{2008}{\natexlab{a}}).

\bibitem[{\citenamefont{Yokoyama
  et~al.}(2008{\natexlab{b}})\citenamefont{Yokoyama, Onari, and
  Tanaka}}]{YOT08a}
\bibinfo{author}{\bibfnamefont{T.}~\bibnamefont{Yokoyama}},
  \bibinfo{author}{\bibfnamefont{S.}~\bibnamefont{Onari}}, \bibnamefont{and}
  \bibinfo{author}{\bibfnamefont{Y.}~\bibnamefont{Tanaka}},
  \bibinfo{journal}{J. Phys. Soc. Jpn.} \textbf{\bibinfo{volume}{77}},
  \bibinfo{pages}{064711} (\bibinfo{year}{2008}{\natexlab{b}}).

\bibitem[{\citenamefont{Yokoyama
  et~al.}(2008{\natexlab{c}})\citenamefont{Yokoyama, Onari, and
  Tanaka}}]{YOT08b}
\bibinfo{author}{\bibfnamefont{T.}~\bibnamefont{Yokoyama}},
  \bibinfo{author}{\bibfnamefont{S.}~\bibnamefont{Onari}}, \bibnamefont{and}
  \bibinfo{author}{\bibfnamefont{Y.}~\bibnamefont{Tanaka}},
  \bibinfo{journal}{J. Phys. Soc. Jpn.} \textbf{\bibinfo{volume}{77}},
  \bibinfo{pages}{088001(E)} (\bibinfo{year}{2008}{\natexlab{c}}).

\bibitem[{\citenamefont{Yada et~al.}(2009)\citenamefont{Yada, Onari, Tanaka,
  and Inoue}}]{YOTI09}
\bibinfo{author}{\bibfnamefont{K.}~\bibnamefont{Yada}},
  \bibinfo{author}{\bibfnamefont{S.}~\bibnamefont{Onari}},
  \bibinfo{author}{\bibfnamefont{Y.}~\bibnamefont{Tanaka}}, \bibnamefont{and}
  \bibinfo{author}{\bibfnamefont{J.}~\bibnamefont{Inoue}},
  \bibinfo{journal}{Phys. Rev. B} \textbf{\bibinfo{volume}{80}},
  \bibinfo{pages}{140509} (\bibinfo{year}{2009}).

\bibitem[{dxy()}]{dxy}
\bibinfo{note}{When the symmetry of the singlet component of pair potential is
  $d_{xy}$-wave ($d_{x^2-y^2}$-wave), the number of the sign change of the real
  or imaginary part of triplet one on the Fermi surface is two (six). Thus, we
  call the mixed pair potential $d_{xy}+p$-wave ($d_{x^{2}-y^{2}} + f$-wave).}

\bibitem[{\citenamefont{Schnyder and Ryu}()}]{SR10}
\bibinfo{author}{\bibfnamefont{A.~P.} \bibnamefont{Schnyder}} \bibnamefont{and}
  \bibinfo{author}{\bibfnamefont{S.}~\bibnamefont{Ryu}},
  \eprint{arXiv:1011.1438}.

\bibitem[{\citenamefont{Thouless et~al.}(1982)\citenamefont{Thouless, Kohmoto,
  Nightingale, and den Nijs}}]{TKNN82}
\bibinfo{author}{\bibfnamefont{D.~J.} \bibnamefont{Thouless}},
  \bibinfo{author}{\bibfnamefont{M.}~\bibnamefont{Kohmoto}},
  \bibinfo{author}{\bibfnamefont{M.~P.} \bibnamefont{Nightingale}},
  \bibnamefont{and} \bibinfo{author}{\bibfnamefont{M.}~\bibnamefont{den Nijs}},
  \bibinfo{journal}{Phys. Rev. Lett.} \textbf{\bibinfo{volume}{49}},
  \bibinfo{pages}{405} (\bibinfo{year}{1982}).

\bibitem[{\citenamefont{Kohmoto}(1985)}]{Kohmoto85}
\bibinfo{author}{\bibfnamefont{M.}~\bibnamefont{Kohmoto}},
  \bibinfo{journal}{Ann. Phys.} \textbf{\bibinfo{volume}{160}},
  \bibinfo{pages}{343} (\bibinfo{year}{1985}).

\bibitem[{\citenamefont{Hatsugai}(1993)}]{Hatsugai93}
\bibinfo{author}{\bibfnamefont{Y.}~\bibnamefont{Hatsugai}},
  \bibinfo{journal}{Phys. Rev. Lett.} \textbf{\bibinfo{volume}{71}},
  \bibinfo{pages}{3697} (\bibinfo{year}{1993}).

\bibitem[{\citenamefont{Fu and Kane}(2006)}]{FK06}
\bibinfo{author}{\bibfnamefont{L.}~\bibnamefont{Fu}} \bibnamefont{and}
  \bibinfo{author}{\bibfnamefont{C.~L.} \bibnamefont{Kane}},
  \bibinfo{journal}{Phys. Rev. B} \textbf{\bibinfo{volume}{74}},
  \bibinfo{pages}{195312} (\bibinfo{year}{2006}).

\bibitem[{\citenamefont{Qi et~al.}(2009)\citenamefont{Qi, Hughes, Raghu, and
  Zhang}}]{QHRZ09}
\bibinfo{author}{\bibfnamefont{X.~L.} \bibnamefont{Qi}},
  \bibinfo{author}{\bibfnamefont{T.~L.} \bibnamefont{Hughes}},
  \bibinfo{author}{\bibfnamefont{S.}~\bibnamefont{Raghu}}, \bibnamefont{and}
  \bibinfo{author}{\bibfnamefont{S.~C.} \bibnamefont{Zhang}},
  \bibinfo{journal}{Phys. Rev. Lett.} \textbf{\bibinfo{volume}{102}},
  \bibinfo{pages}{187001} (\bibinfo{year}{2009}).

\bibitem[{\citenamefont{Roy}()}]{Roy08}
\bibinfo{author}{\bibfnamefont{R.}~\bibnamefont{Roy}},
  \eprint{arXiv:0803.2881}.

\bibitem[{\citenamefont{Schnyder et~al.}(2008)\citenamefont{Schnyder, Ryu,
  Furusaki, and Ludwig}}]{SRFL08}
\bibinfo{author}{\bibfnamefont{A.~P.} \bibnamefont{Schnyder}},
  \bibinfo{author}{\bibfnamefont{S.}~\bibnamefont{Ryu}},
  \bibinfo{author}{\bibfnamefont{A.}~\bibnamefont{Furusaki}}, \bibnamefont{and}
  \bibinfo{author}{\bibfnamefont{A.~W.~W.} \bibnamefont{Ludwig}},
  \bibinfo{journal}{Phys. Rev. B} \textbf{\bibinfo{volume}{78}},
  \bibinfo{pages}{195125} (\bibinfo{year}{2008}).

\bibitem[{\citenamefont{Sato}(2009)}]{Sato09}
\bibinfo{author}{\bibfnamefont{M.}~\bibnamefont{Sato}}, \bibinfo{journal}{Phys.
  Rev. B} \textbf{\bibinfo{volume}{79}}, \bibinfo{pages}{214526}
  (\bibinfo{year}{2009}).

\bibitem[{\citenamefont{Fukui and Fujiwara}(2010)}]{FF10}
\bibinfo{author}{\bibfnamefont{T.}~\bibnamefont{Fukui}} \bibnamefont{and}
  \bibinfo{author}{\bibfnamefont{T.}~\bibnamefont{Fujiwara}},
  \bibinfo{journal}{J. Phys. Soc. Jpn.} \textbf{\bibinfo{volume}{79}},
  \bibinfo{pages}{033701} (\bibinfo{year}{2010}).

\bibitem[{\citenamefont{Fu and Berg}(2010)}]{FB10}
\bibinfo{author}{\bibfnamefont{L.}~\bibnamefont{Fu}} \bibnamefont{and}
  \bibinfo{author}{\bibfnamefont{E.}~\bibnamefont{Berg}},
  \bibinfo{journal}{Phys. Rev. Lett.} \textbf{\bibinfo{volume}{105}},
  \bibinfo{pages}{097001} (\bibinfo{year}{2010}).

\bibitem[{\citenamefont{Sato}(2010{\natexlab{b}})}]{Sato10b}
\bibinfo{author}{\bibfnamefont{M.}~\bibnamefont{Sato}},
  \bibinfo{journal}{Bussei Kenkyu} \textbf{\bibinfo{volume}{94}},
  \bibinfo{pages}{311} (\bibinfo{year}{2010}{\natexlab{b}}).

\bibitem[{\citenamefont{Berezinskii}(1974)}]{Berezinskii74}
\bibinfo{author}{\bibfnamefont{V.~L.} \bibnamefont{Berezinskii}},
  \bibinfo{journal}{JETP Lett.} \textbf{\bibinfo{volume}{20}},
  \bibinfo{pages}{287} (\bibinfo{year}{1974}).

\bibitem[{\citenamefont{Fujimoto}(2007{\natexlab{a}})}]{Fujimoto07a}
\bibinfo{author}{\bibfnamefont{S.}~\bibnamefont{Fujimoto}},
  \bibinfo{journal}{J. Phys. Soc. Jpn.} \textbf{\bibinfo{volume}{76}},
  \bibinfo{pages}{034712} (\bibinfo{year}{2007}{\natexlab{a}}).

\bibitem[{\citenamefont{Fujimoto}(2007{\natexlab{b}})}]{Fujimoto07b}
\bibinfo{author}{\bibfnamefont{S.}~\bibnamefont{Fujimoto}},
  \bibinfo{journal}{J. Phys. Soc. Jpn.} \textbf{\bibinfo{volume}{76}},
  \bibinfo{pages}{051008} (\bibinfo{year}{2007}{\natexlab{b}}).

\bibitem[{\citenamefont{Ryu and Hatsugai}(2002)}]{RH02}
\bibinfo{author}{\bibfnamefont{S.}~\bibnamefont{Ryu}} \bibnamefont{and}
  \bibinfo{author}{\bibfnamefont{Y.}~\bibnamefont{Hatsugai}},
  \bibinfo{journal}{Phys. Rev. Lett.} \textbf{\bibinfo{volume}{89}},
  \bibinfo{pages}{077002} (\bibinfo{year}{2002}).

\bibitem[{\citenamefont{Tanaka et~al.}(2008)\citenamefont{Tanaka, Asano, and
  Golubov}}]{TAG08}
\bibinfo{author}{\bibfnamefont{Y.}~\bibnamefont{Tanaka}},
  \bibinfo{author}{\bibfnamefont{Y.}~\bibnamefont{Asano}}, \bibnamefont{and}
  \bibinfo{author}{\bibfnamefont{A.~A.} \bibnamefont{Golubov}},
  \bibinfo{journal}{Phys. Rev. B} \textbf{\bibinfo{volume}{77}},
  \bibinfo{pages}{220504(R)} (\bibinfo{year}{2008}).

\bibitem[{\citenamefont{Yokoyama
  et~al.}(2008{\natexlab{d}})\citenamefont{Yokoyama, Tanaka, and
  Golubov}}]{Yokoyama08}
\bibinfo{author}{\bibfnamefont{T.}~\bibnamefont{Yokoyama}},
  \bibinfo{author}{\bibfnamefont{Y.}~\bibnamefont{Tanaka}}, \bibnamefont{and}
  \bibinfo{author}{\bibfnamefont{A.~A.} \bibnamefont{Golubov}},
  \bibinfo{journal}{Phys. Rev. B} \textbf{\bibinfo{volume}{78}},
  \bibinfo{pages}{012508} (\bibinfo{year}{2008}{\natexlab{d}}).

\bibitem[{\citenamefont{Tanuma et~al.}(2009)\citenamefont{Tanuma, Hayashi,
  Tanaka, and Golubov}}]{Tanuma09}
\bibinfo{author}{\bibfnamefont{Y.}~\bibnamefont{Tanuma}},
  \bibinfo{author}{\bibfnamefont{N.}~\bibnamefont{Hayashi}},
  \bibinfo{author}{\bibfnamefont{Y.}~\bibnamefont{Tanaka}}, \bibnamefont{and}
  \bibinfo{author}{\bibfnamefont{A.~A.} \bibnamefont{Golubov}},
  \bibinfo{journal}{Phys. Rev. Lett.} \textbf{\bibinfo{volume}{102}},
  \bibinfo{pages}{117003} (\bibinfo{year}{2009}).

\bibitem[{\citenamefont{Yokoyama et~al.}(2010)\citenamefont{Yokoyama, Ichioka,
  and Tanaka}}]{Yokoyama10}
\bibinfo{author}{\bibfnamefont{T.}~\bibnamefont{Yokoyama}},
  \bibinfo{author}{\bibfnamefont{M.}~\bibnamefont{Ichioka}}, \bibnamefont{and}
  \bibinfo{author}{\bibfnamefont{Y.}~\bibnamefont{Tanaka}},
  \bibinfo{journal}{J. Phys. Soc. Jpn.} \textbf{\bibinfo{volume}{79}},
  \bibinfo{pages}{034702} (\bibinfo{year}{2010}).

\bibitem[{\citenamefont{Serene and Rainer}(1983)}]{Serene}
\bibinfo{author}{\bibfnamefont{J.~W.} \bibnamefont{Serene}} \bibnamefont{and}
  \bibinfo{author}{\bibfnamefont{D.}~\bibnamefont{Rainer}},
  \bibinfo{journal}{Phys. Rep.} \textbf{\bibinfo{volume}{101}},
  \bibinfo{pages}{221} (\bibinfo{year}{1983}).

\bibitem[{\citenamefont{Rammer and Smith}(1986)}]{Rammer}
\bibinfo{author}{\bibfnamefont{J.}~\bibnamefont{Rammer}} \bibnamefont{and}
  \bibinfo{author}{\bibfnamefont{H.}~\bibnamefont{Smith}},
  \bibinfo{journal}{Rev. Mod. Phys.} \textbf{\bibinfo{volume}{58}},
  \bibinfo{pages}{323} (\bibinfo{year}{1986}).

\bibitem[{\citenamefont{Kopnin}(2001)}]{Kopnin}
\bibinfo{author}{\bibfnamefont{N.}~\bibnamefont{Kopnin}},
  \emph{\bibinfo{title}{Theory of Nonequilibrium Superconductivity}}
  (\bibinfo{publisher}{Oxford University Press}, \bibinfo{address}{New York},
  \bibinfo{year}{2001}).

\bibitem[{\citenamefont{Chandrasekhar}()}]{Chandrasekhar}
\bibinfo{author}{\bibfnamefont{V.}~\bibnamefont{Chandrasekhar}},
  \bibinfo{note}{in \textit{The Physics of Superconductors}, edited by K.-H.
  Bennemann and J. B. Ketterson (Springer-Verlag, Berlin, 2004), Vol II;
  arXiv:cond-mat/0312507.}

\bibitem[{\citenamefont{Eschrig}(2000)}]{Eschrig00}
\bibinfo{author}{\bibfnamefont{M.}~\bibnamefont{Eschrig}},
  \bibinfo{journal}{Phys. Rev. B} \textbf{\bibinfo{volume}{61}},
  \bibinfo{pages}{9061} (\bibinfo{year}{2000}).

\bibitem[{\citenamefont{Shelankov and Ozana}(2000)}]{Shelankov00}
\bibinfo{author}{\bibfnamefont{A.}~\bibnamefont{Shelankov}} \bibnamefont{and}
  \bibinfo{author}{\bibfnamefont{M.}~\bibnamefont{Ozana}},
  \bibinfo{journal}{Phys. Rev. B} \textbf{\bibinfo{volume}{61}},
  \bibinfo{pages}{7077} (\bibinfo{year}{2000}).

\bibitem[{\citenamefont{Schopohl and Maki}(1995)}]{Schopohl}
\bibinfo{author}{\bibfnamefont{N.}~\bibnamefont{Schopohl}} \bibnamefont{and}
  \bibinfo{author}{\bibfnamefont{K.}~\bibnamefont{Maki}},
  \bibinfo{journal}{Phys. Rev. B} \textbf{\bibinfo{volume}{52}},
  \bibinfo{pages}{490} (\bibinfo{year}{1995}).

\bibitem[{\citenamefont{Ashida et~al.}(1989)\citenamefont{Ashida, S.~Aoyama,
  and Nagai}}]{Ashida}
\bibinfo{author}{\bibfnamefont{M.}~\bibnamefont{Ashida}},
  \bibinfo{author}{\bibfnamefont{J.~H.} \bibnamefont{S.~Aoyama}},
  \bibnamefont{and} \bibinfo{author}{\bibfnamefont{K.}~\bibnamefont{Nagai}},
  \bibinfo{journal}{Phys. Rev. B} \textbf{\bibinfo{volume}{40}},
  \bibinfo{pages}{8673} (\bibinfo{year}{1989}).

\bibitem[{\citenamefont{Nagato et~al.}(1993)\citenamefont{Nagato, Nagai, and
  Hara}}]{Nagato93}
\bibinfo{author}{\bibfnamefont{Y.}~\bibnamefont{Nagato}},
  \bibinfo{author}{\bibfnamefont{K.}~\bibnamefont{Nagai}}, \bibnamefont{and}
  \bibinfo{author}{\bibfnamefont{J.}~\bibnamefont{Hara}}, \bibinfo{journal}{J.
  Low Temp. Phys.} \textbf{\bibinfo{volume}{33}}, \bibinfo{pages}{1993}
  (\bibinfo{year}{1993}).

\end{thebibliography}

\end{document}